\documentclass[12pt,a4paper]{article}
\usepackage{amsmath,amsthm,amsfonts,amssymb,bbm}
\usepackage{graphicx,psfrag,subfigure}
\usepackage{cite}
\usepackage{hyperref}

\numberwithin{equation}{section}
\numberwithin{figure}{section}

\newcommand{\Or}{\mathcal{O}}
\newcommand{\Ai}{\mathrm{Ai}}
\newcommand{\Pb}{\mathbbm{P}}
\newcommand{\E}{\mathbbm{E}}
\newcommand{\Id}{\mathbbm{1}}
\newcommand{\e}{\varepsilon}
\newcommand{\I}{{\rm i}}
\newcommand{\C}{\mathbb{C}}
\newcommand{\R}{\mathbb{R}}
\newcommand{\N}{\mathbb{N}}
\newcommand{\Z}{\mathbb{Z}}
\newcommand{\dx}{\mathrm{d}}

\newcommand{\cte}{\mathrm{const\,}}
\renewcommand{\S}{{\cal S}}
\newcommand{\tht}{\thetag}
\renewcommand{\mid}{\,|\,}
\newcommand{\X}{\frak X}
\renewcommand{\P}{{\mathcal P}}
\renewcommand{\H}{\mathbb{H}}
\newcommand{\W}{\includegraphics[height=0.8em]{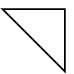}\,}
\newcommand{\B}{\includegraphics[height=0.8em]{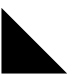}\,}
\newcommand{\TypeI}{\includegraphics[width=0.8em]{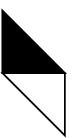}\,}
\newcommand{\TypeIbis}{\includegraphics[width=0.8em]{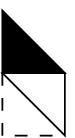}\,}
\newcommand{\TypeIter}{\includegraphics[width=1.6em]{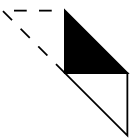}\,}
\newcommand{\TypeII}{\includegraphics[width=1.6em]{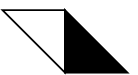}\,}
\newcommand{\TypeIIbis}{\includegraphics[width=1.6em]{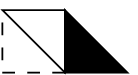}\,}
\newcommand{\TypeIII}{\includegraphics[width=0.8em]{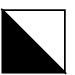}\,}
\newcommand{\TypeIIIter}{\includegraphics[width=1.6em]{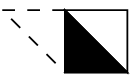}\,}

\DeclareMathOperator*{\virt}{virt}
\DeclareMathOperator*{\sgn}{sgn}
\DeclareMathOperator*{\Var}{Var}

\DeclareMathOperator*{\GFF}{GFF}

\renewcommand{\Re}{\mathrm{Re}}
\renewcommand{\Im}{\mathrm{Im}}

\newtheorem{prop}{Proposition}[section]
\newtheorem{thm}[prop]{Theorem}
\newtheorem{lem}[prop]{Lemma}
\newtheorem{defin}[prop]{Definition}
\newtheorem{cor}[prop]{Corollary}
\newtheorem{conj}[prop]{Conjecture}
\newtheorem{cla}[prop]{Claim}
\newenvironment{claim}{\begin{cla}\normalfont}{\end{cla}}
\newtheorem{rem}[prop]{Remark}
\newenvironment{remark}{\begin{rem}\normalfont}{\end{rem}}

\newenvironment{proofOF}[2]{\removelastskip\vspace{6pt}\noindent {\it Proof of #1.}~\rm#2}{\qed \par\vspace{6pt}}

\title{Anisotropic growth of random\\ surfaces in $2+1$ dimensions}
\author{Alexei Borodin\thanks{California Institute of Technology and Institute for Information Transmission Problems, Moscow, e-mail: \texttt{borodin@caltech.edu}},
Patrik L. Ferrari\thanks{Weierstrass Institute of Applied Analysis and Stochastics (WIAS), Berlin. Now at the Institute for Applied Mathematics, Bonn University. \newline
e-mail: \texttt{ferrari@wiener.iam.uni-bonn.de}}}

\date{November 4, 2008}

\begin{document}
\maketitle \sloppy

\begin{abstract}
We construct a family of stochastic growth models in $2+1$
dimensions, that belong to the anisotropic KPZ class. Appropriate
projections of these models yield $1+1$ dimensional growth models in
the KPZ class and random tiling models. We show that correlation
functions associated to our models have determinantal structure, and
we study large time asymptotics for one of the models.

The main asymptotic results are: (1) The growing surface has a limit
shape that consists of facets interpolated by a curved piece. (2)
The one-point fluctuations of the height function in the curved part
are asymptotically normal with variance of order $\ln(t)$ for time
$t\gg 1$. (3) There is a map of the $(2+1)$-dimensional space-time
to the upper half-plane $\H$ such that on space-like submanifolds
the multi-point fluctuations of the height function are
asymptotically equal to those of the pullback of the Gaussian free
(massless) field on $\H$.
\end{abstract}

\tableofcontents

\section{Introduction}\label{sectIntroduction}
In recent years there has been a lot of progress in understanding
large time fluctuations of driven interacting particle systems on
the one-dimensional lattice, see
e.g.~\cite{BKS85,BS06,BJ08,QV06,Jo00b,SI07,NS04,PS01,FS05a,Sas05,PS02b,BFP06,BFPS06,BFS07,BF07,BFS07b}.
Evolution of such systems is commonly interpreted as random growth
of a one-dimensional interface, and if one views the time as an
extra variable, the evolution produces a random surface (see e.g.\
Figure 4.5 in~\cite{Pra03} for a nice illustration). In a different
direction, substantial progress have also been achieved in studying
the asymptotics of random surfaces arising from dimers on planar
bipartite graphs, see the review~\cite{KenLectures} and references
therein.

Although random surfaces of these two kinds were shown to share
certain asymptotic properties (also common to random matrix models),
no direct connection between them was known. One goal of this paper
is to establish such a connection.

We construct a class of two-dimensional random growth models (that
is, the principal object is a randomly growing surface, embedded in the
four-dimensional space-time). In two different projections these
models yield random surfaces of the two kinds mentioned above (one
reduces the spatial dimension by one, the second projection
is fixing time). We partially compute the correlation functions of
an associated (three-dimensional) random point process and show that
they have determinantal form that is typical for determinantal point
processes.

For one specific growth model we compute the correlation kernel
explicitly, and use it to establish Gaussian fluctuations of the
growing random surface. We then determine the covariance structure.

Let us describe our results in more detail.

\subsection{A two-dimensional growth model}
\begin{figure}
\begin{center}
\psfrag{x}[b]{$x$}
\psfrag{n}[b]{$n$}
\psfrag{h}[b]{$h$}
\psfrag{x11}[cl]{$x_1^1$}
\psfrag{x12}[cl]{$x_1^2$}
\psfrag{x13}[cl]{$x_1^3$}
\psfrag{x14}[cl]{$x_1^4$}
\psfrag{x15}[cl]{$x_1^5$}
\psfrag{x22}[cl]{$x_2^2$}
\psfrag{x23}[cl]{$x_2^3$}
\psfrag{x24}[cl]{$x_2^4$}
\psfrag{x25}[cl]{$x_2^5$}
\psfrag{x33}[cl]{$x_3^3$}
\psfrag{x34}[cl]{$x_3^4$}
\psfrag{x35}[cl]{$x_3^5$}
\psfrag{x44}[cl]{$x_4^4$}
\psfrag{x45}[cl]{$x_4^5$}
\psfrag{x55}[cl]{$x_5^5$}
\includegraphics[height=5cm]{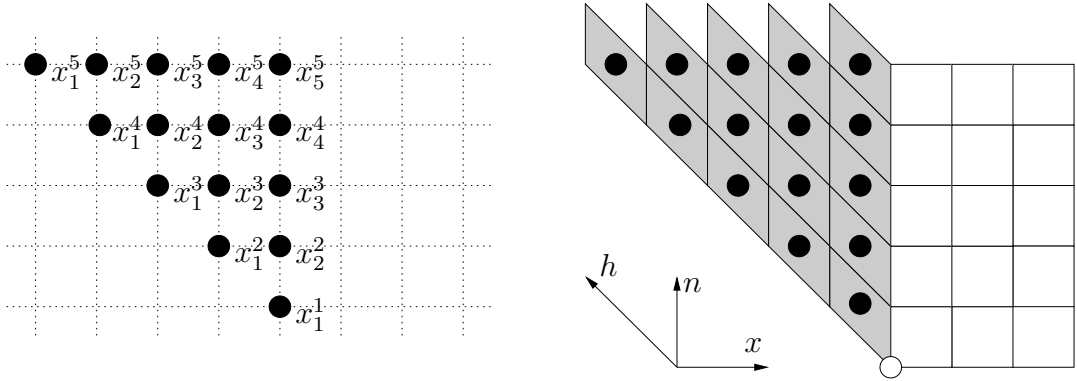}
\caption{Illustration of the initial conditions for the particles system and the corresponding lozenge tilings. In the height function picture, the white circle has coordinates $(x,n,h)=(-1/2,0,0)$.}
\label{FigureInitialConditions}
\end{center}
\end{figure}

Consider a continuous time Markov chain on the state space of interlacing variables
\begin{equation}
{\cal S}^{(n)}=\Big\{\{x_k^m\}_{\begin{subarray}{ll} k=1,\dots,m\\m=1,\dots,n\end{subarray}}
\subset \Z^{\frac{n(n+1)}2}\mid x^m_{k-1}<x_{k-1}^{m-1}\leq x_k^m\Big\},\quad n=1,2,\ldots.
\end{equation}
$x_k^m$ can be interpreted as the position of particle with label $(k,m)$, but we will also refer to a given particle as $x_k^m$. As initial condition, we consider the fully-packed one, namely at time moment $t=0$ we have $x_k^m(0)=k-m-1$ for all $k,m$, see Figure~\ref{FigureInitialConditions}.

The particles evolve according to the following dynamics. Each of the particles $x_k^m$ has an independent exponential clock of rate one, and when the $x_k^m$-clock rings the particle attempts to jump to the right by one. If at that moment $x_k^m=x_k^{m-1}-1$ then the jump is blocked. If that is not the case, we find the largest $c\geq 1$ such that
$x_k^m=x_{k+1}^{m+1}=\dots=x_{k+c-1}^{m+c-1}$, and all $c$ particles in this string jump to the right by one. For any $t\geq 0$ denote by ${\cal M}^{(n)}(t)$ the resulting measure on ${\cal S}^{(n)}$ at time moment $t$.

Informally speaking, the particles with smaller upper indices are heavier than those with larger upper indices, so that the heavier particles block and push the lighter ones in order for the interlacing conditions to be preserved. This anisotropy is essential, see more details in Section~\ref{sc:AKPZ}.

\begin{figure}
\begin{center}
\psfrag{x}[b]{$x$}
\psfrag{n}[b]{$n$}
\psfrag{h}[b]{$h$}
\psfrag{x11}[cl]{$x_1^1$}
\psfrag{x12}[cl]{$x_1^2$}
\psfrag{x13}[cl]{$x_1^3$}
\psfrag{x14}[cl]{$x_1^4$}
\psfrag{x15}[cl]{$x_1^5$}
\psfrag{x22}[cl]{$x_2^2$}
\psfrag{x23}[cl]{$x_2^3$}
\psfrag{x24}[cl]{$x_2^4$}
\psfrag{x25}[cl]{$x_2^5$}
\psfrag{x33}[cl]{$x_3^3$}
\psfrag{x34}[cl]{$x_3^4$}
\psfrag{x35}[cl]{$x_3^5$}
\psfrag{x44}[cl]{$x_4^4$}
\psfrag{x45}[cl]{$x_4^5$}
\psfrag{x55}[cl]{$x_5^5$}
\psfrag{tasep}[c]{\textbf{TASEP}}
\psfrag{pushasep}[c]{\textbf{PushASEP}}
\psfrag{charlier}[c]{\textbf{Charlier process}}
\includegraphics[height=5cm]{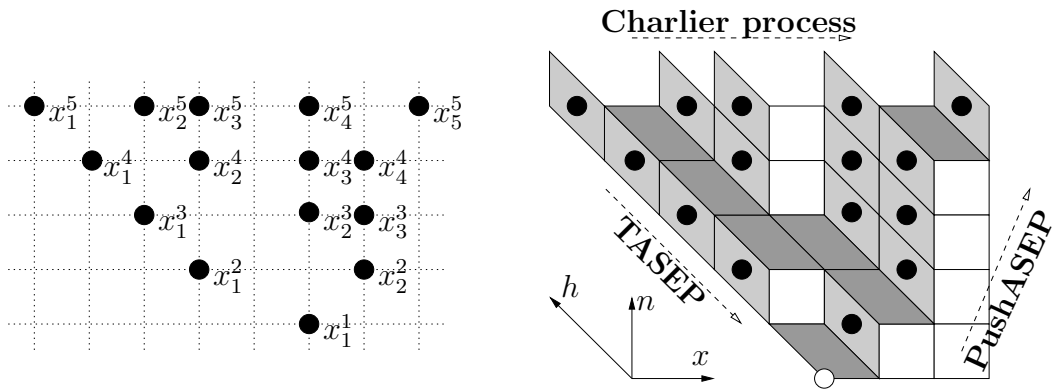}
\caption{From particle configurations (left) to 3d visualization via lozenge tilings (right). The corner with the white circle has coordinates $(x,n,h)=(-1/2,0,0)$.}
\label{FigureIntro}
\end{center}
\end{figure}

Let us illustrate the dynamics using Figure~\ref{FigureIntro},
which shows a possible configuration of particles obtained from our
initial condition. If in this state of the system the $x_1^3$-clock
rings, then particle $x_1^3$ does not move, because it is blocked by
particle $x_1^2$. If it is the $x_2^2$-clock that rings, then
particle $x_2^2$ moves to the right by one unit, but to keep the
interlacing property satisfied, also particles $x_3^3$ and $x_4^4$
move by one unit at the same time. This aspect of the dynamics is
called ``pushing''.

Observe that ${\cal S}^{(n_1)}\subset {\cal S}^{(n_2)}$ for $n_1\leq n_2$, and the definition of the evolution implies that ${\cal M}^{(n_1)}(t)$ is a marginal of ${\cal M}^{(n_2)}(t)$ for any $t\geq 0$.
Thus, we can think of ${\cal M}^{(n)}$'s as marginals of the measure
${\cal M}=\lim\limits_{\longleftarrow}{\cal M}^{(n)}$ on
\mbox{${\cal S}=\lim\limits_{\longleftarrow}{\cal S}^{(n)}$}. In other words,
${\cal M}(t)$ are measures on the space ${\cal S}$ of infinite point
configurations ${\{x_k^m\}}_{k=1,\dots,m,\,m\geq 1}$.

Before stating the main results, it is interesting to notice that the Markov chain has different interpretations. Also, some projections of the Markov chain to subsets of ${\cal S}^{(n)}$ are still Markov chains.
\begin{enumerate}
\item The evolution of $x_1^1$ is the one-dimensional Poisson process of rate one.
\item The row $\{x_1^m\}_{m\geq 1}$ evolves as a Markov chain on $\Z$ known as the {\it Totally Asymmetric Simple Exclusion Process} (TASEP), and the initial condition $x_1^m(0)=-m$ is commonly referred to as {\it step initial condition}. In this case, particle $x_1^k$ jumps to its right with unit rate, provided the arrival site is empty (exclusion constraint).
\item The row $\{x^m_m\}_{m\geq 1}$ also evolves as a Markov chain on $\Z$ that is sometimes
called ``long range TASEP''; it was also called PushASEP
in~\cite{BF07}. It is convenient to view $\{x_m^m+m\}_{m\ge 1}$ as
particle locations in $\Z$. Then, when the $x_k^k$-clock rings, the
particle $x_k^k+k$ jumps to its right and pushes by one unit the
(maybe empty) block of particles sitting next to it. If one
disregards the particle labeling, one can think of particles as
independently jumping to the next free site on their right with unit
rate.
\item For our initial condition, the evolution of each row $\{x_k^m\}_{k=1,\dots,m}$, \mbox{$m=1,2,\dots$}, is also a Markov chain. It was called {\it Charlier process} in~\cite{KOR02} because of its relation to the classical orthogonal Charlier polynomials. It can be defined as Doob $h$-transform for $m$ independent rate one Poisson processes with the harmonic function $h$ equal to the Vandermonde determinant.
\item Infinite point configurations
$\{x_k^m\}\in {\cal S}$ can be viewed as {\it Gelfand-Tsetlin schemes}. Then ${\cal M}(t)$ is the ``Fourier transform'' of a suitable irreducible character of the infinite-dimensional unitary group $U(\infty)$, see~\cite{BK07}. Interestingly enough, increasing $t$ corresponds to a deterministic flow on the space of irreducible characters of $U(\infty)$.
\item Elements of $\cal S$ can also be viewed as lozenge tiling of a sector in the plane. To see that one surrounds each particle location by a rhombus of one type and draws edges through locations where there are no particles, see Figure~\ref{FigureIntro}. Our initial condition corresponds to a perfectly regular tiling, see Figure~\ref{FigureInitialConditions}.
\item The random tiling defined by ${\cal M}(t)$ is the limit of the
uniformly distributed lozenge tilings of hexagons with side lengths
$(a,b,c)$, when $a,b,c\to\infty$ so that $ab/c\to t$, and we observe
the hexagon tiling at finite distances from the corner between sides
of lengths $a$ and $b$.
\item Finally, Figure~\ref{FigureIntro} has a clear three-dimensional connotation. Given
the random configuration $\{x_k^n(t)\}\in {\cal S}$ at time moment $t$, define the random
{\it height function}
\begin{equation}\label{eqDefinHeight}
\begin{aligned}
&h:(\Z+\tfrac 12)\times \Z_{>0}\times \R_{\geq 0}\to \Z_{\geq 0},\\
&h(x,n,t)=\#\{k\in\{1,\dots,n\}\mid x_k^n(t)> x\}.
\end{aligned}
\end{equation}
In terms of the tiling on Figure~\ref{FigureIntro}, the height
function is defined at the vertices of rhombi, and it counts the
number of particles to the right from a given vertex. (This definition
differs by a simple linear function of $(x,n)$ from the standard
definition of the height function for lozenge tilings, see e.g.
\cite{Ken04,KenLectures}.) The initial condition corresponds to
starting with perfectly flat facets.
\end{enumerate}

Thus, our Markov chain can be viewed as a random growth model of the
surface given by the height function. In terms of the step surface
of Figure~\ref{FigureIntro}, the evolution consists of removing all
columns of $(x,n,h)$-dimensions $(1,*,1)$ that could be removed,
independently with exponential waiting times of rate one. For
example, if $x_2^2$ jumps to its right, then three consecutive cubes
(associated to $x_2^2,x_3^3,x_4^4$) are removed. Clearly, in this
dynamics the directions $x$ and $n$ do not play symmetric roles.
Indeed, this model belongs to the $2+1$ anisotropic KPZ class of
stochastic growth models, see Section~\ref{sc:AKPZ}.

\subsection{Determinantal formula, limit shape and one-point fluctuations}
The first result about the Markov chain ${\cal M}(t)$ that we prove is
the (partial) determinantal structure of the correlation functions. Introduce the notation
\begin{equation}\label{eqPartialOrder}
(n_1,t_1)\prec (n_2,t_2) \quad\text{iff} \quad n_1\leq n_2,
t_1\geq t_2,\text{ and }(n_1,t_1)\neq (n_2,t_2).
\end{equation}

\begin{figure}[t!]
\begin{center}
\includegraphics[angle=90,width=14cm]{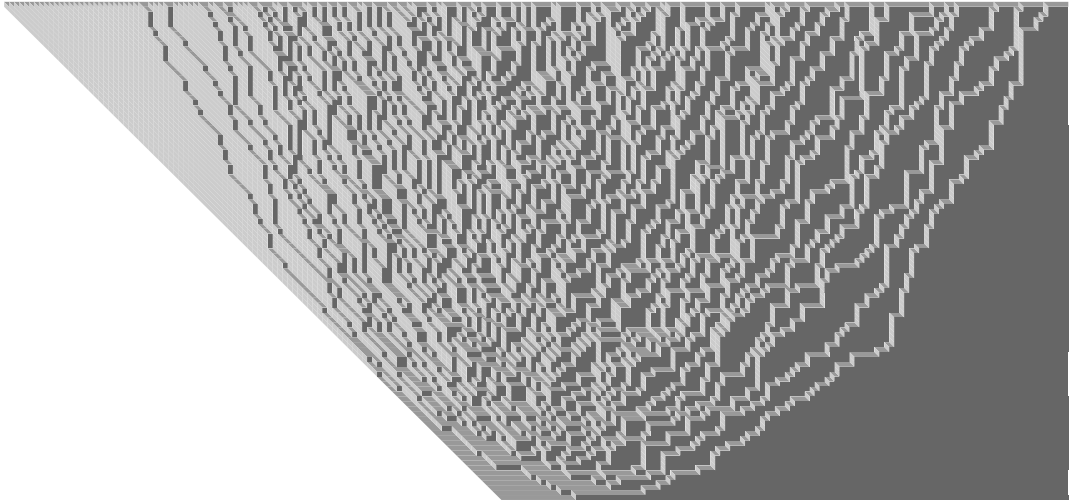}
\caption{A configuration of the model analyzed with $N=100$ particles at time $t=25$, using the same representation as in Figure~\ref{FigureIntro}. In~\cite{FerAKPZ} there is a Java animation of the model.}
\label{FigureSimulazione}
\end{center}
\end{figure}

\begin{thm}\label{ThmDetStructure}
For any $N=1,2,\dots$, pick $N$ triples
\begin{equation*}
\varkappa_j=(x_j,n_j,t_j)\in \Z\times\Z_{>0}\times \R_{\geq 0}
\end{equation*}
such that
\begin{equation}
t_1\leq t_2\leq \dots\leq t_N,\qquad n_1\geq n_2\geq \dots\geq n_N.
\end{equation}
Then
\begin{multline}
\Pb\{\textrm{For each }j=1,\dots,N \textrm{ there exists a } k_j, \\
 1\leq k_j\leq n_j\textrm{ such that }x^{n_j}_{k_j}(t_j)=x_j\}=\det{[{\cal K}(\varkappa_i,\varkappa_j)]}_{i,j=1}^N,
\end{multline}
where
\begin{multline}\label{StartingKernel}
{\cal K}(x_1,n_1,t_1;x_2,n_2,t_2) =-\frac{1}{2\pi\I}\oint_{\Gamma_0}
\frac{\dx w}{w^{x_2-x_1+1}}\,\frac{e^{(t_1-t_2)/w}}
{(1-w)^{n_2-n_1}}\,\Id_{[(n_1,t_1)\prec (n_2,t_2)]} \\
+\frac{1}{(2\pi i)^2}\oint_{\Gamma_0}\dx w\oint_{\Gamma_1}\dx z
\frac{e^{t_1/w}}{e^{t_2/z}} \frac{(1-w)^{n_1}}{(1-z)^{n_2}}
\frac{w^{x_1}}{z^{x_2+1}}\,\frac{1}{w-z},
\end{multline}
the contours $\Gamma_0$, $\Gamma_{1}$ are simple positively oriented
closed paths that include the poles $0$ and $1$, respectively, and
no other poles (hence, they are disjoint).
\end{thm}
This result is proved at the end of Section~\ref{sect8DetStruct}.
The above kernel has in fact already appeared in~\cite{BF07} in
connection with PushASEP. The determinantal structure makes it
possible to study the asymptotics. On a macroscopic scale (large
time limit and hydrodynamic scaling) the model has a limit shape,
which we now describe, see Figure~\ref{FigureSimulazione}. Since we
look at heights at different times, we cannot use time as a large
parameter. Instead, we introduce a large parameter $L$ and consider
space and time coordinates that are comparable to $L$. The limit
shape consists of three facets interpolated by a curved piece. To
describe it, consider the set
\begin{equation}
{\cal D}=\{(\nu,\eta,\tau)\in \R^3_{>0}\mid
(\sqrt{\eta}-\sqrt{\tau})^2<\nu<(\sqrt{\eta}+\sqrt{\tau})^2\}.
\end{equation}
It is exactly the set of triples $(\nu,\eta,\tau)\in \R^3_{>0}$ for
which there exists a nondegenerate triangle with side lengths
$(\sqrt{\nu},\sqrt{\eta},\sqrt{\tau})$. Denote by
$(\pi_\nu,\pi_\eta,\pi_\tau)$ the angles of this triangle that are
opposite to the corresponding sides (see Figure~\ref{FigGeometry} too).

Our second result concerns the limit shape and the Gaussian fluctuations in the curved region, living on a $\sqrt{\ln{L}}$ scale.
\begin{thm}\label{thmLogfluct}
For any $(\nu,\eta,\tau)\in {\cal D}$ we have the
moment convergence of random variables
\begin{equation}
\lim_{L\to\infty}\frac{h([\bigl(\nu-\eta)L]+\tfrac12,[\eta L],\tau L\bigr)-
\E h([\bigl(\nu-\eta)L]+\tfrac 12,[\eta L],\tau L\bigr)}{\sqrt{\kappa\ln{L}}}=\xi\sim {\cal N}(0,1),
\end{equation}
with $\kappa=(2\pi^2)^{-1}$.
\end{thm}

We also give an explicit formula for the limit shape:
\begin{multline}\label{eq1.9}
\lim_{L\to\infty}\frac{\E h([\bigl(\nu-\eta)L]+\tfrac 12,[\eta L],\tau L\bigr)}{L}=:\textit{\textbf{h}}(\nu,\eta,\tau) \\
=\frac{1}{\pi}\left(-\nu\pi_\eta+\eta(\pi-\pi_\nu)+\tau\,\frac{\sin\pi_\nu \sin\pi_\eta}{\sin\pi_\tau}\right).
\end{multline}

Theorem~\ref{thmLogfluct} describes the limit shape $\textit{\textbf{h}}$ of our
growing surface, and the domain $\cal D$ describes the points where
this limit shape is {\it curved}. The logarithmic fluctuations is
essentially a consequence of the local asymptotic behavior being
governed by the discrete sine kernel (this local behavior occurs
also in tiling models~\cite{Jo02b,FS03,OR01}). Using the connection
with the Charlier ensembles, see above, the formula (\ref{eq1.9})
for the limit shape can be read off the formulas of~\cite{Bia00}.

Using Theorem~\ref{ThmDetStructure} it is not hard to verify (see
Proposition \ref{propBulkLimit} below) that near every point of the
limit shape in the curved region, at any fixed time moment the
random lozenge tiling approaches the unique translation invariant
measure $M_{\pi_\nu,\pi_\eta,\pi_\tau}$ on lozenge tilings of the
plane with prescribed slope (see~\cite{KOS03,KenLectures,BS08} and
references therein for discussions of these measures). The slope is
exactly the slope of the tangent plane to the limit shape, given by
\begin{equation}\label{eqIntro1}
\frac{\partial \textit{\textbf{h}}}{\partial \nu}=-\frac{\pi_\eta}\pi,\qquad
\frac{\partial \textit{\textbf{h}}}{\partial \eta}=1-\frac{\pi_\nu}\pi.
\end{equation}
This implies in particular, that
$(\pi_\nu/\pi,\pi_\eta/\pi,\pi_\tau/\pi)$ are the asymptotic
proportions of lozenges of three different types in the neighborhood
of the point of the limit shape. One also computes the growth
velocity (see (\ref{Omega}) for the definition of $\Omega$)
\begin{equation}\label{eqIntro2}
\frac{\partial \textit{\textbf{h}}}{\partial \tau} =\frac{1}{\pi}\frac{\sin\pi_\nu\sin\pi_\eta}{\sin\pi_\tau}=\frac{\Im(\Omega(\nu,\eta,\tau))}{\pi}.
\end{equation}
Since the right-hand side depends only on the slope of the tangent
plane, this suggest that it should be possible to extend the
definition of our surface evolution to the random surfaces
distributed according to measures $M_{\pi_\nu,\pi_\eta,\pi_\tau}$;
these measures have to remain invariant under evolution, and the
speed of the height growth should be given by the right-hand side of (\ref{eqIntro2}).
This is an interesting open problem that we do not address in this paper.

\subsection{Complex structure and multipoint fluctuations}
To describe the correlations of the interface, we first need to introduce a complex structure. Set $\H=\{z\in\C\mid \Im(z)> 0\}$ and define the map \mbox{$\Omega:{\cal D}\to \H$} by
\begin{equation}\label{Omega}
|\Omega(\nu,\eta,\tau)|=\sqrt{\eta/\tau},\quad
|1-\Omega(\nu,\eta,\tau)|=\sqrt{\nu/\tau}.
\end{equation}
Observe that $\arg\Omega=\pi_\nu$ and $\arg(1-\Omega)=-\pi_\eta$.
The preimage of any $\Omega\in\H$ is a ray in $\cal D$ that consists
of triples $(\nu,\eta,\tau)$ with constant ratios $(\nu:\eta:\tau)$.
Denote this ray by $R_\Omega$. One sees that $R_\Omega$'s are also
the level sets of the slope of the tangent plane to the limit shape.
Since $\textit{\textbf{h}}(\alpha \nu, \alpha\eta,\alpha\tau)=\alpha \textit{\textbf{h}}(\nu,\eta,\tau)$ for any $\alpha>0$, the height function grows
linearly in time along each $R_\Omega$.
Note also that the map $\Omega$ satisfies
\begin{equation}
(1-\Omega)\frac{\partial \Omega}{\partial\nu}=\Omega\frac{\partial
\Omega}{\partial\eta}=-\frac{\partial \Omega}{\partial\tau},
\end{equation}
and the first of these relations is the complex Burgers equation,
cf.~\cite{KO07}.

From Theorem~\ref{thmLogfluct} one might think that to get non-trivial correlations we need to consider $(h-\E(h))/\sqrt{\ln{L}}$. However, this is not true and the division by $\sqrt{\ln{L}}$ is not needed. To state the precise result, denote by
\begin{equation}\label{eqGreen}
{\cal G}(z,w)=-\frac 1{2\pi}\ln\left|\frac{z-w}{z-\bar w}\right|
\end{equation}
the Green function of the Laplace operator on $\H$ with Dirichlet boundary conditions.

\begin{thm}\label{TheoremCorr}
For any $N=1,2,\dots$, let $\varkappa_j=(\nu_j,\eta_j,\tau_j)\in
{\cal D}$ be any distinct $N$ triples such that
\begin{equation}\label{eqIntro3}
\tau_1\leq \tau_2\leq \dots\leq \tau_N,\qquad \eta_1\geq\eta_2\geq\dots\geq\eta_N.
\end{equation}
Denote
\begin{equation}\label{HL}
H_L(\nu,\eta,\tau):=\sqrt{\pi}\left(h([(\nu-\eta)L]+\tfrac 12,[\eta
L],\tau L)-\E h([(\nu-\eta)L]+\tfrac 12,[\eta L],\tau L)\right),
\end{equation}
and $\Omega_j=\Omega(\nu_j,\eta_j,\tau_j)$. Then
\begin{equation}\label{eq1.17}
\lim_{L\to\infty}\E\left(H_L(\varkappa_1)\cdots H_L(\varkappa_N)\right)
=\left\{
\begin{array}{ll}
\sum\limits_{\sigma\in {\cal F}_N}\prod\limits_{j=1}^{N/2} {\cal G}(\Omega_{\sigma(2j-1)},\Omega_{\sigma(2j)}),& N\text{ is even},\\[1em]
 0,&N\text{ is odd},
\end{array}
\right.
\end{equation}
where the summation is taken over all fixed point free involutions $\sigma$ on $\{1,\dots,N\}$.
\end{thm}
The result of the theorem means that as $L\to\infty$, $H_L(\Omega^{-1}(z))$ is a Gaussian process with covariance given by ${\cal G}$, i.e., it has correlation of the Gaussian Free Field on $\H$. We can make this statement more precise. Indeed, in addition to Theorem~\ref{TheoremCorr}, a simple consequence of Theorem~\ref{thmLogfluct} gives (see Lemma~\ref{LemmaShortDistance}),
\begin{equation}\label{eqShortDistance}
\E\left(H_L(\varkappa_1)\cdots H_L(\varkappa_N)\right)=O(L^\epsilon),\quad L\to\infty,
\end{equation}
for any $\varkappa_j\in{\cal D}$ and any $\epsilon>0$. This bounds the moments of $H_L(\varkappa_j)$ for infinitesimally close points $\varkappa_j$. A small extension of Theorem~\ref{TheoremCorr} together with this estimate immediately implies that on suitable surfaces in $\cal D$, the random function $H_L(\nu,\eta,\tau)$ converges to the $\Omega$-pullback of the Gaussian free field on $\H$, see Theorem~\ref{ThmGFF2d} and Theorem~\ref{ThmGFF1d} in Section~\ref{SectGFF} for more details.

\begin{conj}\label{Conjecture}
The statement of Theorem~\ref{TheoremCorr} holds without the
assumption (\ref{eqIntro3}), provided that $\Omega$-images of all
the triples are pairwise distinct.
\end{conj}

Theorem~\ref{TheoremCorr} and Conjecture~\ref{Conjecture} indicate that the fluctuations of the
height function along the rays $R_\Omega$ vary slower than in any
other space-time direction. This statement can be rephrased more
generally: the height function has smaller fluctuations along the
curves where the slope of the limit shape remains constant. We have
been able to find evidence for such a claim in one-dimensional
random growth models as well~\cite{Fer08}.

\subsection{Universality class}\label{sc:AKPZ}
In the terminology of physics literature, see e.g.~\cite{BS95}, our Markov
chain falls into the class of local growth models with relaxation
and lateral growth, described by the Kardar-Parisi-Zhang (KPZ) equation
\begin{equation}\label{eqIntro4}
{\partial_t h}=\Delta h +Q(\partial_x h,\partial_y h)+\text{white noise},
\end{equation}
where $Q$ is a quadratic form. Relations (\ref{eqIntro1}) and
(\ref{eqIntro2}) imply that for our growth model the determinant of
the Hessian of $\partial_th$, viewed as a function of the slope, is
strictly negative, which means that the form $Q$ in our case has
signature $(-1,1)$. In such a situation the equation
(\ref{eqIntro4}) is called {\it anisotropic} KPZ or AKPZ equation.

An example of such system is growth of vicinal surfaces, which are
naturally anisotropic because the tilt direction of the surface is
special. Using non-rigorous renormalization group analysis based on
one-loop expansion, Wolf~\cite{Wol91} predicted that large time
fluctuations (the roughness) of the growth models described by AKPZ equation should
be similar to those of linear models described by the
Edwards-Wilkinson equation (heat equation with random term)
\begin{equation}
{\partial_t h}=\Delta h +\text{white noise}.
\end{equation}

Our results can be viewed as the first rigorous analysis of a
non-equilibrium growth model in the AKPZ class. (Some results, like logarithmic fluctuations,
for an AKPZ model in a steady state were obtained in \cite{PS97}. Some numerical numerical results are described in~\cite{JKK93,KKK98,HHA92}). Indeed, Wolf's prediction correctly identifies the logarithmic
behavior of height fluctuations. However, it does not (at least explicitly) predict the appearance of the Gaussian
free field, and in particular the complete structure (map $\Omega$) of the fluctuations described in the previous
section.

On the other hand, universality considerations imply that analogs of
Theorems~\ref{thmLogfluct} and~\ref{TheoremCorr}, as well as possibly Conjecture~\ref{Conjecture}, should hold in
any AKPZ growth model.

\subsection{More general growth models}
It turns out that the determinantal structure of the correlations
functions stated in Theorem~\ref{ThmDetStructure} holds for a much more general class of
two-dimensional growth models. In the first part of the paper we
develop an algebraic formalism needed to show that. At least three
examples where this formalism applies, other than the Markov chain
considered above, are worth mentioning.

\begin{enumerate}
\item In the Markov chain considered above one can make the particle
jump rates depend on the upper index $m$ in an arbitrary way. One
can also allow the particles jump both right and left, with ratio of
left and right jump rates possibly changing in time~\cite{BF07}.
\item The shuffling algorithm for domino tilings of Aztec diamonds
introduced in~\cite{EKLP92} also fits into our formalism. The
corresponding discrete time Markov chain is described in Section~\ref{Sect2dDynamics}
below, and its equivalence to domino shuffling is established in the
recent paper~\cite{Nor08}.
\item A shuffling algorithm for lozenge tilings of the hexagon (also known as \emph{boxed plane partitions}) has
been constructed in~\cite{BG08} using the formalism developed in
this paper, see~\cite{BG08} for details.
\end{enumerate}

Our original Markov chain is a suitable degeneration of each of
these examples.

We expect our asymptotic methods to be applicable to many other
two-dimensional growth models produced by the general formalism,
and we plan to return to this discussion in a later publication.

\subsection{Other connections}\label{sc:other}
We have so far discussed the global asymptotic behavior of our growing surface, and its bulk
properties (measures $M_{\pi_\nu,\pi_\eta,\pi_\tau}$), but have not
discussed the edge asymptotics. As was mentioned above, rows
$\{x_1^m\}_{m\geq 1}$ and $\{x_m^m\}_{m\geq 1}$ can be viewed as
one-dimensional growth models on their own, and their asymptotic
behavior was studied in~\cite{BF07} using essentially the same Theorem~\ref{ThmDetStructure}. This is exactly the edge behavior of our two-dimensional growth model.

Of course, the successive projections to $\{x_1^m\}_{m\geq 1}$ and
then to a fixed (large) time commute. In the first ordering, this
can be seen as the large time interface associated to the TASEP. In
the second ordering, it corresponds to considering a tiling problem
of a large region and focusing on the border of the facet.

Interestingly enough, an analog of Theorem~\ref{ThmDetStructure} remains useful for the
edge computations even in the cases when the measure on the space
$\cal S$ is no longer positive (but its projections to
$\{x_1^m\}_{m\geq 1}$ and $\{x_m^m\}_{m\geq 1}$ remain positive).
These computations lead to the asymptotic results of
\cite{Sas05,BFPS06,BFP06,BF07,BFS07,BFS07b} for one-dimensional growth models with more general
types of initial conditions.

Another natural asymptotic question that was not discussed is the limiting behavior of ${\cal M}^{(n)}(t)$ when $t\to\infty$ but $n$ remains fixed. After proper normalization, in the limit one obtains the Markov chain investigated in \cite{War07}.

Two of the four one-dimensional growth models constructed in~\cite{DW07} (namely, ``Bernoulli with blocking'' and
``Bernoulli with pushing'') are projections to $\{x_1^m\}_{m\geq 1}$ and $\{x_m^m\}_{m\geq 1}$ of one of our two-dimensional growth models, see Section~\ref{Sect2dDynamics} below. It remains unclear however, how to interpret the other two models of~\cite{DW07} in a similar fashion.

Finally, let us mention that our proof of
Theorem~\ref{ThmDetStructure} is based on the argument
of~\cite{CL95} and~\cite{Sos04}, the proof of
Theorem~\ref{TheoremCorr} uses several ideas from~\cite{Ken04}, and
the algebraic formalism for two-dimensional growth models employs a
crucial idea of constructing bivariate Markov chains out of
commuting univariate ones from \cite{DF90}.

\vspace{1em} \textbf{Outline.} The rest of the paper is organized as
follows. It has essentially two main parts. The first part is
Section~\ref{Sect2dDynamics}. It contains the construction of the
Markov chains, with the final result being the determinantal
structure and the associated kernel (Theorem~\ref{ThmAB18}). Its
continuous time analogue is Corollary~\ref{corAB19}, whose further
specialization to particle-independent jump rate leads to
Theorem~\ref{ThmDetStructure}. The second main part concerns the
limit results for the continuous time model that we analyze. We
start by collecting various geometric identities in
Section~\ref{SectGeometry}. We also shortly discuss why our model is
in the AKPZ class. In Section~\ref{sectGaussFluct} we first give a shifted
version of the kernel, whose asymptotic analysis is the content of
Section~\ref{SectAsymptAnalysis}. These results then allow us to
prove Theorem~\ref{thmLogfluct} in Section~\ref{sectGaussFluct} and
Theorem~\ref{TheoremCorr} in Section~\ref{SectCorrelations}.

\vspace{1em} \textbf{Acknowledgments.} The authors are very grateful
to P.~Diaconis, E.~Rains, and H.~Spohn for numerous illuminating
discussions. The first named author (A. B.) was partially supported
by the NSF grant DMS-0707163.

\section{Two dimensional dynamics}\label{Sect2dDynamics}

All the constructions below are based on the following
basic idea. Consider two Markov operators $P$ and $P^*$ on state
spaces $\S$ and $\S^*$, and a Markov link $\Lambda:\S^*\to \S$ that
intertwines $P$ and $P^*$, that is $\Lambda P=P^* \Lambda$. Then one
can construct Markov chains on (subsets of) $\S^*\times \S$ that in
some sense has both $P$ and $P^*$ as their projections. There is
more than one way to realize this idea, and in this paper we discuss
two variants.

In one of them the image $(y^*,y)$ of $(x^*,x)\in \S^*\times \S$
under the Markov operator is determined by {\it sequential
update\/}: One first chooses $y$ according to $P(x,y)$, and then one
chooses $y^*$ so that the needed projection properties are
satisfied. A characteristic feature of the construction is that $x$
and $y^*$ are independent, given $x^*$ and $y$. This bivariate
Markov chain is denoted $P_\Lambda$; its construction is borrowed
from \cite{DF90}.

In the second variant, the images $y^*$ and $y$ are independent,
given $(x,x^*)$, and we say that they are obtained by {\it parallel
update\/}. The distribution of $y$ is still $P(x,y)$, independently
of what $x^*$ is. This Markov chain is denoted $P_\Delta$ for the
operator $\Delta=\Lambda P=P^*\Lambda$ that plays an important role.

By induction, one constructs multivariate Markov chains out of
finitely many univariate ones and links that intertwine them. Again,
we use two variants of the construction --- with sequential and
parallel updates.

The key property that makes these constructions useful is the
following: If the chains $P$, $P^*$, and $\Lambda$, are $h$-Doob
transforms of some (simpler) Markov chains, and the harmonic
functions $h$ used are consistent, then the transition probabilities
of the multivariate Markov chains do not depend on $h$. Thus,
participating multivariate Markov chains may be fairly complex, while
the transition probabilities of the univariate Markov chains remain
simple.

Below we first explain the abstract construction of $P_\Lambda$,
$P_\Delta$, and their multivariate extensions. Then we exhibit a
class of examples that are of interest to us. Finally, we show how
the knowledge of certain averages (correlation functions) for the
univariate Markov chains allows one to compute similar averages for
the multivariate chains.

\subsection{Bivariate Markov chains}\label{sect1MarkovChains}

Let $\S$ and $\S^*$ be discrete sets, and let $P$ and $P^*$ be
stochastic matrices on these sets:
\begin{equation}
\sum_{y\in \S}P(x,y)=1,\quad x\in \S;\qquad \sum_{y^*\in \S^*}P^*(x^*,y^*)=1, \quad x^*\in \S^*.
\end{equation}

Assume that there exists a third stochastic matrix
\mbox{$\Lambda={\Vert\Lambda(x^*,x)\Vert}_{x^*\in \S^*,\, x\in \S}$} such
that for any $x^*\in\S^*$ and $y\in\S$
\begin{equation}\label{eqAB1}
\sum_{x\in \S}\Lambda(x^*,x)
P(x,y)=\sum_{y^*\in\S^*}P^*(x^*,y^*)\Lambda(y^*,y).
\end{equation}
Let us denote the above quantity by $\Delta(x^*,y)$. In matrix notation
\begin{equation}\label{1.3}
\Delta=\Lambda P=P^*\Lambda.
\end{equation}
Set
\begin{eqnarray*} \S_\Lambda=\{(x^*,x)\in\S^*\times\S\mid
\Lambda(x^*,x)>0\}, \\
\S_\Delta=\{(x^*,x)\in\S^*\times\S\mid \Delta(x^*,x)>0\}.
\end{eqnarray*}
Define bivariate Markov chains on $\S_\Lambda$ and $S_\Delta$ by
their corresponding transition probabilities
\begin{equation}
P_\Lambda((x^*,x),(y^*,y))=
\begin{cases}
\dfrac{P(x,y)P^*(x^*,y^*)\Lambda(y^*,y)}{\Delta(x^*,y)}\,,&\Delta(x^*,y)>0,
\\
0,&\text{otherwise},
\end{cases}
\end{equation}
\begin{equation}P_{\Delta}((x^*,x),(y^*,y))= \frac{P(x,y)P^*(x^*,y^*)\Lambda(y^*,x)}
{\Delta(x^*,x)}\,.
\end{equation}
It is immediately verified that both matrices $P_\Lambda$ and
$P_\Delta$ are stochastic.

The chain $P_\Lambda$ was introduced by Diaconis-Fill in \cite{DF90},
and we are using the notation of that paper.

One could think of $P_\Lambda$ and $P_\Delta$ as follows.

For $P_\Lambda$, starting from $(x^*,x)$ we first choose $y$
according to the transition matrix $P(x,y)$, and then choose $y^*$
using $\frac{P^*(x^*,y^*)\Lambda(y^*,y)}{\Delta(x^*,y)}$, which is
the conditional distribution of the middle point in the successive
application of $P^*$ and $\Lambda$ provided that we start at $x^*$
and finish at $y$.

For $P_\Delta$, starting from $(x^*,x)$ we independently choose $y$
according to $P(x,y)$ and $y^*$ according to
$\frac{P^*(x^*,y^*)\Lambda(y^*,x)} {\Delta(x^*,x)}$, which is the
conditional distribution of the middle point in the successive
application of $P^*$ and $\Lambda$ provided that we start at $x^*$
and finish at $x$.

\begin{lem}\label{LemmaAB1}
For any $(x^*,x)\in \S_\Lambda,\ y\in \S$, we have
\begin{equation}
\begin{aligned} \sum_{y^*\in\S^*:(y^*,y)\in \S_\Lambda}
P_\Lambda((x^*,x),(y^*,y))&=P(x,y), \\ \sum_{y^*\in S^*:(y^*,y)\in
\S_\Delta} P_\Delta((x^*,x),(y^*,y))&=P(x,y),
\end{aligned}
\end{equation}
and for any $x^*\in \S^*,\ (y^*,y)\in \S_\Lambda$,
\begin{equation}
\begin{aligned}
\sum_{x\in\S:(x^*,x)\in\S_\Lambda}\Lambda(x^*,x)P_\Lambda((x^*,x),(y^*,y))
&=P^*(x^*,y^*)\Lambda(y^*,y),\\
\sum_{x\in\S:(x^*,x)\in\S_\Delta}\Delta(x^*,x)P_\Delta((x^*,x),(y^*,y)) &=P^*(x^*,y^*)\Delta(y^*,y).
\end{aligned}
\end{equation}
\end{lem}
\begin{proofOF}{Lemma~\ref{LemmaAB1}}
Straightforward computation using the relation
\mbox{$\Delta=\Lambda P=P^*\Lambda$}.
\end{proofOF}

\begin{prop}\label{PropAB2}
Let $m^*(x^*)$ be a probability measure on
$\S^*$. Consider the evolution of the measure $m(x^*)\Lambda(x^*,x)$
on $\S_\Lambda$ under the Markov chain $P_\Lambda$ and denote by
$(x^*(j),x(j))$ the result after $j=0,1,2,\dots$ steps. Then for any
$k,l= 0,1,\dots$ the joint distribution of
\begin{equation}\label{1.6}
(x^*(0),x^*(1),\dots,x^*(k),x(k),x(k+1),\dots,x(k+l))
\end{equation}
coincides with the stochastic evolution of $m^*$ under transition
matrices
\begin{equation}(\underbrace{P^*,\dots,P^*}_k, \Lambda, \underbrace{P,\dots,P}_l).
\end{equation}

Exactly the same statement holds for the Markov chain $P_\Delta$ and
the initial condition $m^*(x^*)\Delta(x^*,x)$ with $\Lambda$
replaced by $\Delta$ in the above sequence of matrices.
\end{prop}
\begin{proofOF}{Proposition~\ref{PropAB2}}
Successive application of the first relations of Lemma~\ref{LemmaAB1} to evaluate the sums over $x^*(k+l),\dots,x^*(k+1)$, and of the
second relations to evaluate the sums over $x(1),\dots,x(k-1)$.
\end{proofOF}

Note that Proposition~\ref{PropAB2} also implies that the joint distribution of
$x^*(k)$ and $x(k)$ has the form
$m^*_k(x^*(k))\Lambda(x^*(k),x(k))$, where $m^*_k$ is the result of
$k$-fold application of $P^*$ to $m^*$.

The above constructions can be generalized to the nonautonomous
situation.

Assume that we have a time variable $t\in \Z$, and our state spaces
as well as transition matrices depend on $t$, which we will indicate
as follows:
\begin{equation}\S(t), \quad \S^*(t),\quad P(x,y\mid t),\quad P^*(x^*,y^*\mid t),
\quad \Lambda(x^*,x\mid t),\quad P(t),\quad P^*(t),\quad
\Lambda(t).
\end{equation}
The commutation relation \tht{1.3} is replaced by
$\Lambda(t)P(t)=P^*(t)\Lambda(t+1)$ or
\begin{equation}\label{eqAB1prime}
\Delta(x^*,y\mid t):=\sum_{x\in \S(t)}\Lambda(x^*,x\mid t) P(x,y\mid
t)=\sum_{y^*\in\S^*(t+1)}P^*(x^*,y^*\mid t)\,\Lambda(y^*,y\mid t+1).
\end{equation}

Further, we set
\begin{equation}
\begin{aligned}
\S_\Lambda(t)&=\{(x^*,x)\in\S^*(t)\times\S(t)\mid
\Lambda(x^*,x\mid t)>0\}, \\
\S_\Delta(t)&=\{(x^*,x)\in\S^*(t)\times\S(t+1)\mid \Delta(x^*,x\mid
t)>0\},
\end{aligned}
\end{equation}
and
\begin{equation}P_\Lambda((x^*,x),(y^*,y)\mid t)=\begin{cases} \dfrac{P(x,y\mid
t)P^*(x^*,y^*\mid t)\Lambda(y^*,y\mid t+1)}{\Delta(x^*,y\mid
t)}\,,&\Delta(x^*,y\mid t)>0,
\\
0,&\text{otherwise},
\end{cases}
\end{equation}
\begin{equation}P_{\Delta}((x^*,x),(y^*,y)\mid t)= \frac{P(x,y\mid
t+1)P^*(x^*,y^*\mid t)\Lambda(y^*,x\mid t+1)} {\Delta(x^*,x\mid
t)}\,.
\end{equation}

The nonautonomous generalization of Proposition~\ref{PropAB2} is proved in
exactly the same way as Proposition~\ref{PropAB2}. Let us state it.

\begin{prop}\label{PropAB2prime}
Fix $t_0\in\Z$, and let $m^*(x^*)$ be a
probability measure on $\S^*(t_0)$. Consider the evolution of the
measure $m(x^*)\Lambda(x^*,x\mid t_0)$ on $\S_\Lambda(t_0)$ under
the Mar\-kov chain $P_\Lambda(t)$, and denote by
$(x^*(t_0+j),x(t_0+j))\in \S_\Lambda(t_0+j)$ the result after
$j=0,1,2,\dots$ steps. Then for any $k,l= 0,1,\dots$ the joint
distribution of
\begin{equation}(x^*(t_0),x^*(t_0+1),\dots,x^*(t_0+k),x(t_0+k),x(t_0+k+1),\dots,x(t_0+k+l))
\end{equation}
coincides with the stochastic evolution of $m^*$ under transition
matrices
\begin{equation}{P^*(t_0),\dots,P^*(t_0+k-1)}, \Lambda(t_0+k),
P(t_0+k),\dots,P(t_0+k+l-1)
\end{equation}
(for $k=l=0$ only $\Lambda(t_0)$ remains in this string).

A similar statement holds for the Markov chain $P_\Delta(t)$ and the
initial condition $m^*(x^*)\Delta(x^*,x\mid t_0)$: For any $k,l=
0,1,\dots$ the joint distribution of
\begin{equation}\label{1.14}
(x^*(t_0),x^*(t_0+1),\dots,x^*(t_0+k),x(t_0+k+1),x(t_0+k+2),\dots,x(t_0+k+l+1))
\end{equation}
coincides with the stochastic evolution of $m^*$ under transition
matrices
\begin{equation}{P^*(t_0),\dots,P^*(t_0+k-1)}, \Delta(t_0+k),
P(t_0+k+1),\dots,P(t_0+k+l).
\end{equation}
\end{prop}

\begin{remark} Observe that there is a difference in the sequences of
times used in (\ref{1.6}) and (\ref{1.14}). The reason is that for
nonautonomous $P_\Delta$, the state space at time $t$ is a subset of
$\S^*(t)\times \S(t+1)$, and we denote its elements as
$(x^*(t),x(t+1))$. In the autonomous case, an element of the state
space $\S_\Delta$ at time $t$ was denoted as $(x^*(t),x(t))$.
\end{remark}

\subsection{Multivariate Markov chains}\label{sect2multiMC}

We now aim at generalizing the constructions of Section~\ref{sect1MarkovChains} to more
than two state spaces.

Let $\S_1,\dots,\S_n$ be discrete sets, $P_1,\dots,P_n$ be
stochastic matrices defining Markov chains on them, and let
$\Lambda_1^2,\dots,\Lambda_{n-1}^n$ be stochastic links between
these sets:
\begin{equation}
\begin{aligned}
P_k:\S_k\times\S_k\to [0,1],&\quad
\sum_{y\in\S_k}P_k(x,y)=1,\quad x\in \S_k,\quad k=1,\dots,n;\\
\Lambda_{k-1}^k:\S_k\times\S_{k-1}\to [0,1],&\quad
\sum_{y\in\S_{k-1}}\Lambda_{k-1}^k(x,y)=1,\quad x\in \S_k,\quad
k=2,\dots,n.
\end{aligned}
\end{equation}

Assume that these matrices satisfy the commutation relations
\begin{equation}\Delta^k_{k-1}:=\Lambda^k_{k-1}P_{k-1}=P_k\Lambda^k_{k-1},\qquad k=2,\dots,n.
\end{equation}

The state spaces for our multivariate Markov chains are defined as
follows
\begin{equation}
\begin{aligned}
\S^{(n)}_\Lambda&=\Bigl\{(x_1,\dots,x_n)\in\S_1\times\cdots\times\S_n\mid
\prod_{k=2}^n\Lambda_{k-1}^k(x_k,x_{k-1})\ne 0\Bigr\},\\
\S^{(n)}_\Delta&=\Bigl\{(x_1,\dots,x_n)\in\S_1\times\cdots\times\S_n\mid
\prod_{k=2}^n\Delta_{k-1}^k(x_k,x_{k-1})\ne 0\Bigr\}.
\end{aligned}
\end{equation}
The transition probabilities for the Markov chains $P^{(n)}_\Lambda$
and $P^{(n)}_\Delta$ are defined as (we use the notation
$X_n=(x_1,\dots,x_n)$, $Y_n=(y_1,\dots,y_n)$)
\begin{equation}\label{1.17}P_\Lambda^{(n)}(X_n,Y_n)=\begin{cases}
P_1(x_1,y_1)\prod\limits_{k=2}^n\dfrac{P_k(x_k,y_k)\Lambda_{k-1}^k(y_k,y_{k-1})}
{\Delta^k_{k-1}(x_k,y_{k-1})}\,,&\prod\limits_{k=2}^n\Delta^k_{k-1}(x_k,y_{k-1})>0,
\\
0,&\text{otherwise},
\end{cases}
\end{equation}
\begin{equation}\label{1.18}P_{\Delta}^{(n)}(X_n,Y_n)=P(x_1,y_1)
\prod_{k=2}^n\frac{P_k(x_k,y_k)\Lambda^k_{k-1}(y_k,x_{k-1})}
{\Delta^k_{k-1}(x_k,x_{k-1})}\,.
\end{equation}

One way to think of $P_\Lambda^{(n)}$ and $P_\Delta^{(n)}$ is as
follows. For $P_\Lambda^{(n)}$, starting from $X_n=(x_1,\dots,x_n)$, we first
choose $y_1$ according to the transition matrix $P(x_1,y_1)$, then
choose $y_2$ using
$\frac{P_2(x_2,y_2)\Lambda_{1}^2(y_2,y_1)}{\Delta^2_{1}(x_2,y_1)}$,
which is the conditional distribution of the middle point in the
successive application of $P_2$ and $\Lambda^2_1$ provided that we
start at $x_2$ and finish at $y_1$, after that we choose $y_3$ using
the conditional distribution of the middle point in the successive
application of $P_3$ and $\Lambda^3_2$ provided that we start at
$x_3$ and finish at $y_2$, and so on. Thus, one could say that $Y_n$
is obtained by the {\it sequential update\/}.

For $P_\Delta^{(n)}$, starting from $X_n=(x_1,\dots,x_n)$ we {\it
independently\/} choose $y_1,\dots,y_n$ according to $P_1(x_1,y_1)$
for $y_1$ and $\frac{P_k(x_k,y_k)\Lambda^k_{k-1}(y_k,x_{k-1})}
{\Delta^k_{k-1}(x_k,x_{k-1})}$, for $y_k$, $k=2,\dots,n$. The latter
formula is the conditional distribution of the middle point in the
successive application of $P_k$ and $\Lambda^k_{k-1}$ provided that
we start at $x_{k}$ and finish at $x_{k-1}$. Thus, it is natural to
say that this Markov chains corresponds to the {\it parallel
update\/}.

We aim at proving the following generalization of Proposition~\ref{PropAB2}.

\begin{prop}\label{PropAB3} Let $m_n(x_n)$ be a probability measure on
$\S_n$. Consider the evolution of the measure
\begin{equation}m_n(x_n)\Lambda^n_{n-1}(x_n,x_{n-1})\cdots \Lambda^2_1(x_2,x_1)
\end{equation}
on $\S_\Lambda^{(n)}$ under the Markov chain $P_\Lambda^{(n)}$, and
denote by $(x_1(j),\dots,x_n(j))$ the result after $j=0,1,2,\dots$
steps. Then for any $ k_1\geq k_{2}\geq\dots\geq k_n\geq 0$ the joint
distribution of
\begin{multline*}
(x_n(0),\dots,x_n(k_n),x_{n-1}(k_n),
x_{n-1}(k_n+1),\dots,x_{n-1}(k_{n-1}),\\x_{n-2}(k_{n-1}),\dots,
x_2(k_{2}),x_1(k_2),\dots,x_1(k_1))
\end{multline*}
coincides with the stochastic evolution of $m_n$ under transition
matrices
\begin{equation}
(\underbrace{P_n,\dots,P_n}_{k_n},
\Lambda^n_{n-1},\underbrace{P_{n-1},\dots,P_{n-1}}_{k_{n-1}-k_n},
\Lambda^{n-1}_{n-2},\dots,
\Lambda^2_1,\underbrace{P_1,\dots,P_1}_{k_1-k_2}).
\end{equation}

Exactly the same statement holds for the Markov chain
$P_\Delta^{(n)}$ and the initial condition
\begin{equation}
m(x_n)\Delta^n_{n-1}(x_n,x_{n-1})\cdots \Delta^2_1(x_2,x_1)
\end{equation}
with $\Lambda$'s replaced by $\Delta$'s in the above sequence of
matrices.
\end{prop}
The following lemma is useful.

\begin{lem}\label{LemmaAB4}
Consider the matrix $\Lambda:\S_n\times \S_\Lambda^{(n-1)}\to [0,1]$
given by
\begin{equation}\Lambda(x_n,(x_1,\dots,x_{n-1})):=\Lambda^n_{n-1}(x_n,x_{n-1})\cdots
\Lambda^2_1(x_2,x_1).
\end{equation}
Then $\Lambda P_\Lambda^{(n-1)}=P_n\Lambda$. If we denote this
matrix by $\Delta$ then
\begin{equation}
P_\Lambda^{(n)}(X_n,Y_n)= \begin{cases}
\dfrac{P_\Lambda^{(n-1)}(X_{n-1},Y_{n-1})
P_n(x_n,y_n)\Lambda(y_n,Y_{n-1})}{\Delta(x_n,Y_{n-1})}\,,
&\Delta(x_n,Y_{n-1})>0,
\\
0,&\text{otherwise}.
\end{cases}
\end{equation}
Also, using the same notation,
\begin{equation}P_{\Delta}^{(n)}(X_n,Y_n)=
\frac{P_\Delta^{(n-1)}(X_{n-1},Y_{n-1})P_n(x_n,y_n)
\Lambda(y_n,X_{n-1})} {\Delta(x_n,X_{n-1})}\,.
\end{equation}
\end{lem}

\begin{proofOF}{Lemma~\ref{LemmaAB4}}
Let us check the commutation relation $\Lambda
P_\Lambda^{(n-1)}=P_n\Lambda$. We have
\begin{multline}
\Lambda
P_\Lambda^{(n-1)}(x_n,Y_{n-1})=\sum_{x_1,\dots,x_{n-1}}\Lambda^n_{n-1}(x_n,x_{n-1})\cdots
\Lambda^2_1(x_2,x_1)\\ \times
P_1(x_1,y_1)\prod\limits_{k=2}^{n-1}\dfrac{P_k(x_k,y_k)\Lambda_{k-1}^k(y_k,y_{k-1})}
{\Delta^k_{k-1}(x_k,y_{k-1})},
\end{multline}
where the sum is taken over all $x_1,\dots,x_{n-1}$ such that
\mbox{$\prod_{k=2}^{n-1}\Delta^k_{k-1}(x_k,y_{k-1})>0$}. Computing the sum
over $x_1$ and using the relation $\Lambda^2_1P_1=\Delta^2_1$ we
obtain
\begin{multline}
 \Lambda P_\Lambda^{(n-1)}(x_n,Y_{n-1})=
\sum_{x_2,\dots,x_{n-1}}\Lambda^n_{n-1}(x_n,x_{n-1})\cdots
\Lambda^3_2(x_3,x_2)\\
\times P_2(x_2,y_2)\Lambda^2_1(y_2,y_1)\prod\limits_{k=3}^{n-1}\dfrac{P_k(x_k,y_k)\Lambda_{k-1}^k(y_k,y_{k-1})}
{\Delta^k_{k-1}(x_k,y_{k-1})}\,.
\end{multline}
Now we need to compute the sum over $x_2$. If
$\Delta^2_1(x_2,y_1)=0$ then \mbox{$P_2(x_2,y_2)=0$} because otherwise the
relation $\Delta^2_1=P_2\Lambda^2_1$ implies that
$\Lambda^2_1(y_2,y_1)=0$, which contradicts the hypothesis that
$Y_{n-1}\in \S_\Lambda^{(n-1)}$. Thus, we can extend the sum to all
$x_2\in \S_2$, and the relation $\Lambda^3_2P_2=\Delta^3_2$ gives
\begin{multline}
 \Lambda
P_\Lambda^{(n-1)}(x_n,Y_{n-1})
=\sum_{x_3,\dots,x_{n-1}}\Lambda^n_{n-1}(x_n,x_{n-1})\cdots
\Lambda^4_3(x_4,x_3)\\
\times
P_3(x_3,y_3)\Lambda^3_2(y_3,y_2)\Lambda^2_1(y_2,y_1)\prod\limits_{k=4}^{n-1}\dfrac{P_k(x_k,y_k)\Lambda_{k-1}^k(y_k,y_{k-1})}
{\Delta^k_{k-1}(x_k,y_{k-1})}\,.
\end{multline}
Continuing like that we end up with
\begin{equation}\Lambda^{n-1}_{n-2}(y_{n-1},y_{n-2}) \cdots
\Lambda^2_1(y_2,y_1)\sum_{x_{n-1}}\Lambda^n_{n-1}(x_n,x_{n-1})
P_{n-1}(x_{n-1},y_{n-1}),
\end{equation}
which, by $\Lambda^n_{n-1}P_{n-1}=P_n\Lambda^n_{n-1}$ is exactly
$P_n\Lambda(x_n,Y_{n-1})$. Let us also note that
\begin{equation}\Delta(x_n,Y_{n-1})=\Delta^n_{n-1}(x_n,y_{n-1})\Lambda^{n-1}_{n-2}(y_{n-1},y_{n-2})
\cdots \Lambda^2_1(y_2,y_1).
\end{equation}
The needed formulas for $P_\Lambda^{(n)}$ and $P_\Delta^{(n)}$ are
now verified by straightforward substitution.
\end{proofOF}

\begin{proofOF}{Proposition~\ref{PropAB3}}
Let us give the argument for
$P_\Lambda^{(n)}$; for $P_\Delta^{(n)}$ the proof is literally the
same. By virtue of Lemma~\ref{LemmaAB4}, we can apply Proposition~\ref{PropAB2} by taking
\begin{equation}
\S^*=\S_n, \quad \S=\S_\Lambda^{(n-1)},\quad P^*=P_n,\quad
P=P_\Lambda^{(n-1)},\quad k=k_n, \quad l=k_1-k_n,
\end{equation}
and $\Lambda(x_n,X_{n-1})$ as in Lemma~\ref{LemmaAB4}. Proposition~\ref{PropAB2} says that
the joint distribution
\begin{equation}(x_n(0),x_n(1),\dots,x_n(k_n),X_{n-1}(k_n),X_{n-1}(k_n+1),
\dots,X_{n-1}(k_1))
\end{equation}
is the evolution of $m_n$ under
\begin{equation}(\underbrace{P_n,\dots,P_n}_{k_n}, \Lambda,
\underbrace{P_\Lambda^{(n-1)},\dots,P_\Lambda^{(n-1)}}_{k_1-k_n}).
\end{equation}
Induction on $n$ completes the proof.
\end{proofOF}

As in the previous section, Proposition~\ref{PropAB3} can be also proved in the
nonautonomous situation. Let us give the necessary definitions.

We now have a time variable $t\in \Z$, and our state spaces as well
as transition matrices depend on $t$:
\begin{equation}
\S_k(t), \quad P_k(x,y\mid t),\quad k=1,\dots,n, \qquad
\Lambda^k_{k-1}(x_k,x_{k-1}\mid t),\quad k=2,\dots,n.
\end{equation}
The commutation relations are
\begin{equation}
\Delta^k_{k-1}(t):=\Lambda^k_{k-1}(t)P_{k-1}(t)=P_k(t)\Lambda^k_{k-1}(t+1),\qquad
k=2,\dots,n.
\end{equation}

The multivariate state spaces are defined as
\begin{eqnarray*}
\S^{(n)}_\Lambda&=&\Bigl\{(x_1,\dots,x_n)\in\S_1(t)\times\cdots\times\S_n(t)\mid
\prod_{k=2}^n\Lambda_{k-1}^k(x_k,x_{k-1}\mid t)\ne 0\Bigr\},\\
\S^{(n)}_\Delta&=&\Bigl\{(x_1,\dots,x_n)
\in\S_1(t+n-1)\times\cdots\times\S_n(t)\mid
\prod_{k=2}^n\Delta_{k-1}^k(x_k,x_{k-1}\mid t+n-k)\ne 0\Bigr\}.
\end{eqnarray*}

Then the transition matrices for $P^{(n)}_\Lambda$ and
$P^{(n)}_\Delta$ are defined as
\begin{equation}
P_\Lambda^{(n)}(X_n,Y_n\mid t)=P_1(x_1,y_1\mid t)\prod\limits_{k=2}^n\dfrac{P_k(x_k,y_k\mid t)\Lambda_{k-1}^k(y_k,y_{k-1}\mid t+1)}{\Delta^k_{k-1}(x_k,y_{k-1}\mid t)}
\end{equation}
if $\prod_{k=2}^n\Delta^k_{k-1}(x_k,y_{k-1}\mid t)>0$ and $0$
otherwise; and
\begin{multline}P_{\Delta}^{(n)}(X_n,Y_n)=P(x_1,y_1\mid t+n-1) \\
\times\prod_{k=2}^n\frac{P_k(x_k,y_k\mid
t+n-k)\Lambda^k_{k-1}(y_k,x_{k-1}\mid t+n-k+1)}
{\Delta^k_{k-1}(x_k,x_{k-1}\mid t+n-k)}.
\end{multline}

\begin{prop}\label{PropAB3prime}
Fix $t_0\in\Z$, and let $m_n(x_n)$ be a
probability measure on $\S_n(t_0)$. Consider the evolution of the
measure
\begin{equation}m_n(x_n)\Lambda^n_{n-1}(x_n,x_{n-1}\mid t_0)\cdots
\Lambda^2_1(x_2,x_1\mid t_0)
\end{equation}
on $\S_\Lambda^{(n)}(t_0)$ under $P_\Lambda^{(n)}(t)$. Denote by
$(x_1(t_0+j),\dots,x_n(t_0+j))$ the result after $j=0,1,2,\dots$
steps. Then for any $ k_1\geq k_{2}\geq\dots\geq k_n\geq t_0$ the
joint distribution of
\begin{multline*} (x_n(t_0),\dots,x_n(k_n),x_{n-1}(k_n),
x_{n-1}(k_n+1),\dots,x_{n-1}(k_{n-1}),\\
x_{n-2}(k_{n-1}),\dots, x_2(k_{2}),x_1(k_2),\dots,x_1(k_1))
\end{multline*}
coincides with the stochastic evolution of $m_n$ under transition
matrices
\begin{multline*} P_n(t_0),\dots,P_n(k_n-1),
\Lambda^n_{n-1}(k_n),P_{n-1}(k_n),\dots,P_{n-1}(k_{n-1}-1),\\
\Lambda^{n-1}_{n-2}(k_{n-1}),\dots,
\Lambda^2_1(k_2),P_1(k_2),\dots,P_1(k_1-1).
\end{multline*}

A similar statement holds for the Markov chain $P_\Delta^{(n)}(t)$
and the initial condition
\begin{equation}\label{icDelta}
m(x_n)\Delta^{n}_{n-1}(x_n,x_{n-1}\mid t_0)\cdots
\Delta^2_1(x_2,x_1\mid t_0+n-2).
\end{equation}
For any $ k_1> k_{2}>\dots> k_n\geq t_0$ the joint distribution of
\begin{multline*} (x_n(t_0),\dots,x_n(k_n),x_{n-1}(k_n+1),
x_{n-1}(k_n+2),\dots,x_{n-1}(k_{n-1}),\\
x_{n-2}(k_{n-1}+1),\dots, x_2(k_{2}),x_1(k_2+1),\dots,x_1(k_1))
\end{multline*}
coincides with the stochastic evolution of $m_n$ under transition
matrices
\begin{multline*} P_n(t_0),\dots,P_n(k_n-1),
\Delta^n_{n-1}(k_n),P_{n-1}(k_n+1),\dots,P_{n-1}(k_{n-1}-1),
\\ \Delta^{n-1}_{n-2}(k_{n-1}),\dots,
\Delta^2_1(k_2),P_1(k_2+1),\dots,P_1(k_1-1).
\end{multline*}
\end{prop}

The proof is very similar to that of Proposition~\ref{PropAB3}.

\subsection{Toeplitz-like transition probabilities}\label{sect3Toeplitz}

The goal of this section is to provide some general recipe on how to
construct commuting stochastic matrices.

\begin{prop}\label{PropAB4}
Let $\alpha_1,\dots,\alpha_n$ be nonzero
complex numbers, and let $F(x)$ be an analytic function in an
annulus $A$ centered at the origin that contains all
$\alpha_j^{-1}$'s. Assume that $F(\alpha_1^{-1})\cdots
F(\alpha_n^{-1})\ne 0$. Then
\begin{equation}\frac 1{F(\alpha_1^{-1})\cdots F(\alpha_n^{-1})}\sum_{
y_1<\dots<y_n\in\Z}
\det{[\alpha_i^{y_j}]}_{i,j=1}^n\det{[f(x_j-y_i)]}_{i,j=1}^n=
\det{[\alpha_i^{x_j}]}_{i,j=1}^n
\end{equation}
where
\begin{equation}f(m)=\frac1{2\pi \I}\oint \frac{F(z)\dx z}{z^{m+1}}\,,
\end{equation}
and the integral is taken over any positively oriented simple loop
in $A$.
\end{prop}
\begin{proofOF}{Proposition~\ref{PropAB4}}
Since the left-hand side is symmetric with respect to
permutations of $y_j$'s and it vanishes when two $y_j$'s are equal,
we can extend the sum to $\Z^n$ and divide the result by $n!$. We
obtain
\begin{equation}\sum_{y_1,\dots,y_n\in\Z} \det{[\alpha_i^{y_j}]}_{i,j=1}^n
\det{[f(x_j-y_i)]}_{i,j=1}^n=n!\det\Bigl[\sum_{y=-\infty}^{+\infty}
\alpha_k^yf(x_j-y)\Bigr]_{k,j=1}^n.
\end{equation}
Further,
\begin{eqnarray*}
& &\sum_{y=-\infty}^{+\infty}
\alpha_k^yf(x_j-y)=\sum_{y=-\infty}^{+\infty}\frac1{2\pi \I}\oint
\frac{\alpha_k^yF(z)\dx z}{z^{x_j-y+1}}\\
&=&\frac 1{2\pi \I}\hspace{-1em}\oint\limits_{|z|=c_1<|\alpha_k|^{-1}}\hspace{-1em}F(z)\dx z
\sum_{y=x_j+1}^{+\infty}\frac{\alpha_k^y}{z^{x_j-y+1}}+
\frac 1{2\pi \I}\hspace{-1em}\oint\limits_{|z|=c_2>|\alpha_k|^{-1}}\hspace{-1em}F(z)\dx z
\sum_{y=-\infty}^{x_j}\frac{\alpha_k^y}{z^{x_j-y+1}}\\
&=&\frac 1{2\pi\I}\hspace{-1em}\oint\limits_{|z|=c_1<|\alpha_k|^{-1}}\hspace{-1em} \frac{\alpha_k^{x_j+1}F(z)}{1-\alpha_kz}-\frac
1{2\pi \I}\hspace{-1em}\oint\limits_{|z|=c_2>|\alpha_k|^{-1}}\hspace{-1em}\frac{\alpha_k^{x_j+1}F(z)}{1-\alpha_kz}
=\alpha_k^{x_j}F(\alpha_k^{-1}).
\end{eqnarray*}
\end{proofOF}

\begin{prop}\label{PropAB5}
In the notation of Proposition~\ref{PropAB4}, assume
that the variable $y_n$ is virtual, $y_n=\virt$, and set
$f(x_k-\virt)=\alpha_n^{x_k}$ for any $k=1,\dots,n$. Then
\begin{equation}\frac 1{F(\alpha_1^{-1})\cdots
F(\alpha_{n-1}^{-1})}\sum_{y_1<\dots<y_{n-1}\in\Z}
\det{[\alpha_i^{y_j}]}_{i,j=1}^{n-1}\det{[f(x_j-y_i)]}_{i,j=1}^n=
\det{[\alpha_i^{x_j}]}_{i,j=1}^n.
\end{equation}
\end{prop}
\begin{proofOF}{Proposition~\ref{PropAB5}}
Expansion of $\det{[f(x_j-y_i)]}_{i,j=1}^n$ along the last row gives
\begin{equation}\det{[f(x_j-y_i)]}_{i,j=1}^n=
\sum_{k=1}^n(-1)^{n-k}\alpha_n^{x_k} \cdot
\det{[f(x_j-y_i)]}_{\begin{subarray}{l} i=1,...,n-1\\j=1,\dots,k-1,
k+1,\dots,n
\end{subarray}}.
\end{equation}
The application of Proposition~\ref{PropAB4} to each of the
resulting summands in the left-hand side of the desired equality
produces the expansion of $\det{[\alpha_i^{x_j}]}_{i,j=1}^n$ along
the last row.
\end{proofOF}

For $n=1,2,\dots$, denote
\begin{equation}\X_n=\{(x_1,\dots,x_n)\in\Z^n\mid x_1<\dots<x_n\}.
\end{equation}

In what follows we assume that the (nonzero) complex parameters
$\alpha_1,\alpha_2,\dots$ are such that the ratios
${\det[\alpha_i^{x_j}]}_{i,j=1}^n/\det{[\alpha_i^{j-1}]}_{i,j=1}^n$
are nonzero for all $n=1,2,\dots$ and all $(x_1\dots,x_n)$ in
$\X^n$. This holds, for example, when all $\alpha_j$'s are positive.
The Vandermonde determinant in the denominator is needed to make
sense of ${\det[\alpha_i^{x_j}]}_{i,j=1}^n$ when some of the
$\alpha_j$'s are equal.

Under this assumption, define the matrices $\X_n\times\X_n$ and
$\X_n\times \X_{n-1}$ by
\begin{eqnarray*} T_n(\alpha_1,\dots,\alpha_n;F)(X,Y)&=&
\frac{\det{[\alpha_i^{y_j}]}_{i,j=1}^n}
{\det{[\alpha_i^{x_j}]}_{i,j=1}^n}\,\frac{\det{[f(x_i-y_j)]}_{i,j=1}^n}{\prod_{j=1}^n
F(\alpha_j^{-1})},\quad X,Y\in \X_n,
\\
T^n_{n-1}(\alpha_1,\dots,\alpha_{n};F)(X,Y)&=&
\frac{\det{[\alpha_i^{y_j}]}_{i,j=1}^{n-1}}
{\det{[\alpha_i^{x_j}]}_{i,j=1}^n}\,\frac{\det{[f(x_i-y_j)]}_{i,j=1}^n}
{\prod_{j=1}^{n-1} F(\alpha_j^{-1})},\quad X\in\X_n,\,Y\in \X_{n-1},
\end{eqnarray*}
where in the second formula $y_n=\virt$. By
Propositions~\ref{PropAB4} and~\ref{PropAB5}, the sums of entries of
these matrices along rows are equal to 1. We will often omit the
parameters $\alpha_j$ from the notation so that the above matrices
will be denoted as $T_n(F)$ and $T^n_{n-1}(F)$.

We are interested in these matrices because they have nice
commutation relations, as the following proposition shows.

\begin{prop}\label{PropAB6}
Let $F_1$ and $F_2$ be two functions
holomorphic in an annulus containing $\alpha_j^{-1}$'s, that are
also nonzero at these points. Then
\begin{equation}
\begin{aligned}
&T_n(F_1)T_n(F_2)=T_n(F_2)T_n(F_1)=T_n(F_1F_2),\\
&T_n(F_1)T^n_{n-1}(F_2)=T^n_{n-1}(F_1)T_{n-1}(F_2)=T^n_{n-1}(F_1F_2).
\end{aligned}
\end{equation}
\end{prop}
\begin{proofOF}{Proposition~\ref{PropAB6}}
The first line and the relation
$T^n_{n-1}(F_1)T_{n-1}(F_2)=T_{n-1}^n(F_1F_2)$ are proved by
straightforward computations using the fact the Fourier transform of
$F_1F_2$ is the convolution of those of $F_1$ and $F_2$. The only
additional ingredient in the proof of the relation
$T_n(F_1)T^n_{n-1}(F_2)=T^n_{n-1}(F_1F_2)$ is
\begin{equation}\sum_{y\in\Z} f_1(x-y)f_2(y-\virt)=\sum_{y\in\Z}
f_1(x-y)\alpha_n^y=F_1(\alpha_n^{-1})\alpha_n^x.
\end{equation}
\end{proofOF}

\begin{remark}
In the same way one proves the commutation relation
\begin{equation}T_{n-1}^n(F_1)T^{n-1}_{n-2}(F_2)=T_{n-1}^n(F_2)T^{n-1}_{n-2}(F_1)
\end{equation}
but we will not need it later.
\end{remark}

\subsection{Minors of some simple Toeplitz matrices}\label{sect4MinorsToeplitz}

The goal of the section is to derive explicit formulas for $T_n(F)$
and $T^n_{n-1}(F)$ from the previous section for some simple
functions $F$.

\begin{lem}\label{LemmaAB7}
Consider $F(z)=1+pz$, that is
\begin{equation}f(m)=\begin{cases} p,&m=1,\\ 1,&m=0,\\ 0,& \text{otherwise}.\end{cases}
\end{equation}
Then for integers $x_1<\dots<x_n$ and $y_1<\dots<y_n$
\begin{equation}\label{1.44}\det{[f(x_i-y_j)]}_{i,j=1}^n= \begin{cases} p^{\sum_{i=1}^n(x_i-y_i)},
& \text{if}\ y_i-x_i\in \{-1, 0\}
\ \text{for all}\ 1\leq i\leq n,\\
0,& \text{otherwise}.
\end{cases}
\end{equation}
\end{lem}

\begin{proofOF}{Lemma~\ref{LemmaAB7}}
If $x_i<y_i$ for some $i$ then $x_k<y_l$ for $k\leq i$
and $l\geq i$, which implies that $f(x_k-y_l)=0$ for such $k,l$, and
thus the determinant in question vanishes. If $x_i>y_i+1$ then
$x_k>y_{l}+1$ for $k\geq i$ and $l\leq i$, which means $f(x_k-y_l)=0$,
and the determinant vanishes again. Hence, it remains to consider
the case when $x_i-y_i\in \{0,1\}$ for all $1\leq i\leq n$.

Split $\{x_i\}_{i=1}^n$ into blocks of neighboring integers with
distance between blocks being at least $2$. Then it is easy to see
that $\det{[f(x_i-y_j)]}$ splits into the product of determinants
corresponding to blocks. Let $(x_k,\dots,x_{l-1})$ be such a block.
Then there exists $m$, $k\leq m< l$, such that $x_i=y_i+1$ for $l\leq i< m$, and $x_i=y_i$ for $m\leq i< l$. The determinant corresponding
to this block is the product of determinants of two triangular
matrices, one has size $m-k$ and diagonal entries equal to $p$,
while the other one has size $l-m$ and diagonal entries equal to 1.
Thus, the determinant corresponding to this block is equal to
$p^{m-k}$, and collecting these factors over all blocks yields the
result.
\end{proofOF}

\begin{lem}\label{LemmaAB8}
Consider $F(z)=(1-qz)^{-1}$, that is
\begin{equation}f(m)=\begin{cases} q^m,&m\geq 0,\\ 0,&otherwise.\end{cases}
\end{equation}

(i) For integers $x_1<\dots<x_n$ and $y_1<\dots<y_n$
\begin{equation}\det{[f(x_i-y_j)]}_{i,j=1}^n=\begin{cases}
q^{\sum_{i=1}^n(x_i-y_i)},&
x_{i-1}<y_i\leq x_i,\ 1\leq i\leq n,\\
0,&otherwise.\end{cases}
\end{equation}
(The condition $x_0<y_1$ above is empty.)

(ii) For integers $x_1<\dots<x_n$ and $y_1<\dots<y_{n-1}$, and with
virtual variable $y_n=\virt$ such that $f(x-\virt)=q^x$,
\begin{equation}\label{1.47}\det{[f(x_i-y_j)]}_{i,j=1}^n=\begin{cases} (-1)^{n-1} q^{\sum_{i=1}^n
x_i-\sum_{i=1}^{n-1} y_i},&
x_{i}<y_i\leq x_{i+1},\ 1\leq i\leq n-1,\\
0,&otherwise.\end{cases}
\end{equation}
\end{lem}
\begin{proofOF}{Lemma~\ref{LemmaAB8}}
(i) Let us first show that the needed inequalities are satisfied.
Indeed, if $x_i<y_i$ for some $i$ then $\det{[f(x_i-y_j)]}=0$ by the
same reasoning as in the previous lemma. On the other hand, if
$x_{i-1}\geq y_i$ then $x_k\geq y_l$ for $k\geq i-1$, $l\leq i$. Let
$i$ be the smallest number such that $x_{i-1}\geq y_i$. Then columns
$i$ and $i+1$ have the form
\begin{equation}
\begin{bmatrix} 0&\dots& 0 &q^{x_{i-1}-y_{i-1}}
&q^{x_{i}-y_{i-1}}&*&*&\dots\\ 0&\dots& 0 &q^{x_{i-1}-y_{i}}
&q^{x_{i}-y_{i}}&*&*&\dots
\end{bmatrix}^T,
\end{equation}
where the $2\times 2$ block with powers of $q$ is on the main
diagonal.
 This again implies that the determinant
vanishes. On the other hand, if the interlacing inequalities are
satisfied then the matrix $[f(x_i-y_j)]$ is triangular, and
computing the product of its diagonal entries yields the result.

(ii) The statement follows from (i). Indeed, we just need to
multiply both sides of (i) by $q^{y_1}$, denote $y_1(\leq x_1)$ by
$\virt$, and then cyclically permute $y_j$'s.
\end{proofOF}

\begin{lem}\label{LemmaAB9}
Consider $F(z)=p+qz(1-qz)^{-1}$, that is
\begin{equation}f(m)=\begin{cases} p,&m=0,\\q^m,&m\geq 1,\\ 0,&otherwise.\end{cases}
\end{equation}

(i) For integral $x_1<\dots<x_n$ and $y_1<\dots<y_n$
\begin{equation}\det{[f(x_i-y_j)]}_{i,j=1}^n=q^{\sum_{i=1}^n(x_i-y_i)} p^{\#\{i\mid
x_i=y_{i}\}}(1-p)^{\#\{i\mid x_{i-1}=y_{i}\}}
\end{equation}
if $x_{i-1}\leq y_i\leq x_i$ for all $1\leq i\leq n$, and $0$ otherwise.

(ii) For integral $x_1<\dots<x_n$ and $y_1<\dots<y_{n-1}$, and with
virtual variable $y_n=\virt$ such that $f(x-\virt)=q^x$,
\begin{equation}\det{[f(x_i-y_j)]}_{i,j=1}^n=(-1)^{n-1}q^{\sum_{i=1}^n
x_i-\sum_{i=1}^{n-1}y_i} p^{\#\{i\mid
x_{i+1}=y_{i}\}}(1-p)^{\#\{i\mid x_{i}=y_{i}\}}
\end{equation}
if $x_{i}\leq y_i\leq x_{i+1}$ for all $1\leq i\leq n-1$, and $0$
otherwise.
\end{lem}

\begin{proofOF}{Lemma~\ref{LemmaAB9}}
(i) The interlacing conditions are verified by the same argument as
in the proof of Lemma~\ref{LemmaAB8}(i) (although the conditions
themselves are slightly different). Assuming that they are
satisfied, we observe that the matrix elements of $[f(x_i-y_j)]$ are
zero for $j\geq i+2$ because $x_i\leq y_{i+1}<y_{i+2}$ and $f(m)=0$
for $m<0$. Further, the $(i,i+1)$-element is equal to $p$ if
$x_i=y_{i+1}$ or $0$ if $x_i<y_{i+1}$. Thus, the matrix is
block-diagonal, with blocks being either of size 1 with entry
$f(x_i-y_i)$, or of larger size having the form
\begin{equation}\label{1.51} \left(
 \begin{array}{ccccc}
q^{x_k-y_k}&p&0&\dots&0\\
q^{x_{k+1}-y_{k}}&q^{x_{k+1}-y_{k+1}}&p&\dots&0\\
q^{x_{k+2}-y_{k}}&q^{x_{k+2}-y_{k+1}}&q^{x_{k+2}-y_{k+2}}&\dots&0\\
\dots&\dots&\dots&\dots&\dots\\
q^{x_{l}-y_{k}}&q^{x_{l}-y_{k+1}}&q^{x_{l}-y_{k+2}}&\dots&
q^{x_{l}-y_{l}}
 \end{array}
\right)
\end{equation}
with $x_k=y_{k+1},\dots,x_{l-1}=y_l$, and $x_{k-1}<y_k$,
$x_l<y_{l+1}$. The determinant of (\ref{1.51}) is computable
via Lemma 1.2 of \cite{Bor08}, and it is equal to
\begin{equation}
q^{x_l-y_k}(1-p)^{l-k}=q^{x_k+\dots+x_l-(y_k+\dots+y_l)}(1-p)^{l-k}.
\end{equation}
Collecting all the factors yields the desired formula.

The proof of (ii) is very similar to that of Lemma~\ref{LemmaAB8}(ii).
\end{proofOF}

\subsection{Examples of bivariate Markov chains}\label{sect5Examples}

We now use the formulas from the previous two sections to make the
constructions of the first two sections more explicit.

Let us start with bivariate Markov chains. Set $\S^*=\X_n$ and
$\S=\X_{n-1}$, where the sets $\X_m$, $m=1,2,\dots$, were introduced
in Section~\ref{sect3Toeplitz}. We will also take
\begin{equation}
\Lambda=T^n_{n-1}(\alpha_1,\dots,\alpha_n;(1-\alpha_nz)^{-1})
\end{equation}
for some fixed $\alpha_1,\dots,\alpha_n>0$.

The first case we consider is
\begin{equation}
P=T_{n-1}(\alpha_1,\dots,\alpha_{n-1};1+\beta z),\quad
P^*=T_{n}(\alpha_1,\dots,\alpha_{n};1+\beta z),\qquad \beta>0.
\end{equation}
Then Proposition~\ref{PropAB6} implies that
\begin{equation}
\Delta=\Lambda
P=P^*\Lambda=T^n_{n-1}(\alpha_1,\dots,\alpha_n;(1+\beta
z)/(1-\alpha_nz)).
\end{equation}

According to (\ref{1.17}), (\ref{1.18}), we have to compute
expressions of the form
\begin{equation*}
\frac{P^*(x^*,y^*)\Lambda(y^*,y)}{\Delta(x^*,y)}\,,\qquad
\frac{P^*(x^*,y^*)\Lambda(y^*,x)} {\Delta(x^*,x)}\,
\end{equation*}
for the sequential and parallel updates, respectively.

We start with the condition probability needed for the Markov chain
$P_\Lambda$.

\begin{prop}\label{PropAB11}
Assume that $x^*\in \S^*$ and $y\in \S$ are such that \mbox{$\Delta(x^*,y)>0$}, that is, $x_{k}^*\leq y_k\leq x_{k+1}^*$ for all $1\leq k\leq n-1$. Then the probability
distribution
\begin{equation}
\frac{P^*(x^*,y^*)\Lambda(y^*,y)}{\Delta(x^*,y)},\qquad y^*\in\S^*,
\end{equation}
has nonzero weights iff
\begin{equation}
y_k^*-x_k^*\in\{-1,0\}, \qquad y_{k-1}\leq y_k^*< y_{k},\qquad
k=1,\dots,n,
\end{equation}
(equivalently, $\max(x_k^*-1,y_{k-1})\leq y_k^*\leq \min(x_k^*,y_k-1)$
for all $k$), and these weights are equal to
\begin{equation}
\prod_{\begin{subarray}{c} \max(x_k^*-1,y_{k-1})< \min(x_k^*,y_k-1)\\
k=1,\dots,n\end{subarray}}
\left(\frac{\beta}{\alpha_n+\beta}\right)^{x^*_k-y^*_k}\left(\frac
{\alpha_n}{\alpha_n+\beta}\right)^{1-x^*_k+y^*_k}
\end{equation}
with empty product equal to 1.
\end{prop}

\begin{remark}
One way to think about the distribution of $y^*\in
\S^*$ is as follows. For each $k$ there are two possibilities for
$y_k^*$: Either $\max(x_k^*-1,y_{k-1})=\min(x_k^*,y_k-1)$, in which
case $y_k^*$ is forced to be equal to this number, or
$\max(x_k^*-1,y_{k-1})=x_k^*-1$ and $\min(x_k^*,y_k-1)=x_k^*$, in
which case $y_k^*$ is allowed to take one of the two values $x_k^*$
or $x_k^*-1$. Then in the latter case, $x_k^*-y_k^*$ are i.~i.~d.
Bernoulli random variables with the probability of the value $0$
equal to $\alpha_n/(\alpha_n+\beta)$.
\end{remark}

\begin{proofOF}{Proposition~\ref{PropAB11}}
The conditions for non-vanishing of the weights follow from those of
Lemmas~\ref{LemmaAB7} and \ref{LemmaAB8}, namely from (\ref{1.44})
and (\ref{1.47}). Using these formulas we extract the factors of
${P^*(x^*,y^*)\Lambda(y^*,y)}$ that depend on $y^*$. This yields
$(\alpha_n/\beta)^{\sum_{i=1}^n y_i^*}$. Normalizing these weights
so that they provide a probability distribution leads to the desired
formula.
\end{proofOF}

Let us now look at the conditional distribution involved in the
definition of the Markov chain $P_\Delta$. The following statement
is a direct consequence of Proposition~\ref{PropAB11}.

\begin{cor}\label{CorAB12}
Assume that $x^*\in \S^*$ and $x\in \S$ are
such that $\Delta(x^*,x)>0$, that is, $x_{k}^*\leq x_k\leq x_{k+1}^*$
for all $1\leq k\leq n-1$. Then the probability distribution
\begin{equation}\frac{P^*(x^*,y^*)\Lambda(y^*,x)} {\Delta(x^*,x)},\qquad y^*\in
\S^*,
\end{equation}
has nonzero weights iff $ \max(x_k^*-1,x_{k-1})\leq y_k^*\leq \min(x_k^*,x_{k}-1)$, and these weights are equal to
\begin{equation}
\prod_{\begin{subarray}{c}
\max(x_k^*-1,x_{k-1})< \min(x_k^*,x_{k}-1)\\
k=1,\dots,n\end{subarray}}\left(\frac {\beta
}{\alpha_n+\beta}\right)^{x^*_k-y^*_k}\left(\frac
{\alpha_n}{\alpha_n+\beta}\right)^{1-x^*_k+y^*_k}.
\end{equation}
\end{cor}

Let us now proceed to the case
\begin{equation}
P=T_{n-1}(\alpha_1,\dots,\alpha_{n-1};(1-\gamma z)^{-1}),\quad
P^*=T_{n}(\alpha_1,\dots,\alpha_{n};(1-\gamma z)^{-1}).
\end{equation}
We assume that $0<\gamma<\min\{\alpha_1,\dots,\alpha_n\}$.

By Proposition~\ref{PropAB6}
\begin{equation}
\Delta=\Lambda P=P^*\Lambda=T^n_{n-1}\bigl(\alpha_1,\dots,\alpha_n;1/((1-\alpha_nz)(1-\gamma z))\bigr).
\end{equation}
Again, let us start with $P_\Lambda$.

\begin{prop}\label{PropAB13}
Assume that $x^*\in \S^*$ and $y\in \S$ are such that \mbox{$\Delta(x^*,y)>0$}, that is,
$x_{k-1}^*<y_k-1<x_{k+1}^*$ for all $k$. Then the probability
distribution
\begin{equation}
\frac{P^*(x^*,y^*)\Lambda(y^*,y)}{\Delta(x^*,y)},\qquad y^*\in\S^*,
\end{equation}
has nonzero weights iff
\begin{equation}
x_{k-1}^*< y_k^*\leq x^*_{k},\qquad y_{k-1}\leq y_{k}^*< y_k,\qquad
k=1,\dots,n-1,
\end{equation}
(equivalently, $\max(x_{k-1}^*+1,y_{k-1})\leq y_k^*\le\min(x_{k}^*,y_{k}-1)$ for all $k$), and these weights are
equal to
\begin{equation}
\prod_{k=1}^n\frac{{(\alpha_n/\gamma)}^{y_k^*}}
{\sum\limits_{l=\max(x_{k-1}^*+1,y_{k-1})}^{\min(x^*_{k},y_k-1)}
{(\alpha_n/\gamma)}^{l}}\,.
\end{equation}
Here $\max(x_{0}^*+1,y_{0})$ is assumed to denote $-\infty$.
\end{prop}

\begin{remark}
Less formally, these formulas state the following:
Each $y_k^*$ has to belong to the segment
$[\max(x_{k-1}^*+1,y_{k-1}),\min(x_{k}^*,y_{k}-1)]$, and the
restriction that $\Delta(x^*,y)>0$ guarantees that these segments
are nonempty. Then the claim is that $y_k^*$'s are independent, and
the distribution of $y_k^*$ in the corresponding segment is
proportional to the weights ${(\alpha_n/\gamma)}^{y_k^*}$. In other
words, this is the geometric distribution with ratio
$\alpha_n/\gamma$ conditioned to live in the prescribed segment.
\end{remark}

\begin{proofOF}{Proposition~\ref{PropAB13}}
Similarly to the proof of Proposition~\ref{PropAB11}, we use
Lemmas~\ref{LemmaAB8} to derive the needed inequalities and to
single out the part of the ratio
${P^*(x^*,y^*)\Lambda(y^*,y)}/{\Delta(x^*,y)}$ that depends on
$y^*$. One readily sees that it is equal to
${(\alpha_n/\gamma)}^{\sum_{k=1}^n y_k^*}$, and this concludes the
proof.
\end{proofOF}

Let us state what this computation means in terms of the conditional
distribution used in the construction of $P_\Delta$.

\begin{cor}\label{CorAB14}
Assume that $x^*\in \S^*$ and $x\in \S$ are
such that $\Delta(x^*,x)>0$, that is, $x_{k-1}^*<x_k-1<x_{k+1}^*$
for all $k$. Then the probability distribution
\begin{equation}
\frac{P^*(x^*,y^*)\Lambda(y^*,x)} {\Delta(x^*,x)},\qquad y^*\in \S^*,
\end{equation}
has nonzero weights iff $\max(x_{k-1}^*+1,x_{k-1})\leq y_k^*\le\min(x_{k}^*,x_{k}-1)$ for all $k$, and these weights are
equal to
\begin{equation}\prod_{k=1}^n\frac{{(\alpha_n/\gamma)}^{y_k^*}}
{\sum\limits_{l=\max(x_{k-1}^*+1,x_{k-1})}^{\min(x^*_{k},x_k-1)}
{(\alpha_n/\gamma)}^{l}}\,.
\end{equation}
\end{cor}

In the four statements above we computed the ingredients needed for
the constructions of the bivariate Markov chains for the simplest
possible Toeplitz-like transition matrices. In these examples we
always had \mbox{$x_k^*\geq y_k^*$}, or, informally speaking, ``particles
jump to the left''. Because of the previous works on the subject, it
is more convenient to deal with the case when particles ``jump to
the right''. The arguments are very similar, so let us just state
the results.

Consider
\begin{equation}
P=T_{n-1}(\alpha_1,\dots,\alpha_{n-1};1+\beta z^{-1}),\quad
P^*=T_{n}(\alpha_1,\dots,\alpha_{n};1+\beta z^{-1}),\quad \beta>0.
\end{equation}

\noindent $\bullet$ \quad For $P_\Lambda$, we have
$\max(x_k^*,y_{k-1})\leq y_k^*\leq \min(x_k^*+1,y_k-1)$. This segment
consists of either 1 or 2 points, in the latter case $y_k^*-x_k^*$
are i.~i.~d. Bernoulli random variables with the probability of $0$
equal to $(1+\alpha_n\beta)^{-1}$.

\noindent $\bullet$ \quad For $P_\Delta$, we have
$\max(x_k^*,x_{k-1})\leq y_k^*\leq \min(x_k^*+1,x_k-1)$, and the rest
is the same as for $P_\Lambda$.

Now consider
\begin{equation}
P=T_{n-1}(\alpha_1,\dots,\alpha_{n-1};(1-\gamma z^{-1})^{-1}),\quad
P^*=T_{n}(\alpha_1,\dots,\alpha_{n};(1-\gamma z^{-1})^{-1}),
\end{equation}
for $0<\gamma<\min\{\alpha_1^{-1},\dots,\alpha_{n}^{-1}\}$.

\noindent $\bullet$ \quad For $P_\Lambda$, we have
$\max(x_{k}^*,y_{k-1})\leq y_k^*\le\min(x_{k+1}^*,y_{k})-1$, and
$y_k^*$ are independent geometrically distributed with ratio
$(\alpha_n\gamma)$ random variables conditioned to stay in these
segments.

\noindent $\bullet$ \quad For $P_\Delta$, we have
$\max(x_{k}^*,x_{k-1})\leq y_k^*\le\min(x_{k+1}^*,x_{k})-1$, and the
rest is the same as for $P_\Lambda$.

Thus, we have so far considered eight bivariate Markov chains. It is
natural to denote them as
\begin{equation}P_\Lambda(1+\beta z^{\pm 1}), \quad P_\Delta(1+\beta z^{\pm
1}),\quad P_\Lambda((1-\gamma z^{\pm 1})^{-1}), \quad
P_\Delta((1-\gamma z^{\pm 1})^{-1}).
\end{equation}

Observe that although all four chains of type $P_\Lambda$ live on
one and the same state space, all four chains of type $P_\Delta$
live on different state spaces. For the sake of completeness, let us
list those state spaces:
\begin{eqnarray*}
S_\Lambda&=&\{(x^*,x)\in \X_n\times\X_{n-1}\mid x_{k}^*+1\leq x_k\leq x_{k+1}^*\text{ for all $k$}\}\\
S_{\Delta}(1+\beta z)&=&\{(x^*,x)\in \X_n\times\X_{n-1}\mid
x_{k}^*\leq x_k\leq x_{k+1}^*\text{ for all $k$}\}\\
S_{\Delta}(1+\beta z^{-1})&=&\{(x^*,x)\in \X_n\times\X_{n-1}\mid
x_{k}^*+1\leq x_k\leq x_{k+1}^*+1\text{ for all $k$}\}
\\
S_{\Delta}((1-\gamma z)^{-1})&=&\{(x^*,x)\in \X_n\times\X_{n-1}\mid
x_{k-1}^*+2\leq x_k\leq x_{k+1}^*\text{ for all $k$}\}
\\
S_{\Delta}((1-\gamma z^{-1})^{-1})&=&\{(x^*,x)\in
\X_n\times\X_{n-1}\mid x_{k}^*+1\leq x_k\leq x_{k+2}^*-1\text{ for
all $k$}\}
\end{eqnarray*}
In the above formulas we always use the convention that if an
inequality involves a nonexistent variable (like $x_0$ or
$x^*_{n+1}$), it is omitted.

\subsection{Examples of multivariate Markov chains}\label{sect6ExampleMulti}

Let us now use some of the examples of the bivariate Markov chains
from the previous section to construct explicit examples of
multivariate (not necessarily autonomous) Markov chains following
the recipe of Section~\ref{sect1MarkovChains}.

For any $m\geq 0$ we set $\S_m=\X_m$, which is the set of strictly
increasing $m$-tuples of integers. In this section we will denote
these integers by \mbox{$x^m_1<\dots<x^m_m$}.

Fix an integer $n\geq 1$, and choose $n$ positive real numbers
$\alpha_1,\dots,\alpha_n$. We take the maps $\Lambda^k_{k-1}$ to be
\begin{equation}\Lambda^k_{k-1}=T^k_{k-1}(\alpha_1,\dots,\alpha_k;(1-\alpha_kz)^{-1}),\qquad
k=2,\dots,n.
\end{equation}

We consider the Markov chain $S_\Lambda^{(n)}$, i.e., the sequential
update, first. Its state space has the form
\begin{equation}
\begin{aligned}
S_\Lambda^{(n)}&=\Bigl\{(x^1,\dots,x^n)\in
\S_1\times\dots\times\S_n\mid \prod_{m=2}^n
\Lambda^m_{m-1}(x^m,x^{m-1})>0\Bigr\}\\
&=\Bigl\{\{x_k^m\}_{\begin{subarray}{c} m=1,\dots,n\\k=1,\dots,m\end{subarray}}\subset
\Z^{\frac{n(n+1)}2}\mid x_k^{m+1}< x_k^{m}\leq x_{k+1}^{m+1}\text{
 for all $k,m$}\Bigr\}.
\end{aligned}
\end{equation}
In other words, this is the space of $n$ interlacing integer
sequences of length $1,\dots,n$.

Let $t$ be an integer time variable. We now need to choose the
transition probabilities $P_m(t)$, $m=1,\dots,n$.

Let $\{F_t(z)\}_{t\geq t_0}$ be a sequence of functions each of which
has one of the four possibilities:
\begin{equation}F_t(z)=(1+\beta^+_tz)\ \text{ or }\ (1+\beta^-_t/z) \ \text{ or }\
(1-\gamma^+_tz)^{-1}\ \text{ or }\ (1-\gamma^-_t/z)^{-1}.
\end{equation}
Here we assume that
\begin{equation}\beta^{\pm}_t,\gamma^{\pm}_t>0,\qquad
\gamma^+_t<\min\{\alpha_1,\dots,\alpha_n\},\quad
\gamma^-_t<\min\{\alpha_1^{-1},\dots,\alpha_n^{-1}\}.
\end{equation}
We set
\begin{equation}P_m(t)=T_m(\alpha_1,\dots,\alpha_m;F_t(z)),\qquad m=1,\dots,n.
\end{equation}
Then all needed commutation relations are satisfied, thanks to
Proposition~\ref{PropAB6}.

The results of Section~\ref{sect5Examples} enable us to describe the resulting Markov
chain on $\S^{(n)}_\Lambda$ as follows.

At time moment $t$ we observe a (random) point $\{x_k^m(t)\}\in
S^{(n)}_\Lambda$. In order to obtain $\{x_k^m(t+1)\}$, we perform
the sequential update from level $1$ to level $n$. When we are at
level $m$, $1\leq m\leq n$, the new positions of the particles
$x_1^m<\dots<x_m^m$ are decided independently.

\textbf{(1)}\quad For $F_t(z)=1+\beta^+_tz$, the particle $x_k^m$ is
either forced to stay where it is if $x^{m-1}_{k-1}(t+1)=x_k^m(t)$,
or it is forced to jump to the left by $1$ if
\mbox{$x^{m-1}_{k}(t+1)=x_k^m(t)$}, or it chooses between staying put or
jumping to the left by $1$ with probability of staying equal to
$1/(1+\beta^+_t\alpha_m^{-1})$. This follows from Proposition
\ref{PropAB11}.

\textbf{(2)}\quad For $F_t(z)=1+\beta^-_t/z$, the particle $x_k^m$
is either forced to stay where it is if
$x^{m-1}_{k}(t+1)=x_k^m(t)+1$, or it is forced to jump to the right
by 1 if $x^{m-1}_{k-1}(t+1)=x_k^m(t)+1$, or it chooses between
staying put or jumping to the right by 1 with probability of staying
equal to $1/(1+\beta^-_t\alpha_m)$.

\textbf{(3)}\quad For $F_t(z)=(1-\gamma^+_t z)^{-1}$, the particle
$x_k^m$ chooses its new position according to a geometric random
variable with ratio $\alpha_m/\gamma^+_t$ conditioned to stay in the
segment
\begin{equation}[\max(x_{k-1}^m(t)+1,x_{k-1}^{m-1}(t+1)),\min(x_{k}^m(t),x_{k}^{m-1}(t+1)-1)].
\end{equation}
In other words, it tries to jump to the left using the geometric
distribution of jump length, but it is conditioned not to overcome
$x_{k-1}^m(t)+1$ (in order not to ``interact'' with the jump of
$x_{k-1}^m$), and it is also conditioned to obey the interlacing
inequalities with the updated particles on level $m-1$. This follows
from Proposition \ref{PropAB13}.

\textbf{(4)}\quad For $F_t(z)=(1-\gamma^-_t/z)^{-1}$, the particle
$x_k^m$ chooses its new position according to a geometric random
variable with ratio $\alpha_m\gamma^-_t$ conditioned to stay in the
segment
\begin{equation}[\max(x_{k}^m(t),x_{k-1}^{m-1}(t+1)),\min(x_{k+1}^m(t),x_{k}^{m-1}(t+1))-1].
\end{equation}
In other words, it tries to jump to the right using the geometric
distribution of jump length, but it is conditioned not to overcome
$x_{k+1}^m(t)-1$ (so that it does not interact with jumps of
$x_{k+1}^m$), and it is also conditioned to obey the interlacing
inequalities with the updated particles on level $m-1$.
\medskip

\textbf{Projection to $\{x^m_1\}_{m\geq 1}$}. A remarkable property
of the Markov chain $P_\Lambda^{(n)}$ with steps of the first three
types is that its projection onto the \mbox{$n$-dimensional} subspace
$\{x_1^1>x_1^2>\dots>x_1^n\}$ (the smallest coordinates on each
level) is also a Markov chain. Moreover, since these are the
leftmost particles on each level, they have no interlacing condition
on their left to be satisfied, which makes the evolution simpler.
Let us describe these Markov chains.

At time moment $t$ we observe
$\{x_1^1(t)>x_1^2(t)>\dots>x_1^n(t)\}$. In order to obtain
$\{x_1^m(t+1)\}_{m=1}^n$, we perform the sequential update from
$x_1^1$ to $x_1^n$.

\textbf{(1)}\quad For $F_t(z)=1+\beta^+_tz$, the particle $x_1^m$ is
either forced to jump (it is being {\it pushed\/}) to the left by 1
if $x^{m-1}_{1}(t+1)=x_1^m(t)$, or it chooses between not moving at
all or jumping to the left by 1 with probability of not moving equal
to $1/(1+\beta^+_t\alpha_m^{-1})$.

\textbf{(2)}\quad For $F_t(z)=1+\beta^-_t/z$, the particle $x_1^m$
is either forced to stay where it is if
$x^{m-1}_{1}(t+1)=x_1^m(t)+1$, or it chooses between staying put or
jumping to the right by 1 with probability of staying equal to
$1/(1+\beta^-_t\alpha_m)$.

\textbf{(3)}\quad For $F_t(z)=(1-\gamma^+_t z)^{-1}$, the particle
$x_1^m$ chooses its new position according to a geometrically
distributed with ratio $\gamma^+_t/\alpha_n$ jump to the left from
the point $\min(x_{1}^m(t),x_{1}^{m-1}(t+1)-1)$. That is, if
$x_{1}^m(t)< x_{1}^{m-1}(t+1)$ then $x_1^m$ simply jumps to the left
with the geometric distribution of the jump, while if
$x_{1}^m(t)\geq x_{1}^{m-1}(t+1)$ then $x_1^m$ is first being pushed
to the position $x_{1}^{m-1}(t+1)-1$ and then it jumps to the left
using the geometric distribution.

\textbf{(4)}\quad For the transition probability with
$F_t(z)=(1-\gamma^-_t/z)^{-1}$, the particle $x_1^m$ is conditioned
to stay below $\min(x_{2}^m(t),x_{1}^{m-1}(t+1))-1$, which involves
$x_2^m$, thus the projection is \emph{not Markovian}.

The Markov chains on $\{x_1^1>\dots>x_1^n\}$ corresponding to
$1+\beta^+_tz$ and $1+\beta^-_t/z$ are the ``Bernoulli jumps with
pushing'' and ``Bernoulli jumps with blocking'' chains discussed in
\cite{DW07}.

\medskip

\textbf{Projection to $\{x^m_m\}_{m\geq 1}$.} Similarly, the
projection of the ``big'' Markov chain to $\{x_1^1\leq
x_2^2\le\dots\leq x_n^n\}$ is Markovian for the steps of types one,
two, and four, but it is not Markovian for the step of the third
type \mbox{$F_t(z)=(1-\gamma^+_tz)^{-1}$}.

\medskip

Let us now consider the parallel update Markov chain
$P_\Delta^{(n)}$, or rather one of them.

Choose a sequence of functions $G_t(z)=1+\beta_tz^{-1}$ with
$\beta_t\geq 0$, and set
\begin{equation}P_m(t)=T_m(\alpha_1,\dots,\alpha_m;G_t(z)),\qquad m=1,\dots,n.
\end{equation}
In case $\beta_t=0$, $P_m(t)$ is the identity matrix. As before, the
needed commutation relations are satisfied by Proposition~\ref{PropAB6}.

The (time-dependent) state space of our Markov chain is
\begin{multline}
 \S_\Delta^{(n)}(t)=\Bigl\{(x^1,\dots,x^n)\in
\S_1\times\dots\times\S_n\mid \prod\limits_{m=2}^{n}
\Delta^{m}_{m-1}(x^{m},x^{m-1}\mid t+n-m)>0\Bigr\}\\
\hspace{2em}=\Bigl\{\{x_k^m\}_{\begin{subarray}{c}
m=1,\dots,n\\k=1,\dots,m\end{subarray}}\subset
\Z^{\frac{n(n+1)}2}\mid x_k^{m}< x_k^{m-1}\leq x_{k+1}^{m}\text{ \ if\quad $\beta_{t+n-m}=0$},\\
x_k^{m}< x_k^{m-1}\leq x_{k+1}^{m}+1 \text{ \ if\quad $\beta_{t+n-m}>0$}\Bigr\}.
\end{multline}

The update rule follows from the analog of Corollary~\ref{CorAB12}
for $(1+\beta_tz^{-1})$. Namely, assume we have $\{x_k^m(t)\}\in
S_\Delta^{(n)}(t)$. Then we choose $\{x_k^m(t)\}$ {\it independently
of each other\/} as follows. We have
\begin{equation}\max(x_k^m(t),x_{k-1}^{m-1}(t))\leq x_k^{m}(t+1)\leq \min(x_k^m(t)+1,x_k^{m-1}(t)-1).
\end{equation}

 This segment
consists of either 1 or 2 points, and in the latter case
$x_k^{m+1}(t+1)$ has probability of not moving equal to
$(1+\alpha_m\beta_{t-n+m})^{-1}$, and it jumps to the right by 1
with remaining probability. In particular, if $\beta_{t-n+m}=0$ then
$x_k^m(t+1)=x_k^m(t)$ for all $k=1,\dots,m$.

Less formally, each particle $x^m_k$ either stays put or moves to
the right by 1. It is forced to stay put if
$x^m_k(t)=x^{m-1}_k(t)-1$, and it is forced to move by 1 if
$x^m_k(t)=x_{k-1}^{m-1}(t)-1$. Otherwise, it jumps with probability
$1-(1+\alpha_n\beta_{t-n+m})^{-1}$.

\medskip

\textbf{Projection to $\{x^m_1\}_{m\geq 1}$}. Once again, the
projection of this Markov chain to $\{x_1^1>\dots>x_1^n\}$ is also a
Markov chain, and its transition probabilities are as follows: Each
particle $x_1^m$ at time moment $t$ is either forced to stay if
$x_1^{m}(t)=x_1^{m-1}(t)-1$ or it stays with probability
$(1+\alpha_n\beta_{t-n+m})^{-1}$ and jumps to the right by 1 with
complementary probability. This Markov chain has no pushing because
$x_1^m$'s do not have neighbors on the left. This is the ``TASEP
with parallel update'', see e.g.~\cite{BFS07b}.

\medskip

\textbf{Projection to $\{x^m_m\}_{m\geq 1}$}. We can also restrict
our ``big'' Markov chain to the particles $\{x_1^1,
x_2^2,\dots,x_n^n\}$. Then at time moment $t$ they satisfy the
inequalities
\begin{equation}x_{m-1}^{m-1}(t)\leq x_{m}^{m}(t) \quad \text{if}\quad
\beta_{t+n-m}=0,\qquad x_{m-1}^{m-1}(t)\leq x_{m}^{m}(t)+1 \quad
\text{if}\quad \beta_{t+n-m}>0,
\end{equation}
and the update rule is as follows. If
$x_{m-1}^{m-1}(t)=x_{m}^m(t)+1$ then $x_m^m$ moves to the right by
1: $x_m^m(t+1)=x_m^m(t)$. However, if $x_{m-1}^{m-1}(t)\leq
x_{m}^m(t)$ then $x_m^m$ stays put with probability
$(1+\alpha_n\beta_{t-n+m})^{-1}$, and it jumps to the right by 1
with the complementary probability.

In the special case when all $\alpha_j=1$,
\begin{equation}\beta_k=\begin{cases} \beta, &k\geq n-1,\\0, & k<n-1,\end{cases}
\end{equation}
and with the densely packed initial condition $x_k^m(n-m)=k-m-1$,
the Markov chain $P^{(n)}_\Delta$ discussed above is equivalent to
the so-called shuffling algorithm on domino tilings of the Aztec
diamonds that at time $n$ produces a random domino tiling of the
diamond of size $n$ distributed according to the measure that
assigns to a tiling the weight proportional to $\beta$ raised to the
number of vertical tiles, see~\cite{Nor08}.

\subsection{Continuous time multivariate Markov chain}\label{sect7ContTime}

The (discrete time) Markov chains considered above admit
degenerations to continuous time Markov chains. Let us work out one
of the simplest examples.

As in the previous sections, we fix an integer $n\geq 1$ and $n$
positive real numbers $\alpha_1,\dots,\alpha_n$, and take
\begin{equation}\Lambda^k_{k-1}=T^k_{k-1}(\alpha_1,\dots,\alpha_k;(1-\alpha_kz)^{-1}),\qquad
k=2,\dots,n.
\end{equation}

We will consider a limit of the Markov chain $S_\Lambda^{(n)}$, so
our state space is
\begin{equation}S_\Lambda^{(n)}=\Bigl\{\{x_k^m\}_{\begin{subarray}{c}
m=1,\dots,n\\k=1,\dots,m\end{subarray}}\subset \Z^{\frac{n(n+1)}2}\mid
x_k^{m+1}< x_k^{m}\leq x_{k+1}^{m+1}\text{
 for all $k,m$}\Bigr\}.
\end{equation}

In the notation of the previous section, let us take
$F_t(z)=1+\beta^-/z$ for a fixed $\beta_->0$ and $t=1,2,\dots$.
Thus, we obtain an autonomous Markov chain on $\S_\Lambda^{(n)}$,
whose transition probabilities are determined by the following
recipe.

In order to obtain $\{x_k^m(t+1)\}$ from $\{x_k^m(t)\}$, we perform
the sequential update from level $1$ to level $n$. When we are at
level $m$, $1\leq m\leq n$, for each $k=1,\dots,m$ the particle
$x_k^m$ is either forced to stay if $x^{m-1}_{k}(t+1)=x_k^m(t)+1$,
or it is forced to jump to the right by 1 if
$x^{m-1}_{k-1}(t+1)=x_k^m(t)+1$, or it chooses between staying put
or jumping to the right by 1 with probability of staying equal to
$(1+\beta^-\alpha_m)^{-1}$. Note that, since particles can only move
to the right, it is easy to order the elements of the state space so
that the matrix of transition probabilities is triangular.

We are now interested in taking the limit $\beta^-\to 0$.

\begin{lem}\label{LemmaAB15}
Let $A(\epsilon)$ be a (possibly infinite)
triangular matrix, whose matrix elements are polynomials in an
indeterminate $\epsilon>0$:
\begin{equation}
A(\epsilon)=A_0+\epsilon A_1+\epsilon^2 A_2+\dots,
\end{equation}
and assume that $A_0=\Id$. Then for any $\tau\in \Bbb R$,
\begin{equation}
\lim_{\epsilon\to 0} (A(\epsilon))^{[\tau/\epsilon]}=\exp(\tau A_1).
\end{equation}
\end{lem}
\begin{proofOF}{Lemma~\ref{LemmaAB15}}
For the finite size matrix the claim is standard, and
the triangularity assumption reduces the computation of any fixed
matrix element of $(A(\epsilon))^{[\tau/\epsilon]}$ to the finite
matrix case.
\end{proofOF}

This lemma immediately implies that the transition probabilities of
the Markov chain described above converge, in the limit $\beta^-\to
0$ and time rescaling by $\beta^-$, to those of the continuous time
Markov chain on $S_\Lambda^{(n)}$, whose generator is the linear in
$\beta^-$ term of the generator of the discrete time Markov chain.
Let denote this linear term by $L^{(n)}$. Its off-diagonal entries
are not hard to compute:
\begin{equation}L^{(n)}\left(\{x_k^m\}_{\begin{subarray}{c}
m=1,\dots,n\\k=1,\dots,m\end{subarray}},\{y_k^m\}_{\begin{subarray}{c}
m=1,\dots,n\\k=1,\dots,m\end{subarray}}\right)=1
\end{equation}
if there exists $1\leq a\leq b$, $1\leq b\leq n$, $0\leq c\leq n-b$ such
that
\begin{eqnarray*} &x_a^b=x_{a+1}^{b+1}=\dots=x_{a+c}^{b+c}=x,\\
&y_a^b=y_{a+1}^{b+1}=\dots=y_{a+c}^{b+c}=x+1,
\end{eqnarray*}
and $x_k^m=y_k^m$ for all other values of $(k,m)$, and
\begin{equation}L^{(n)}\left(\{x_k^m\}_{\begin{subarray}{c}
m=1,\dots,n\\k=1,\dots,m\end{subarray}},\{y_k^m\}_{\begin{subarray}{c}
m=1,\dots,n\\k=1,\dots,m\end{subarray}}\right)=0
\end{equation}
in all other cases.

Less formally, this continuous time Markov chain can be described as
follows. Each of the particles $x_k^m$ has its own exponential
clock, all clocks are independent. When $x_a^b$-clock rings, the
particle checks if its jump by one to the right would violate the
interlacing condition. If no violation happens, that is, if
\begin{equation}x_a^b<x_a^{b-1}-1 \quad \text{and} \quad x_a^b<x_{a+1}^{b+1},
\end{equation}
then this jump takes place.
 If $x_a^b=x_a^{b-1}-1$ then the jump is blocked. On the other
hand, if $x_a^b=x_{a+1}^{b+1}$ then we find the longest string
$x_a^b=x_{a+1}^{b+1}=\dots=x_{a+c}^{b+c}$ and move all the particles
in this string to the right by one. One could think that the
particle $x_a^b$ has pushed the whole string.

We denote this continuous time Markov chain by $\P^{(n)}$.

Similarly to $P_\Lambda^{(n)}$, each of the Markov chains $P_m$ on
$\S_m$ also has a continuous limit as $\beta^-\to 0$. Indeed, the
transition probabilities of the Markov chain generated by
$T_m(\alpha_1,\dots,\alpha_m;1+\beta^-/z)$ converge to ($x^m,y^m\in
\S_m$)
\begin{multline}
\left(\lim_{\beta^-\to 0}
\bigl(T_m(\alpha_1,\dots,\alpha_m;1+\beta^-/z)\bigr)^{[\tau/\beta^-]}\right)
(x^m,y^m)\\
= \frac{\det{[\alpha_i^{y^m_j}]}_{i,j=1}^m}
{\det{[\alpha_i^{x^m_j}]}_{i,j=1}^m}\,\frac{\det{[\tau^{y_i^m-x_j^m} \Id(y_i^m-x_j^m\geq 0)/(y_i^m-x_j^m)!]}_{i,j=1}^m}
{\exp(\tau\sum_{j=1}^n \alpha_j)}.
\end{multline}
Thus, the limit of $P_m$ is the Doob $h$-transform of $m$
independent Poisson processes by the harmonic function
$h(x_1,\dots,x_m)=\det{[\alpha_i^{x_j}]}_{i,j=1}^m$, cf.
\cite{OCon03}. Let us denote this continuous
time Markov chain by $\P_{m}$, and the above matrix of its
transition probabilities over time $\tau$ by $\P_m(\tau)$.

Taking the same limit $\beta^-$ in Proposition~\ref{PropAB3} leads to the
following statement.

\begin{prop}\label{PropAB16} Let $m_n(x^n)$ be a probability measure on
$\S_n$. Consider the evolution of the measure
\begin{equation}m_n(x^n)\Lambda^n_{n-1}(x^n,x^{n-1})\cdots \Lambda^2_1(x^2,x^1)
\end{equation}
on $\S_\Lambda^{(n)}$ under the Markov chain $\P^{(n)}$, and denote
by $(x^1(t),\dots,x^n(t))$ the result after time $t\geq 0$. Then for
any
\begin{equation}0=t^{0}_n\le\dots\leq t^{c(n)}_n= t^{0}_{n-1}\le\dots\leq t_{c(n-1)}^{n-1}=t^0_{n-2}\leq \dots\leq t_2^{c(2)}=t_1^0\le\dots \leq t_1^{c(1)}
\end{equation}
(here $c(1),\dots,c(n)$ are arbitrary nonnegative integers) the
joint distribution of
\begin{multline*} x^n(t_n^0),\dots,x^n(t_n^{c(n)}),x^{n-1}(t_{n-1}^0),
x^{n-1}(t_{n-1}^1),\dots,x^{n-1}(t_{n-1}^{c(n-1)}),\\x^{n-2}(t_{n-2}^0),
\dots, x^2(t_{2}^{c(2)}),x^1(t_1^0),\dots,x^1(t_1^{c(1)})
\end{multline*}
coincides with the stochastic evolution of $m_n$ under transition
matrices
\begin{eqnarray*}
& &{\P_n(t_n^1-t_n^0),\dots,\P_n\bigl(t_n^{c(n)}-t_n^{c(n)-1}\bigr)},
\Lambda^n_{n-1},\\
& &{\P_{n-1}(t_{n-1}^1-t_{n-1}^0),\dots,
\P_{n-1}\bigl(t_{n-1}^{c(n-1)}-t_{n-1}^{c(n-1)-1}\bigr)},
\Lambda^{n-1}_{n-2},\dots,\\
& &\dots,
\Lambda^2_1,{\P_1(t_1^1-t_1^0),\dots,\P_1\bigl(t_1^{c(1)}-t_1^{c(1)-1}\bigr)}.
\end{eqnarray*}
\end{prop}

\begin{remark}\label{RemarkAB17}
It is not hard to see that if in the
construction of $P_\Lambda^{(n)}$ we used
$F_t(z)=(1-\gamma^-/z)^{-1}$ and took the limit $\gamma^-\to0$ then
the resulting continuous Markov chains would have been exactly the
same. On the other hand, if we used $F_t(z)=(1+\beta^+z)$ or
$F_t(z)=(1-\gamma^+z)^{-1}$ then the limiting continuous Markov
chain would have been similar to $\P^{(n)}$, but with particles
jumping to the left.

It is slightly technically harder to establish the convergence of
Markov chains with alternating steps, for example,
\begin{equation}
F_{2s}(z)=1+\beta^+(s)z,\qquad F_{2s+1}=1+\beta^-(s)/z,
\end{equation}
because the transition matrix is no longer triangular (particles
jump in both directions). It is possible to prove, however, the
following fact:

For any two continuous functions $a(\tau)$ and $b(\tau)$ on $\Bbb
R_+$ with \mbox{$a(0)=b(0)=0$}, consider the limit as $\epsilon\to 0$ of
the Markov chain $P_\Lambda^{(n)}$ with alternating $F_t$'s as
above,
\begin{equation}\beta^-(s)=\epsilon a(\epsilon s),\qquad \beta^+(s)=\epsilon
b(\epsilon s),
\end{equation}
and the time rescaled by $\epsilon$. Then this Markov chain
converges to a continuous time Markov chain, whose generator at time
$\tau$ is equal to $a(\tau)$ times the generator of $\P^{(n)}$ plus
$b(\tau)$ times the generator of the Markov chain similar to
$\P^{(n)}$ but with particles jumping to the left.

The statement of Proposition~\ref{PropAB16} also remains true, but in the
definition of the Markov chains $\P_m$ one needs to replace the
Poisson process by the one-dimensional process whose generator is
$a(\tau)$ times the generator of the Poisson process plus $b(\tau)$
times the generator of the Poisson process jumping to the left.
\end{remark}

\subsection{Determinantal structure of the correlation functions}\label{sect8DetStruct}

The goal of this section is to compute certain averages often called
correlation functions for the Markov chains $P_\Lambda^{(n)}$ and
$P_\Delta^{(n)}$ with $F_t(z)=(1+ \beta_t^\pm z^{\pm 1})$ or $(1-
\gamma_t^\pm z^{\pm 1})^{-1}$, and their continuous time counterpart
$\P^{(n)}$, starting from a certain specific initial condition.

As usual, we begin with $P_\Lambda^{(n)}$. The initial condition
that we will use is natural to call {\it densely packed initial
condition\/}. It is defined by
\begin{equation}
x_k^m(0)=k-m-1,\qquad k=1,\dots,m,\ m=1,\dots,n.
\end{equation}

\begin{defin}
For any $M\geq 1$, pick $M$ points
\begin{equation}
\varkappa_j=(y_j,m_j,t_j)\in \Z\times \{1,\dots,n\}\times\Z_{\geq 0}
\quad \text{or}\quad \Z\times \{1,\dots,n\}\times\Bbb R_{\geq 0},
\end{equation}
$j=1,\dots,M$. The value of the $M$th correlation function $\rho_M$
of $P_\Lambda^{(n)}$ (or $P_\Delta^{(n)}$) at
$(\varkappa_1,\dots,\varkappa_M)$ is defined as
\begin{multline}
\rho_M(\varkappa_1,\dots,\varkappa_M)=\operatorname{Prob}\{\text{For
each $j=1,\dots,M$ there exists a $k_j$,} \\ \text{$1\leq k_j\leq
m_j$, such that $x^{m_j}_{k_j}(t_j)=y_j$}\}.
\end{multline}
\end{defin}

The goal of this section is to partially evaluate the correlation
functions corresponding to the densely packed initial condition.

Introduce a partial order on pairs $(m,t)\in \{1,\dots,n\}\times
\Z_{\geq 0}$ or $\{1,\dots,n\}\times \Bbb R_{\geq 0}$ via
\begin{equation}
(m_1,t_1)\prec (m_2,t_2) \quad\text{iff} \quad m_1\leq m_2,\ t_1\geq
t_2\ \text{and}\ (m_1,t_1)\neq (m_2,t_2).
\end{equation}

In what follows we use positive numbers $\alpha_1,\dots,\alpha_n$
that specify the links $\Lambda^k_{k-1}$ as in Section~\ref{sect6ExampleMulti}, and as
before we assume that
\begin{equation}\beta^{\pm}_t,\gamma^{\pm}_t>0,\qquad
\gamma^+_t<\min\{\alpha_1,\dots,\alpha_n\},\quad
\gamma^-_t<\min\{\alpha_1^{-1},\dots,\alpha_n^{-1}\}.
\end{equation}

\begin{thm}\label{ThmAB18}
Consider the Markov chain $P_\Lambda^{(n)}$ with the densely packed
initial condition and $F_t(z)=(1+ \beta_t^\pm z^{\pm 1})$ or
$(1-\gamma_t^\pm z^{\pm 1})^{-1}$. Assume that triplets
$\varkappa_j=(y_j,m_j,t_j)$, $j=1,\dots,M$, are such that any two
distinct pairs $(m_j,t_j)$, $(m_{j'},t_{j'})$ are comparable with
respect to $\prec$. Then
\begin{equation}
\rho_M(\varkappa_1,\dots,\varkappa_M)=\det{[{\cal K}(\varkappa_i,\varkappa_j)]}_{i,j=1}^M,
\end{equation}
where
\begin{multline*}
{\cal K}(y_1,m_1,t_1;y_2,m_2,t_2)=-\frac 1{2\pi \I}\oint_{\Gamma_0}
\frac{\dx w}{w^{y_2-y_1+1}}\,\frac{\prod_{t=t_2}^{t_1-1}F_t(w)}
{\prod_{l=m_1+1}^{m_2} (1-\alpha_l w)}\,\Id_{[(m_1,t_1)\prec (m_2,t_2)]}\\
+\frac{1}{(2\pi \I)^2}\oint_{\Gamma_0}\dx w\oint_{\Gamma_{\alpha^{-1}}}\dx z\,
\frac{\prod_{t=0}^{t_1-1}F_t(w)}{\prod_{t=0}^{t_2-1}F_t(z)}
\frac{\prod_{l=1}^{m_1}(1-\alpha_lw)}{\prod_{l=1}^{m_2}(1-\alpha_lz)}
\frac{w^{y_1}}{z^{y_2+1}}\,\frac{1}{w-z}\,,
\end{multline*}
the contours $\Gamma_0$, $\Gamma_{\alpha^{-1}}$ are closed and
positively oriented, and they include the poles $0$ and
$\{\alpha_1^{-1},\dots,\alpha_n^{-1}\}$, respectively, and no other
poles.
\end{thm}

This statement obviously implies

\begin{cor}\label{corAB19}
For the Markov chain $\P^{(n)}$, with the
notation of Theorem~\ref{ThmAB18} and densely packed initial condition, the
correlation functions are given by the same determinantal formula
with the kernel
\begin{multline*}
{\cal K}(y_1,m_1,\tau_1;y_2,m_2,\tau_2)=-\frac 1{2\pi \I}\oint_{\Gamma_0} \frac{\dx w}{w^{y_2-y_1+1}}\,\frac{e^{(t_1-t_2)/w}}
{\prod_{l=m_1+1}^{m_2} (1-\alpha_l w)}\,\Id_{[(m_1,t_1)\prec (m_2,t_2)]}\\
+\frac{1}{(2\pi \I)^2}\oint_{\Gamma_0}\dx w\oint_{\Gamma_{\alpha^{-1}}}\dx z\,
\frac{e^{t_1/w}}{e^{t_2/z}}
\frac{\prod_{l=1}^{m_1}(1-\alpha_lw)}{\prod_{l=1}^{m_2}(1-\alpha_lz)}
\frac{w^{y_1}}{z^{y_2+1}}\,\frac{1}{w-z}\,.
\end{multline*}
\end{cor}

\begin{remark}For the more general continuous time Markov chain
described in Remark~\ref{RemarkAB17} a similar to Corollary~\ref{corAB19} result holds true,
where one needs to replace the function $e^{t/w}$ by
$e^{a(t)/w+b(t)w}$.
\end{remark}

\begin{proofOF}{Theorem~\ref{ThmAB18}} The starting point is
Proposition~\ref{PropAB3prime}. The densely packed initial condition
is a measure on $\S_\Lambda^{(n)}$ of the form
$m_n(x^n)\Lambda^n_{n-1}(x^n,x^{n-1})\cdots \Lambda^2_1(x^2,x^1)$
with $m_n$ being the delta-measure at the point
$(-n,-n+1,\dots,-1)\in \S_n$.

This delta-measure can be rewritten (up to a constant) as
$\det[\alpha_j^{x_i^n}]_{i,j=1,\ldots,n}\det[\Psi^n_{n-l}(x^n_{k})]_{k,l=1,\dots,n}$ with
\begin{equation}
\Psi^n_{n-l}(x)=\frac 1{2\pi \I}\oint_{\Gamma_0} \prod_{j=l+1}^n
(1-\alpha_j w)w^{x+l}\,\frac{\dx w}{w}\,, \qquad l=1,\dots,n.
\end{equation}
Indeed, $\operatorname{Span}(\Psi^n_{n-l}\mid l=1,\dots,n)$ is
exactly the space of all functions on $\Z$ supported by
$\{-1,\dots,-n\}$.

We are then in a position to apply Theorem 4.2 of~\cite{BF07}. For convenience of the reader, this theorem can be found in Appendix~\ref{Appendix}. (In fact, the change of notation that facilitates the application
was already used in Proposition~\ref{PropAB16} above.) The computation of the
matrix $M^{-1}$ of that theorem follows verbatim the computation in
the proof of Theorem 3.2 of \cite{BK07}, where $\theta_j$ of
\cite{BK07} have to be replaced by $\alpha_j^{-1}$ for all
$j=1,\dots,n$. Arguing exactly as in that proof we arrive at the
desired integral representation for the correlation kernel.
\end{proofOF}

Finally, one can also derive similar formulas for the Markov chain
$P_\Delta^{(n)}$. As the state space $\S_\Delta^{(n)}$ is now
\begin{equation}
\S_\Delta^{(n)}(t)=\{(x^n(t),x^{n-1}(t+1),\dots,x^1(t+n-1)\},
\end{equation}
we need to define the densely packed initial condition differently,
cf. the end of Section \ref{sect6ExampleMulti}. We set
\begin{equation}
x^{m}_k(n-m)=k-m-1, \qquad k=1,\dots,m,\ m=1,\dots,n,
\end{equation}
and assume that $F_t(z)\equiv 1$ for $t=0,\dots,n-2$. This means
that
\begin{equation}
\Delta_{m-1}^{m}(x^{m},x^{m-1}\mid
n-m)=\Lambda^{m}_{m-1}(x^{m},x^{m-1}), \qquad m=2,\dots,n,
\end{equation}
and our initial condition is of the form (\ref{icDelta}).

\begin{cor}\label{corAB20}
For the Markov chain $P^{(n)}_\Delta$, with the above assumptions,
notation of Theorem~\ref{ThmAB18}, and densely packed initial
condition, under the additional assumption that for any two pairs
$(m_j,t_j)\prec (m_{j'},t_{j'})$ we have
\begin{equation}
t_{j}-t_{j'}\geq m_{j'}-m_j,
\end{equation}
the correlation functions are given by the same determinantal
formula as in Theorem~\ref{ThmAB18}.
\end{cor}
\begin{proofOF}{Corollary \ref{corAB20}} Comparing the formulas for
the joint distributions for $P_\Lambda^{(n)}$ and $P_\Delta^{(n)}$
in Proposition~\ref{PropAB3prime} we see that with the densely
packed initial conditions they simply coincide. Hence, the
correlation functions are the same.
\end{proofOF}

Note that according to the remark at the end of Section
\ref{sect6ExampleMulti}, the correlation functions for the shuffling
algorithm of domino tilings of Aztec diamonds can be obtained from
Theorem~\ref{ThmAB18} and Corollary \ref{corAB20}.

\section{Geometry}\label{SectGeometry}

\subsection{Macroscopic behavior, limit shape}
It is more convenient for us to slightly modify the definition of
the height function (\ref{eqDefinHeight}) by assuming that its first
argument varies over $\Z$, and
\begin{equation}
h(x,n,t)=\left|\{k | x_k^n(t)> x\}\right|.
\end{equation}
Clearly, this modification has no effect on asymptotic statements.

We are interested in large time behavior of the interface. The macroscopic choice of variables is
\begin{equation}
x=[(\nu-\eta)L],\quad n=[\eta L],\quad t=\tau L,
\end{equation}
where $(\nu,\eta,\tau)\in \R_+^3$ and $L\gg 1$ is a large parameter
setting the macroscopic scale. For fixed $\eta$ and $\tau$,
$h(x,n,t)=n$ for $\nu$ small enough (e.g., $\nu=0$) and $h(x,n,t)=0$
for $\nu$ large enough. Define the $x$-density of our system as the
local average number of particles on unit length in the $x$-direction. Then, for large
$L$, one expects that $-L^{-1}\partial h/\partial \nu \simeq
x\textrm{-density}$. Thus, our model has facets when the $x$-density
is constant (equal to $0$ or $1$ in our situation), which are
interpolated by curved pieces of the surface, see
Figure~\ref{FigureSimulazione}.

\begin{claim}\label{Claim1}
The domain ${\cal D}\subset \R_+^3$, where the $x$-density of our system is asymptotically strictly between $0$ and $1$ is given by
\begin{equation}\label{eqClaim1}
|\sqrt{\tau}-\sqrt{\eta}|<\sqrt{\nu}<\sqrt{\tau}+\sqrt{\eta}.
\end{equation}
\end{claim}
Equivalently, $x\textrm{-density}\in (0,1)$ iff there exists a (non-degenerate) triangle with sides $\sqrt{\nu},\sqrt{\eta},\sqrt{\tau}$. Denote by $\pi_\nu$, $\pi_\eta$ and $\pi_\tau$ the angles of this triangle as indicated in Figure~\ref{FigGeometry}. Claim~\ref{Claim1} follows from Proposition~\ref{propBulkLimit} below.
\begin{figure}[h!]
\begin{center}
\psfrag{eta}{$\sqrt\eta$}
\psfrag{tau}{$\sqrt\tau$}
\psfrag{nu}{$\sqrt\nu$}
\psfrag{pieta}{$\pi_\eta$}
\psfrag{pitau}{$\pi_\tau$}
\psfrag{pinu}{$\pi_\nu$}
\psfrag{Omega}{$\Omega$}
\psfrag{0}[c]{$0$}
\psfrag{1}[c]{$1$}
\psfrag{r1}[r]{$\sqrt{\eta/\tau}$}
\psfrag{r2}[l]{$\sqrt{\nu/\tau}$}
\includegraphics[height=4cm]{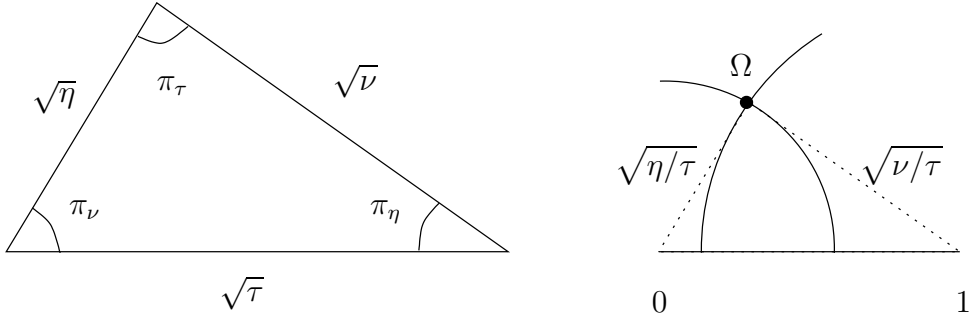}
\caption{The triangle of (\ref{eqClaim1}) on the left and its scaled version defined by intersection of circles on the right.}
\label{FigGeometry}
\end{center}
\end{figure}

The condition (\ref{eqClaim1}) is also equivalent to saying that the
circle centered at $0$ of radius $\sqrt{\eta/\tau}$ has two
disjoint intersections with the circle centered at $1$ of radius
$\sqrt{\nu/\tau}$. In that case, the two intersections are complex
conjugate. Denote by $\Omega(\nu,\eta,\tau)$ the intersection in
\begin{equation}
\H=\{z\in\C\, |\, \Im(z)> 0\}.
\end{equation}
Then, we have the following properties
\begin{equation}\label{eqGeom}
|\Omega|^2=\frac{\eta}{\tau},\quad |1-\Omega|^2=\frac{\nu}{\tau},\quad \arg(\Omega)=\pi_\nu,\quad \arg(1-\Omega)=-\pi_\eta.
\end{equation}
The cosine rule gives the angles $\pi_*$'s in $(0,\pi)$ by
\begin{eqnarray}\label{eqAngles}
\pi_\nu&=&\arccos\left(\frac{\tau+\eta-\nu}{2\sqrt{\tau\eta}}\right), \nonumber \\
\pi_\eta&=&\arccos\left(\frac{\tau+\nu-\eta}{2\sqrt{\tau\nu}}\right), \\
\pi_\tau&=&\arccos\left(\frac{\eta+\nu-\eta}{2\sqrt{\nu\eta}}\right). \nonumber
\end{eqnarray}

\begin{prop}[Bulk scaling limit]\label{propBulkLimit}
For any $k=1,2,\dots$, consider
\begin{equation}
\varkappa_j(L)=(x_j(L),n_j(L),t_j(L)),\qquad j=1,\dots,k,
\end{equation}
such that for any $i\ne j$ and any $L>0$ either
$(n_i(L),t_i(L))\prec(n_j(L),t_j(L))$ or
$(n_j(L),t_j(L))\prec(n_i(L),t_i(L))$ (the notation $\prec$ was defined in (\ref{eqPartialOrder})). Assume that
\begin{equation}
\lim_{L\to\infty} \frac {x_j}L=\nu,\qquad\lim_{L\to\infty} \frac
{n_j}L=\eta,\qquad\lim_{L\to\infty} \frac {t_j}L=\tau,\qquad
j=1,\dots,k;
\end{equation}
we have $(\nu,\eta,\tau)\in\mathcal{D}$; and also all the
differences $x_i-x_j$, $n_i-n_j$, $t_i-t_j$ do not depend on the
large parameter $L$. Then the $k$-point correlation function
$\rho_k(\varkappa_1,\ldots,\varkappa_k)$ converges to the
determinant $\det[K^{\rm bulk}_{ij}]_{1\leq i,j\leq k}$, where
\begin{equation}\label{eq3.9}
K^{\rm
bulk}_{i,j}=\frac{1}{2\pi\I}\int_{1-\Omega(\nu,\eta,\tau)}^{1-\overline
\Omega(\nu,\eta,\tau)} \dx w \frac{(1-w)^{n_i-n_j}
e^{(t_j-t_i)w}}{w^{x_i-x_j+1}},
\end{equation}
where for $(n_i,t_i)\not\prec(n_j,t_j)$ the integration contour
crosses $\R_+$, while for $(n_i,t_i)\prec (n_j,t_j)$ the contour
crosses $\R_-$. On the other hand, if $(\nu,\eta,\tau)\not\in \mathcal{D}$, then
\begin{equation}
\begin{aligned}
\lim_{L\to\infty}\rho_k(\varkappa_1,\ldots,\varkappa_k)=0, &\quad\textrm{if }\sqrt{\nu}>\sqrt{\eta}+\sqrt{\tau}\\
\lim_{L\to\infty}\rho_k(\varkappa_1,\ldots,\varkappa_k)=0, &\quad\textrm{if }\sqrt{\nu}<\sqrt{\tau}-\sqrt{\eta}\\
\lim_{L\to\infty}\rho_k(\varkappa_1,\ldots,\varkappa_k)=1, &\quad\textrm{if }\sqrt{\nu}<\sqrt{\eta}-\sqrt{\tau}.
\end{aligned}
\end{equation}
\end{prop}
\begin{proofOF}{Proposition~\ref{propBulkLimit}}
One follows exactly the same steps as in Section 3.2 of~\cite{Ok03},
replacing the double integral (35) in there by
(\ref{eqDoubleIntRepr}). The deformed paths are then like in
Figure~\ref{FigPrincValue} but with $z_c=w_c$. The degenerate cases when $\varkappa_j\not\in\mathcal{D}$ are treated in the same way with limiting kernel $K_{i,j}$ being either $0$ (no residue in the contour integral computation) or triangular ($K_{i,j}=0$ for $x_i<x_j$) with $K_{i,i}=1$, when the integral in (\ref{eq3.9}) is over a complete circle around the origin.
\end{proofOF}

\begin{cor}
Let $\rho$ denote the asymptotic $x$-density. Then, in $\cal D$, it is given by
\begin{equation}\label{EqDensity}
\rho(\nu,\eta,\tau)=\lim_{L\to\infty}\rho_1([\nu L],[\eta L],\tau L)=\pi_\eta/\pi \in [0,1].
\end{equation}
Consequently,
\begin{equation}\label{eqLimitShape}
\textit{\textbf{h}}(\nu,\eta,\tau):=\lim_{L\to\infty} \frac{\E h([(\nu-\eta)L],[\eta L],\tau L)}{L}=
\frac{1}{\pi}\int_{\nu}^{(\sqrt\tau+\sqrt\eta)^2}
\pi_\eta(\nu',\eta,\tau)\dx \nu'.
\end{equation}
\end{cor}

Below we perform the integral in (\ref{eqLimitShape}) to get an
explicit expression for the limit shape $\textit{\textbf{h}}$. Along the way we
derive some interesting geometric relations. First of all, $\textit{\textbf{h}}$
is homogeneous of degree one (since it is the scaling limit under same scaling in all
directions).
\begin{lem}\label{LemmaLimitShape}
For any $\alpha>0$,
\begin{equation}
\textit{\textbf{h}}(\alpha\nu,\alpha\eta,\alpha\tau)=\alpha \textit{\textbf{h}}(\nu,\eta,\tau),
\end{equation}
from which it follows
\begin{equation}\label{eqLS2}
\left(\nu\frac{\partial}{\partial \nu}+\eta\frac{\partial}{\partial \eta}+\tau\frac{\partial}{\partial \tau}\right) \textit{\textbf{h}}(\nu,\eta,\tau)=\textit{\textbf{h}}(\nu,\eta,\tau).
\end{equation}
\end{lem}
\begin{proofOF}{Lemma~\ref{LemmaLimitShape}}
It follows directly from the geometric property $\pi_\eta(\alpha\nu,\alpha\eta,\alpha\tau)=\pi_\eta(\nu,\eta,\tau)$.
\end{proofOF}
Therefore, we just need to compute the partial derivatives of $\textit{\textbf{h}}$, then the limit shape $\textit{\textbf{h}}$ will be determined by the l.h.s.\ of (\ref{eqLS2}).

\begin{prop}\label{PropLS}
The partial derivatives of the limit shape $\textit{\textbf{h}}$ are given by
\begin{equation}\label{eqSlopes}
\frac{\partial \textit{\textbf{h}}}{\partial \nu}=-\frac{\pi_\eta}{\pi}, \quad
\frac{\partial \textit{\textbf{h}}}{\partial \eta}=1-\frac{\pi_\nu}{\pi}, \quad
\frac{\partial \textit{\textbf{h}}}{\partial \tau}=\frac{\sin(\pi_\nu)\sin(\pi_\eta)}{\pi\sin(\pi_\tau)}.
\end{equation}
\end{prop}

\begin{remark}
Another expression for the growth velocity is
\begin{equation}
\frac{\partial \textit{\textbf{h}}}{\partial \tau}=\frac{1}{\pi}\Im\Omega(\nu,\eta,\tau).
\end{equation}
This can be understood using Proposition~\ref{propBulkLimit}. The macroscopic growth velocity is equal to the average flow of particles, $J$. It is computed in Section~\ref{SectCorrelations}, see (\ref{eq5.11}) with $Q=0$: $\E(J)=-\partial_2 K(x,n,t;x,n,t)$. Then, by (\ref{eq5.18}) we have $\E(J)=K(x,n,t;x+1,n,t)$. Finally, by Proposition~\ref{propBulkLimit} one gets $\E(J)=\Im\Omega/\pi$.
\end{remark}

As a corollary of Lemma~\ref{LemmaLimitShape} and Proposition~\ref{PropLS}, the limit shape is given as follows.
\begin{cor}\label{corLS}
For $(\nu,\eta,\tau)\in {\cal D}$, we have
\begin{equation}
\textit{\textbf{h}}(\nu,\eta,\tau)=\frac{1}{\pi}\left(-\nu \pi_\eta+\eta(\pi-\pi_\nu)+\tau \frac{\sin(\pi_\nu)\sin(\pi_\eta)}{\sin(\pi_\tau)}\right).
\end{equation}
\end{cor}

\begin{proofOF}{Proposition~\ref{PropLS}}
From (\ref{eqLimitShape}) we immediately have the first relation: $\partial \textit{\textbf{h}}/\partial \nu=-\pi_\eta/\pi$. In the derivatives of $\textit{\textbf{h}}$ with respect to $\tau$ and $\eta$ we have one term coming from the boundary term and one from the internal derivative. The boundary terms will actually be zero, since the density at the upper edge is zero. We need to compute
\begin{equation}
\frac{\partial \pi_\eta}{\partial \eta}=\frac{1}{\sqrt{4\eta\tau-(\nu-\eta-\tau)^2}},\quad
\frac{\partial \pi_\eta}{\partial \tau}=\frac{\nu-\eta-\tau}{2\tau\sqrt{4\eta\tau-(\nu-\eta-\tau)^2}}.
\end{equation}
Then, we apply the indefinite integrals
\begin{equation}
\int\frac{\dx x}{a^2-x^2}=\arcsin(x/|a|)+C,\quad \int\frac{x\dx x}{\sqrt{a^2-x^2}}=-\sqrt{a^2-x^2}+C.
\end{equation}
For the derivative with respect to $\eta$,
\begin{eqnarray}
\pi \frac{\partial \textit{\textbf{h}}}{\partial \eta} &=&\int_{\nu}^{(\sqrt{\eta}+\sqrt{\tau})^2}\frac{\partial \pi_\eta}{\partial \eta}\dx \nu'+(1+\sqrt{\tau/\eta})\pi_\eta\left((\sqrt{\eta}+\sqrt{\tau})^2,\eta,\tau\right)\nonumber \\
&=&\pi/2+\arcsin\left(\frac{\eta+\tau-\nu}{2\sqrt{\eta\tau}}\right) =\pi-\arccos\left(\frac{\eta+\tau-\nu}{2\sqrt{\eta\tau}}\right),
\end{eqnarray}
the latter being $\pi_\nu$. Finally,
\begin{eqnarray}
\pi \frac{\partial \textit{\textbf{h}}}{\partial \tau} &=&\int_{\nu}^{(\sqrt{\eta}+\sqrt{\tau})^2}\frac{\partial \pi_\eta}{\partial \tau}\dx \nu'+(1+\sqrt{\eta/\tau})\pi_\eta\left((\sqrt{\eta}+\sqrt{\tau})^2,\eta,\tau\right)\nonumber \\
&=&\frac{\sqrt{4\eta\tau-(\nu-\eta-\tau)^2}}{2\tau}=\sqrt{\eta/\tau}\sin(\pi_\nu),
\end{eqnarray}
and by the sinus theorem for the triangle of Figure~\ref{FigGeometry} we have $\sqrt\eta/\sqrt\tau=\sin(\pi_\eta)/\sin(\pi_\tau)$.
\end{proofOF}

\begin{figure}[t!]
 \subfigure[]{
 \psfrag{eta}[t]{$\eta$}
 \psfrag{nu}[t]{$\nu$}
 \psfrag{rho}[tc]{$\rho(\nu,\eta,1)$}
 \psfrag{0}[tc][0.6]{$0$}
 \psfrag{1}[tc][0.6]{$1$}
 \psfrag{2}[tc][0.6]{$2$}
 \psfrag{3}[tc][0.6]{$3$}
 \psfrag{4}[tc][0.6]{$4$}
 \psfrag{v0}[c][0.6]{$0$}
 \psfrag{v0.2}[c][0.6]{$0.2$}
 \psfrag{v0.4}[c][0.6]{$0.4$}
 \psfrag{v0.6}[c][0.6]{$0.6$}
 \psfrag{v0.8}[c][0.6]{$0.8$}
 \psfrag{v1}[c][0.6]{$1$}
 \hspace{-1em}\includegraphics[width=5.5cm,angle=-90]{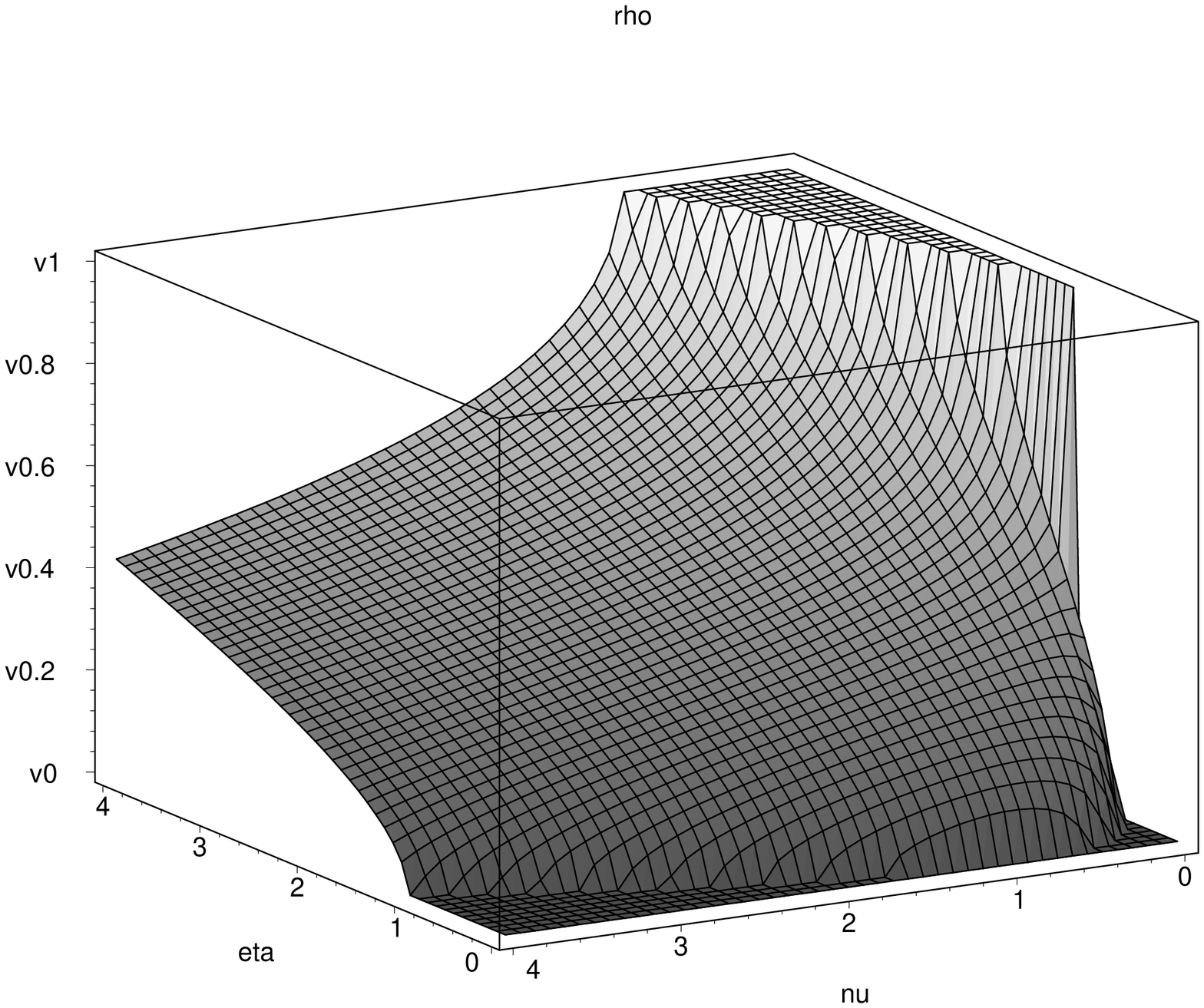}}
 \subfigure[]{
 \psfrag{eta}[t]{$\eta$}
 \psfrag{nu}[t]{$\nu$}
 \psfrag{h}[tc]{$\textit{\textbf{h}}(\nu,\eta,1)$}
 \psfrag{0}[tc][0.6]{$0$}
 \psfrag{1}[tc][0.6]{$1$}
 \psfrag{2}[tc][0.6]{$2$}
 \psfrag{3}[tc][0.6]{$3$}
 \psfrag{4}[tc][0.6]{$4$}
 \psfrag{h0}[tc][0.6]{$0$}
 \psfrag{h1}[tc][0.6]{$1$}
 \psfrag{h2}[tc][0.6]{$2$}
 \psfrag{h3}[tc][0.6]{$3$}
 \psfrag{h4}[tc][0.6]{$4$}
 \psfrag{v1}[c][0.6]{$1$}
 \psfrag{v0}[c][0.6]{$0$}
 \psfrag{v1}[c][0.6]{$1$}
 \psfrag{v2}[c][0.6]{$2$}
 \psfrag{v3}[c][0.6]{$3$}
 \psfrag{v4}[c][0.6]{$4$}
 \includegraphics[width=5.5cm,angle=-90]{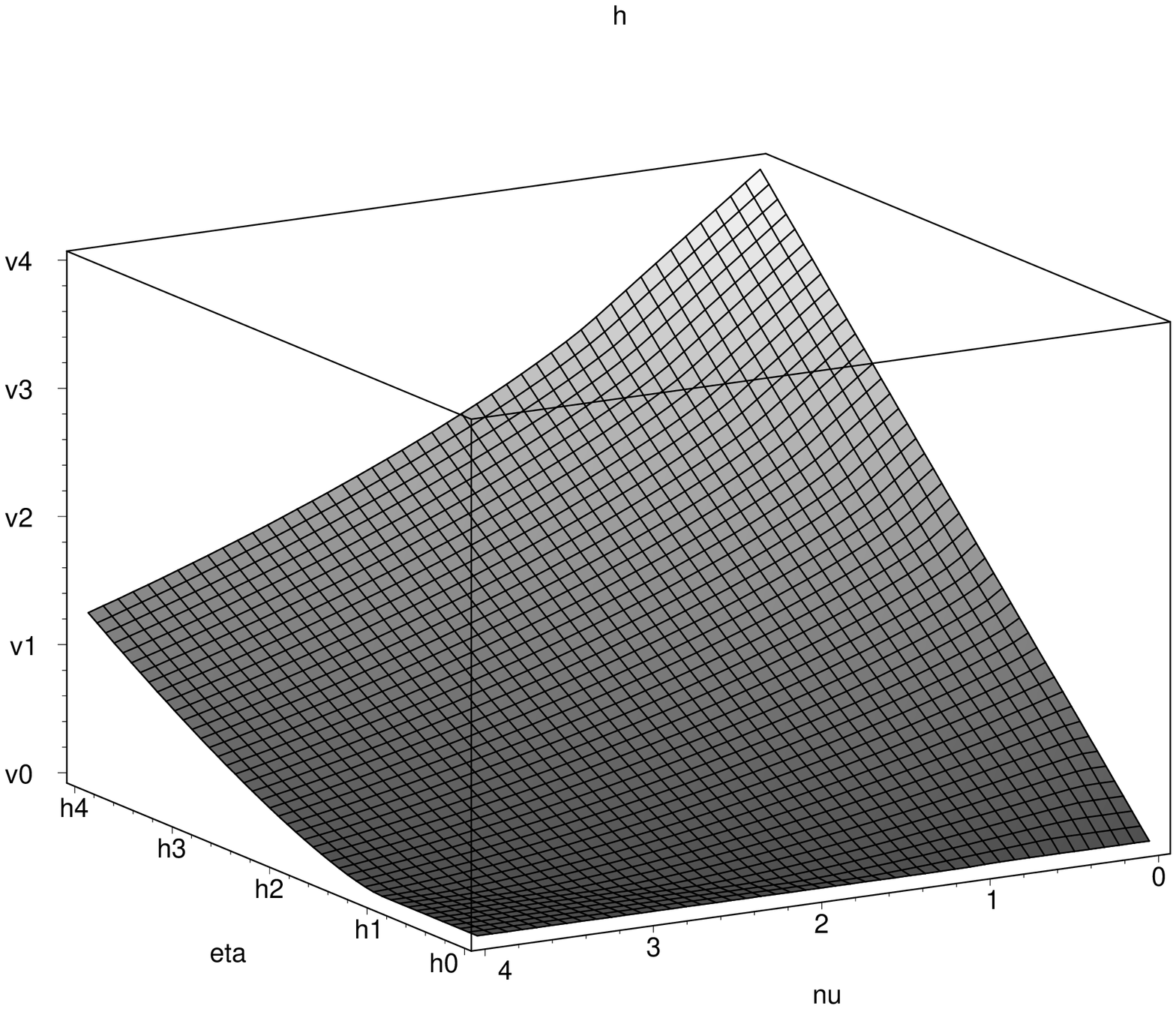}}
 \caption{(a) Limiting density of the particles with $\tau=1$.
 (b) The associated limiting height function. Two facets are visible. }\label{FigDensity}
\end{figure}

\subsection{Growth model in the anisotropic KPZ class}
For fixed $\tau$, the macroscopic slopes of the interface in the $\nu$- and $\eta$-directions are given by $\textit{\textbf{h}}_\nu :=\partial_\nu \textit{\textbf{h}}$ and $\textit{\textbf{h}}_\eta :=\partial_\eta \textit{\textbf{h}}$. The speed of growth of the surface, $\partial_\tau \textit{\textbf{h}}$, depends only on these two slopes. Indeed, by (\ref{eqSlopes}), we can rewrite
\begin{equation}
v=\frac{\partial \textit{\textbf{h}}}{\partial\tau}=-\frac{1}{\pi}\frac{\sin(\pi \textit{\textbf{h}}_\nu )\sin(\pi \textit{\textbf{h}}_\eta )}{\sin(\pi (\textit{\textbf{h}}_\nu +\textit{\textbf{h}}_\eta ))}.
\end{equation}
Remark that the speed of growth is monotonically decreasing with the slope
\begin{equation}
\frac{\partial v(\textit{\textbf{h}}_\nu ,\textit{\textbf{h}}_\eta )}{\partial \textit{\textbf{h}}_\nu }<0,\quad \frac{\partial v(\textit{\textbf{h}}_\nu ,\textit{\textbf{h}}_\eta )}{\partial \textit{\textbf{h}}_\eta }<0
\end{equation}
for $\textit{\textbf{h}}_\nu ,\textit{\textbf{h}}_\eta ,\textit{\textbf{h}}_\nu +\textit{\textbf{h}}_\eta \in (0,1)$.

To see which universality class our model belongs to, we need to compute the determinant of the Hessian of $v=v(\textit{\textbf{h}}_\nu ,\textit{\textbf{h}}_\eta )$. Explicit computations give
\begin{equation}
\left|\begin{array}{cc}
\partial_{\textit{\textbf{h}}_\nu }\partial_{\textit{\textbf{h}}_\nu }v & \partial_{\textit{\textbf{h}}_\nu }\partial_{\textit{\textbf{h}}_\eta }v \\ \partial_{\textit{\textbf{h}}_\eta }\partial_{\textit{\textbf{h}}_\nu }v & \partial_{\textit{\textbf{h}}_\eta }\partial_{\textit{\textbf{h}}_\eta }v
\end{array}\right| = -4\pi^2\frac{\sin(\pi \textit{\textbf{h}}_\nu )^2\sin(\pi \textit{\textbf{h}}_\eta )^2}{\sin(\pi(\textit{\textbf{h}}_\nu +\textit{\textbf{h}}_\eta ))^4} <0
\end{equation}
for $\textit{\textbf{h}}_\nu ,\textit{\textbf{h}}_\eta ,\textit{\textbf{h}}_\nu +\textit{\textbf{h}}_\eta \in (0,1)$, i.e., for $(\nu,\eta,\tau)\in {\cal D}$. Thus, our model belongs to the \emph{anisotropic} KPZ universality class of growth models in $2+1$ dimensions.

\subsection{A few other geometric properties}
During the asymptotic analysis we will use a few more geometric quantities, which we collect in this section. The key function to be analyzed is
\begin{equation}\label{eqDefinG}
G(w)\equiv G(w|\nu,\eta,\tau)=\tau w+\nu\ln(1-w)-\eta\ln(w), \quad w\in \C.
\end{equation}
The critical points of $G$ coincide with $\Omega$ as stated below.
\begin{prop}\label{PropGeom1}
On $\C\setminus\{0,1\}$, the function $G$ has two critical
points (counted with multiplicities). These two points are distinct
and complex conjugate if and only if $(\nu,\eta,\tau)\in {\cal D}$,
in which case the critical points are $\{\Omega,\overline \Omega\}$.
\end{prop}
\begin{proofOF}{Proposition~\ref{PropGeom1}}
The derivative of $G$ gives
\begin{equation}\label{eqGprime}
G'(w)=\frac{\tau}{w(w-1)}\left(\left(w-\frac{\eta+\tau-\nu}{2\tau}\right)^2+\frac{4\eta\tau-(\eta+\tau-\nu)^2}{4\tau^2}\right),
\end{equation}
and we have two distinct complex conjugate solutions iff $4\eta\tau-(\eta+\tau-\nu)^2>0$, i.e., iff $(\nu,\eta,\tau)\in {\cal D}$. Also, from (\ref{eqGeom}) and (\ref{eqAngles}) we get
\begin{equation}
\Re(\Omega)=\frac{\eta+\tau-\nu}{2\tau},\quad \Im(\Omega)=\frac{\sqrt{4\eta\tau-(\eta+\tau-\nu)^2}}{2\tau}.
\end{equation}
Thus, $\Omega$ and $\overline \Omega$ are the two solutions of $G'(w)=0$, i.e., the two critical points.
\end{proofOF}

The main formulas needed later are the partial derivatives of $\Omega$ as well as $G''(\Omega)$.
\begin{prop}\label{PropGeom2}
Denote $\kappa=2\tau\Im(\Omega)=\sqrt{4\eta\tau-(\eta+\tau-\nu)^2}$. Then we have
\begin{equation}
G''(\Omega)=\frac{-\I \kappa}{\Omega(1-\Omega)},
\end{equation}
which implies
\begin{equation}\label{eqG2}
|G''(\Omega)|=\frac{\kappa}{|\Omega(1-\Omega)|},\quad \arg(G''(\Omega))=-\frac{\pi}{2}-\pi_\nu+\pi_\eta.
\end{equation}
Moreover,
\begin{equation}
\frac{\partial \Omega}{\partial \nu}= \frac{\I \Omega}{\kappa}, \quad
\frac{\partial \Omega}{\partial \eta}= \frac{\I (1-\Omega)}{\kappa},\quad
\frac{\partial \Omega}{\partial \tau}= \frac{-\I \Omega(1-\Omega)}{\kappa}.
\end{equation}
\end{prop}
\begin{proofOF}{Proposition~\ref{PropGeom2}}
From (\ref{eqGprime}) we get
\begin{equation}
G''(\Omega)=\frac{2\tau}{\Omega(\Omega-1)}(\Omega-\Re(\Omega))=\frac{2\I\tau\Im(\Omega)}{\Omega(\Omega-1)}.
\end{equation}
The modulus is immediate, while the argument is obtained using (\ref{eqGeom}).

Since $\Omega$ is the intersection point of the circles $|z|=\sqrt{\eta/\tau}$ and \mbox{$|1-z|=\sqrt{\nu/\tau}$}, the direction of $\partial_\nu \Omega$ is orthogonal to the vector $\Omega$ and $\partial_\eta \Omega$ is orthogonal to $1-\Omega$. Therefore, for some $c_1,c_2 \in \R$,
\begin{equation}
\frac{\partial\Omega}{\partial\nu}=c_1 \Omega \I,\quad \frac{\partial\Omega}{\partial\eta}=c_2(1-\Omega)\I.
\end{equation}
Looking at the real part of these equations, we get $\partial_\nu\Re(\Omega)=-c_1 \Im(\Omega)$, and $\partial_\nu\Re(\Omega) = c_2 \Im(\Omega)$. On the other hand,
\begin{equation}
\Re(\Omega)=\frac{\eta+\tau-\nu}{2\tau}\quad\Rightarrow \quad \partial_\nu\Re(\Omega)=-\frac{1}{2\tau},\quad \partial_\eta\Re(\Omega)=\frac{1}{2\tau}.
\end{equation}
From this we conclude that
\begin{equation}
\partial_\nu \Omega=\frac{\I\Omega}{2\tau\Im(\Omega)},\quad \partial_\eta \Omega=\frac{\I(1-\Omega)}{2\tau\Im(\Omega)}.
\end{equation}
To get $\partial_\tau\Omega$, we can use the following property: $\Omega(a\nu,a\eta,a\tau)=\Omega(\nu,\eta,\tau)$ for any $a>0$, which implies
\begin{equation}
\left(\nu\partial_\nu+\eta\partial_\eta+\tau\partial_\tau\right)\Omega=0.
\end{equation}
This equation leads to
\begin{equation}
\partial_\tau \Omega=-\frac{\I}{2\tau\Im(\Omega)}\left(\frac{\nu}{\tau}\Omega+\frac{\eta}{\tau}(1-\Omega)\right) =-\frac{\I \Omega(1-\Omega)}{2\tau\Im(\Omega)},
\end{equation}
using $|\Omega|^2=\eta/\tau$ and $|1-\Omega|^2=\nu/\tau$, see (\ref{eqGeom}).
\end{proofOF}

Another important function appearing in the asymptotics of the kernel is the imaginary part of $G(\Omega)$ (and its derivatives).
\begin{prop}\label{propImPart}
We have
\begin{equation}\label{eqGIm}
\gamma(\nu,\eta,\tau):=\Im(G(\Omega))=\tau \Im(\Omega)-\nu\pi_\eta-\eta\pi_\nu,
\end{equation}
Its derivatives are
\begin{equation}
\frac{\partial \Im(G(\Omega))}{\partial \nu}=-\pi_\eta,\quad \frac{\partial \Im(G(\Omega))}{\partial \eta}=-\pi_\nu,
\end{equation}
and
\begin{equation}
\frac{\partial^2 \Im(G(\Omega))}{\partial \nu\partial\eta}=-\frac{1}{\kappa},\quad \kappa=2\tau\Im(\Omega).
\end{equation}
\end{prop}
\begin{proofOF}{Proposition~\ref{propImPart}}
The relation (\ref{eqGIm}) is a direct consequence of
(\ref{eqGeom}). The rest are just simple computations.
\end{proofOF}

\section{Gaussian fluctuations}\label{sectGaussFluct}
In this section we first state a couple of equivalent forms of the correlation kernel. In particular, the kernel for a fixed $(n,t)$ has a Christoffel-Darboux representation in terms of Charlier polynomials. In the second part of the section we prove Theorem~\ref{thmLogfluct} on the Gaussian fluctuations.

\subsection{Kernel representations}\label{SectKernelRepresentations}
For the analysis of the variance we will use a representation in terms of Charlier polynomials. These polynomials are defined on $\Z_+=\{0,1,2,\ldots\}$, while our particles at level $n$ live on $\{-n,-n+1,\ldots\}$. Thus, it is convenient to shift the position at level $n$ by $n$, i.e., the positions of particles at level $n$ will be denoted by $-n+x$, $x\geq 0$. Finally, we also conjugate by a factor $(-1)^{n_1-n_2}$. More precisely, the relation between the shifted and conjugate kernel $K$ and the kernel $\cal K$ in Theorem~\ref{ThmDetStructure}, is the following,
\begin{equation}\label{eq4.1}
K(x_1,n_1,t_1;x_2,n_2,t_2)=(-1)^{n_1-n_2} {\cal K}(x_1-n_1,n_1,t_1;x_2-n_2,n_2,t_2).
\end{equation}

For later use, we give the explicit double integral representation of $K$ which will be used in the asymptotic analysis.
\begin{cor}\label{CorKernelDoubleInt}
The extended kernel $K$ can be expressed as
\begin{multline}\label{eqDoubleIntRepr}
K(x_1,n_1,t_1;x_2,n_2,t_2)\\
= \begin{cases}
\frac{e^{t_1-t_2}}{(2\pi\I)^2}\oint_{\Gamma_1}\dx z \oint_{\Gamma_0}\dx w \frac{z^{n_1}}{e^{t_1 z} (1-z)^{x_1+1}} \frac{e^{t_2 w} (1-w)^{x_2}}{w^{n_2}} \frac{1}{w-z}, & (n_1,t_1)\not\prec (n_2,t_2)\\[0.5em]
\frac{e^{t_1-t_2}}{(2\pi\I)^2}\oint_{\Gamma_{1}}\dx z \oint_{\Gamma_{0,z}}\dx w \frac{z^{n_1}}{e^{t_1 z} (1-z)^{x_1+1}} \frac{e^{t_2 w} (1-w)^{x_2}}{w^{n_2}} \frac{1}{w-z}, & (n_1,t_1)\prec (n_2,t_2)
\end{cases}
\end{multline}
\end{cor}
\begin{proofOF}{Corollary~\ref{CorKernelDoubleInt}}
The kernel (\ref{eqDoubleIntRepr}) is obtained by substituting into (\ref{eq4.1}) the expression for $\cal K$ from (\ref{StartingKernel}), and applying the change of variables $z\to 1/(1-w)$ and $w\to 1/(1-z)$.
\end{proofOF}

It is instructive to see the structure of the kernel that leads the above integral representation.
\begin{prop}\label{PropExtendenKernel}
The extended kernel $K$ is given by
\begin{equation}\label{eq1.10}
K(x_1,n_1,t_1;x_1,n_2,t_2)=-\phi^{((n_1,t_1),(n_2,t_2))}(x_1,x_2)+\sum_{k=1}^{n_2}\Psi^{n_1,t_1}_{n_1-k}(x_1)\Phi^{n_2,t_2}_{n_2-k}(x_2)
\end{equation}
with
\begin{eqnarray}\label{eq1.11}
\Psi^{n,t}_k(x)&=&\frac{1}{2\pi\I}\oint_{\Gamma_{0,1}}\dx w \frac{e^{tw}(1-w)^k}{w^{x+1}},\nonumber \\
\Phi^{n,t}_k(x)&=&\frac{-1}{2\pi\I}\oint_{\Gamma_1}\dx z \frac{z^x e^{-tz}}{(1-z)^{k+1}}, \\
\phi^{((n_1,t_1),(n_2,t_2))}(x_1,x_2) &=& \frac{1}{2\pi\I}\oint_{\Gamma_{0,1}}\dx w \frac{e^{w(t_1-t_2)}}{w^{x_1-x_2+1}(w-1)^{n_2-n_1}}\Id_{[(n_1,t_1)\prec (n_2,t_2)]}, \nonumber
\end{eqnarray}
where $\Gamma_{0,1}$ and $\Gamma_1$ are any simple anticlockwise oriented contours that include poles $\{0,1\}$ and $\{1\}$, respectively.
\end{prop}
\begin{proofOF}{Proposition~\ref{PropExtendenKernel}}
Using the integral representations for $\Psi$ and $\Phi$ one checks that
\begin{equation}
\sum_{k\geq 0} \Psi^{n_1,t_1}_k(x)\Phi^{n_2,t_2}_k(y)=\phi^{((n_1,t_1),(n_2,t_2))}(x,y).
\end{equation}
Thus (\ref{eq1.10}) becomes
\begin{equation}\label{eqExtKernel}
K(x_1,n_1,t_1;x_2,n_2,t_2)= \left\{\begin{array}{ll}
\sum_{k=1}^{n_2}\Psi^{n_1,t_1}_{n_1-k}(x_1)\Phi^{n_2,t_2}_{n_2-k}(x_2), & (n_1,t_1)\not\prec (n_2,t_2)\\[0.5em]
-\sum_{l=0}^{\infty}\Psi^{n_1,t_1}_{n_1+l}(x_1)\Phi^{n_2,t_2}_{n_2+l}(x_2), & (n_1,t_1)\prec (n_2,t_2)
\end{array}\right.
\end{equation}
This new expression is good because in (\ref{eq1.11}) we never have the case when the pole at $w=1$ in $\Psi^{n,t}_k$ survives. Then, one has just to take the sums inside the integral. For example, for $(n_1,t_1)\not\prec (n_2,t_2)$, we first take the sum inside the integrals and then we extend it to $k=\infty$. This can be done provided $|1-w|>|1-z|$. Then, to get the formula (\ref{eqDoubleIntRepr}), one just has to rename the variables $z\to 1-w$ and $w\to 1-z$.
\end{proofOF}

For the computation of the variance, we will need only the kernel at fixed $(n,t)$. It is given in terms of the Charlier polynomials, $C_n(x,t)$, given by
\begin{equation}\label{eq5.1}
C_n(x,t)= \frac{n!}{t^n}\frac{1}{2\pi\I}\oint_{\Gamma_0}\dx w \frac{(1-w)^x e^{wt}}{w^{n+1}},
\end{equation}
which satisfy $C_n(x,t)=C_x(n,t)$, and are orthogonal with respect to the weight $w_t(x)=\frac{e^{-t}t^x}{x!}$, namely
\begin{equation}\label{eq5.2}
\sum_{x\geq 0} C_n(x,t) C_m(x,t) w_t(x)=\frac{n!}{t^n}\delta_{n,m},\quad t>0.
\end{equation}

\begin{cor}\label{CorKernelCharlier}
The kernel $K(x,n,t;y,n,t)$ is equivalent (conjugate) to the kernel $K_{n,t}(x,y)$, given by
\begin{equation}\label{eqKernelCharlier}
K_{n,t}(x,y)=\sqrt{n t} \frac{q_{n-1}(x,t) q_{n}(y,t)-q_{n}(x,t) q_{n-1}(y,t)}{x-y},
\end{equation}
where
\begin{equation} \label{eqCharlier}
q_n(x,t) =w_t(x)^{1/2} \frac{t^{n/2}}{\sqrt{n!}} C_n(x,t),
\end{equation}
\end{cor}
\begin{proofOF}{Corollary~\ref{CorKernelCharlier}}
Consider $n_1=n_2=n$ and $t_1=t_2=t$ in (\ref{eq1.10}). Then,
\begin{equation}
K(x,n,t;y,n,t)=\sum_{k=0}^{n-1} \Psi^{n,t}_k(x)\Phi^{n,t}_k(y).
\end{equation}
For all $k\geq 0$, $w=1$ is not a pole in the integral representation of $\Psi^{n,t}_k$. Using (\ref{eq5.1}) and $C_n(x,t)=C_x(n,t)$, we get $\Psi^{n,t}_k(x)=\frac{t^x}{x!} C_k(x,t)$. Also, by the change of variable $z\to 1-w$ in the integral representation $\Phi^{n,t}_k$ we obtain $\Phi^{n,t}_k(x)=e^{-t}\frac{t^k}{k!} C_k(x,t)$. Thus the kernel writes
\begin{equation}
K(x,n,t;y,n,t)=w_t(x) \sum_{k=0}^{n-1} \frac{t^k}{k!} C_k(x,t) C_k(y,t),
\end{equation}
which is conjugate to the kernel
\begin{equation}
K_{n,t}(x,y)=\sum_{k=0}^{n-1} q_k(x,t) q_k(y,t).
\end{equation}
From $C_n(x,t)=u_n x^n+\cdots$ with $u_n =1/(-t)^n$, we have \mbox{$q_n(x,t)=v_n x^n+\cdots$} with $v_n=(-1)^n/\sqrt{t^n n!}$. Then, (\ref{eqKernelCharlier}) follows from Christoffel-Darboux formula.
\end{proofOF}

\begin{remark}
For later use, we rewrite $q_n$ as
\begin{equation}\label{eqQn}
q_n(x,t)=B_{n,t}(x) I_{n,t}(x),\quad B_{n,t}(x)=\frac{e^{-t/2} t^{x/2}}{\sqrt{x!}}\frac{\sqrt{n!}}{t^{n/2}},
\end{equation}
and
\begin{equation}\label{eqIn}
I_{n,t}(x)=\frac{1}{2\pi\I}\oint_{\Gamma_0}\dx w \frac{(1-w)^x e^{wt}}{w^{n+1}}.
\end{equation}
\end{remark}

\subsection{Proof of Theorem~\ref{thmLogfluct}}
In this section we look only at the height function at a given time. Therefore, it is convenient to set $\lambda=\nu/\tau$ and $c=\eta/\tau$ so that we have $n=[\eta L]= [ct]$ and $x=[\nu L]=[\lambda t]$. In these variables, the equation for the bulk region given by (\ref{eqClaim1}) rewrites as
\begin{equation}
(1-\sqrt{c})^2 < \lambda < (1+\sqrt{c})^2.
\end{equation}

First we compute the variance of the height.
\begin{prop}\label{PropVariance}
For any $\lambda\in ((1-\sqrt{c})^2,(\sqrt{c}+1)^2)$,
\begin{equation}\label{eqPropVar}
\lim_{t\to\infty}\frac{\Var(h([(\lambda-c)t],[ct],t))}{\ln(t)}=\frac{1}{2\pi^2}.
\end{equation}
\end{prop}
With this we can prove Theorem~\ref{thmLogfluct}.
\begin{proofOF}{Theorem~\ref{thmLogfluct}}
It is a consequence of Proposition~\ref{PropVariance}
and~\cite{Sos04}. More precisely, in Section 2 of~\cite{Sos04} the
convergence in distribution (a generalization of the result for the
sine kernel of~\cite{CL95}) is stated. However, following the proof
of the theorem, one realizes that it is done by controlling the
cumulants, i.e., also the moments converge.
\end{proofOF}

\begin{proofOF}{Proposition~\ref{PropVariance}}
The variance can be written in terms of the one and two point correlation functions $\rho_1$ and $\rho_2$. Namely,
\begin{equation}
\Var(h([(\lambda-c)t],[ct],t))=\sum_{x,y>[\lambda t]} \rho_2(x,y)+\sum_{x>[\lambda t]}\rho_1(x)-\Big(\sum_{x>[\lambda t]}\rho_1(x)\Big)^2,
\end{equation}
where $\rho_2(x,y)=K_{n,t}(x,x)K_{n,t}(y,y)-K_{n,t}(x,y)K_{n,t}(y,x)$ and $\rho_1(x)=K_{n,t}(x,x)$. Using $K_{n,t}^2=K_{n,t}$ on $\ell_2(\Z_+)$, we have
\begin{equation}\label{eqVariance}
\begin{aligned}
&\Var(h([(\lambda-c)t],[ct],t))= \sum_{x>[\lambda t]} K_{n,t}(x,x)-\sum_{x,y>[\lambda t]} K_{n,t}(x,y)K_{n,t}(y,x)\\
&= \sum_{x>[\lambda t]} \sum_{y=0}^{\infty} K_{n,t}(x,y)K_{n,t}(y,x)-\sum_{x,y>[\lambda t]} K_{n,t}(x,y)K_{n,t}(y,x) \\ &=\sum_{x>[\lambda t]} \sum_{y\leq [\lambda t]} \left(K_{n,t}(x,y)\right)^2,\qquad n=[ct].
\end{aligned}
\end{equation}

We use the expression (\ref{eqKernelCharlier}) for the kernel $K_{n,t}$. We decompose the sum in (\ref{eqVariance}) into the following three sets:
\begin{eqnarray}
M&=&\{x,y\in \Z_+^2 | x>[\lambda t], y\leq [\lambda t], y-x\leq \e_1 t\}, \nonumber \\
R_1&=&\{x,y \in \Z_+^2| x>[\lambda t], y\leq [\lambda t], \e_1 t< y-x< \e_2 t\}, \\
R_2&=&\{x,y \in \Z_+^2| x>[\lambda t], y\leq [\lambda t], \e_2 t\leq y-x\},\nonumber
\end{eqnarray}
where the parameter $\e_2=\frac12 \min\{(1+\sqrt{c})^2-\lambda,\lambda-(1-\sqrt{c})^2\}$ is chosen so that $R_1$ is a subset of the bulk. Thus
\begin{equation}
\Var(h([(\lambda-c)t],[ct],t))=M_t+R_{t,1}+R_{t,2},
\end{equation}
with
\begin{equation}
M_t=\sum_{x,y\in M} \left|K_{n,t}(x,y)\right|^2,\quad R_{t,k}=\sum_{x,y\in R_k} \left|K_{n,t}(x,y)\right|^2.
\end{equation}

\noindent\emph{Remark:} The parameter $\e_1$, small, will be chosen $t$-dependent in the end.

\medskip

\noindent \textbf{(1) Bound on $R_{t,2}$.} For $x,y\in R_2$, we use
$y-x\geq \e_2 t$, and extend the sum to infinities
\begin{eqnarray}
R_{t,2}&\leq& \frac{1}{\e_2^2} \sum_{x\geq \lambda t}\sum_{y\leq \lambda t} \big(|q_{[ct]}(x,t)|^2 |q_{[ct]-1}(y,t)|^2+ |q_{[ct]-1}(x,t)|^2 |q_{[ct]}(y,t)|^2| \nonumber \\
& &+2 |q_{[ct]-1}(x,t)q_{[ct]}(x,t)| |q_{[ct]-1}(y,t)q_{[ct]}(y,t)|\big) \leq \frac{4}{\e_2^2}.
\end{eqnarray}
The last inequality follows from Cauchy-Schwarz and the property
\begin{equation}\label{eqCompleteness}
\sum_{x\geq 0}|q_k(x,t)|^2=\langle \Psi^{n,t}_k,\Phi^{n,t}_k\rangle=1,\quad \textrm{for all }k.
\end{equation}

\medskip

\noindent \textbf{(2) Bound on $R_{t,1}$.} Since this time $x,y\in R_1$ are always in the bulk, we just use the bound of Lemma~\ref{Lemma6} and get
\begin{eqnarray}
R_{t,1} &\leq & \cte \sum_{x,y\in R_1} \frac{1}{(x-y)^2} =\cte \sum_{z=[\e_1 t]}^{[\e_2 t]} \frac{1}{z}\nonumber \\
&=& \Psi([\e_2 t]+1)-\Psi([\e_1 t]),
\end{eqnarray}
where $\Psi(x)$ is the \emph{digamma} function, which has the Taylor expansion at infinity given by
\begin{equation}
\Psi(x)=\ln(x)-1/(2x)+\Or(1/x^2).
\end{equation}
Thus
\begin{equation}
R_{t,1} \leq \cte \ln(1/\e_1),
\end{equation}
with $\cte$ $t$-independent, as long as $z,t\to\infty$ as $t\to\infty$.

\medskip

\noindent \textbf{(3) Limit value for $M_t$.} This time we need more than just a bound. Recall that $n=ct$ and set $x=[\lambda t]+\xi_1$, $y=[\lambda t]-\xi_2$. We have $1\leq \xi_1+\xi_2\leq \e_1 t$. Lemma~\ref{Lemma4} gives
\begin{multline}\label{eq3.11}
q_{[ct]-\ell}(\lambda t+\xi,t)=\frac{1}{\sqrt{\pi}} \frac{t^{-1/2}}{\sqrt[4]{c-\frac{(1+c-\lambda)^2}{4}}} \Bigg[\Or(t^{-1/2})+\Or(\e_1)\\
 + \cos\Big[t\alpha(c,\lambda+\xi/t)+\beta(c,\lambda)-\ell \partial_{c}\alpha(c,\lambda)\Big]\Bigg].
\end{multline}
We use it with $\ell=0,1$, together with the trigonometric identity
\begin{equation}
\cos(b_1+\delta)\cos(b_2)-\cos(b_1)\cos(b_2+\delta)=\sin(\delta)\sin(b_2-b_1),
\end{equation}
with $\delta=-\partial_c\alpha(c,\lambda)$, $b_1=t\alpha(c,\lambda+\xi_1/t)+\beta(c,\lambda)$, $b_2=t\alpha(c,\lambda-\xi_1/t)+\beta(c,\lambda)$. The factor $\sin^2(\delta)$ cancels the $\sqrt[4]{\cdots}$ term exactly. We obtain (using (\ref{eqKernelCharlier}))
\begin{multline}
M_t=\sum_{\xi_1=1}^{[\e_1 t]}\sum_{\xi_2=0}^{\xi_1-1} \frac{1}{\pi^2}\frac{1}{(\xi_1+\xi_2)^2}\Bigg[\Or(t^{-1/2})+\Or(\e_1) \\
 +\sin^2\Big[t(\alpha(c,\lambda-\xi_2/t)-\alpha(c,\lambda+\xi_1/t))\Big]\Bigg]
\end{multline}
The contribution of the error terms can be bounded by $\ln(\e_1 t)\Or(t^{-1/2},\e_1)$, and the remainder is
\begin{equation}\label{eq1.56}
\sum_{\xi_1=1}^{[\e_1 t]}\sum_{\xi_2=0}^{\xi_1-1} \frac{1}{\pi^2}\frac{1}{(\xi_1+\xi_2)^2}\sin^2\Big[t(\alpha(c,\lambda-\xi_2/t)-\alpha(c,\lambda+\xi_1/t))\Big].
\end{equation}
Let $b(\lambda)=-\alpha(c,\lambda)$, then
\begin{equation}
b'(\lambda)=\arccos\left(\frac{1+\lambda-c}{2\sqrt{\lambda}}\right) \in (0,\pi),\quad \textrm{for }(1-\sqrt{c})^2< \lambda < (1+\sqrt{c})^2.
\end{equation}
By Lemma~\ref{LemmaSinus} below (one needs to shift the argument of $b(\lambda\pm \xi_j/t)$ by $\lambda$ to apply it), for large $t$ the leading term in the sum is identical to the one where $\sin^2(\cdots)$ is replaced by its mean, i.e., $1/2$. Thus
\begin{eqnarray}
(\ref{eq1.56})&=&(1+\Or(\e_1,(\e_1^2 \sqrt{t})^{-1})\sum_{\xi_1=1}^{[\e_1 t]}\sum_{\xi_2=0}^{\xi_1-1} \frac{1}{2\pi^2}\frac{1}{(\xi_1+\xi_2)^2}\nonumber \\
&=& \frac{1}{2\pi^2}\ln(\e_1 t)(1+\Or(\e_1,(\e_1^2 \sqrt{t})^{-1})).
\end{eqnarray}
Thus,
\begin{equation}
M_t=\ln(\e_1 t) \left(\frac{1}{2\pi^2}+\Or(t^{-1/2},\e_1,(\e_1^2 \sqrt{t})^{-1})\right).
\end{equation}
Now we choose $\e_1=1/\ln(t)$. Then,
\begin{equation}
\Var(h([(\lambda-c)t],[ct],t))= \frac{1}{2\pi^2}\ln(t) + \Or\big(1,\ln(\ln(t)), (\ln(t))^3/\sqrt{t}\big),
\end{equation}
which implies (\ref{eqPropVar}). Modulo Lemma~\ref{LemmaSinus}, the proof of Theorem~\ref{thmLogfluct} is complete.
\end{proofOF}

\begin{lem}\label{LemmaSinus}
Let $b(x)$ be a smooth function ($C^2$ is enough) on a neighborhood of the origin with $b'(0)\in (0,\pi)$. Then
\begin{equation}\label{eq1.62}
\sum_{\xi_1=1}^{[\e t]}\sum_{\xi_2=0}^{\xi_1-1} \frac{\sin^2\left[t b(\xi_1/t)-t b(-\xi_2/t)\right]}{(\xi_1+\xi_2)^2} = \sum_{\xi_1=1}^{[\e t]}\sum_{\xi_2=0}^{[\e t]-1} \frac{1}{2(\xi_1+\xi_2)^2} \left(1+\Or\Big(\e, \frac{1}{\e^2 \sqrt{t}}\Big)\right)
\end{equation}
uniformly for $\e>0$ small enough.
\end{lem}
\begin{proofOF}{Lemma~\ref{LemmaSinus}}
We divide the sum into two regions
\begin{eqnarray}
I_1&=&\{\xi_1\geq 1,\xi_2\geq 0 | 1\leq \xi_1+\xi_2 \leq \e \sqrt{t}\},\\
I_2&=&\{\xi_1\geq 1,\xi_2\geq 0 | \e\sqrt{t} < \xi_1+\xi_2 \leq \e t\}. \nonumber
\end{eqnarray}
Let us evaluate the contribution to (\ref{eq1.62}) of $(\xi_1,\xi_2)\in I_1$. We set $z=\xi_1+\xi_2$ and get
\begin{equation}\label{eq1.65}
\sum_{z=1}^{[\e\sqrt{t}]}\sum_{\xi_1=1}^{z}\frac{1}{z^2}\sin^2\left[t b(\xi_1/t)-t b((\xi_1-z)/t)\right].
\end{equation}
Taylor expansion around zero leads to
\begin{equation}
t b(\xi_1/t)-t b((\xi_1-z)/t) = z b'(0)+\Or(\e^2).
\end{equation}
Thus
\begin{equation}\label{eq1.67}
(\ref{eq1.65}) = \sum_{z}^{[\e\sqrt{t}]}\frac{1}{z}\left(\sin^2\left[z b'(0)\right]+\Or(\e^2)\right).
\end{equation}
The sum with the sine squared can be explicitly evaluated:
\begin{equation}
\sum_{z=1}^P\frac{\sin^2(\sigma z)}{z}=\frac12\ln(P)+\Or(1),\quad\textrm{ as } P\to\infty
\end{equation}
provided $0<\sigma<\pi$. Since $\sum_{z=1}^P 1/z=\ln(P)/2+\Or(1/P)$, we have
\begin{equation}
\sum_{z=1}^P\frac{\sin^2(\sigma z)}{z}=\sum_{z=1}^P\frac{1}{2z}\left(1+\Or(1/P)\right).
\end{equation}
Using $P=[\e\sqrt{t}]$ and going back to the original variables $(\xi_1,\xi_2)$ we have
\begin{equation}\label{eq1.69}
\sum_{(\xi_1,\xi_2)\in I_1} \frac{\sin^2\left[t b(\xi_1/t)-t b(-\xi_2/t)\right]}{(\xi_1+\xi_2)^2} =\sum_{(\xi_1,\xi_2)\in I_1}\frac{1}{2(\xi_1+\xi_2)^2}\left(1+\Or\Big(\frac{1}{\e\sqrt{t}},\e^2\Big)\right).
\end{equation}

Now we evaluate the contribution to (\ref{eq1.62}) of $(\xi_1,\xi_2)\in I_2$. Let \mbox{$(X,Y)\in I_2$}, then we have $X+Y\geq \e\sqrt{t}$. We consider a neighborhood of size $M=[\e^2\sqrt{t}]$ around $(X,Y)$, namely the contribution
\begin{equation}\label{eq1.70}
\sum_{x,y=0}^M\frac{1}{(X+Y+x+y)^2}\sin^2\left[t b((X+x)/t)-t b(-(Y+y)/t)\right].
\end{equation}
Since $\sin^2(\cdots)\geq 0$ and $\frac{1}{(X+Y)^2}-\frac{1}{(X+Y+x+y)^2}\geq 0$, if we replace $\frac{1}{(X+Y+x+y)^2}$ by $\frac{1}{(X+Y)^2}$ in (\ref{eq1.70}) the error made is bounded by
\begin{eqnarray}\label{eq1.71}
& &\sum_{x,y=0}^M\left(\frac{1}{(X+Y)^2}-\frac{1}{(X+Y+x+y)^2}\right) \\
&=&\sum_{x,y=0}^M\frac{1}{(X+Y)^2}\left(1-\frac{1}{(1+\Or(\e))^2}\right)
= \sum_{x,y=0}^M\frac{1}{(X+Y)^2}\Or(\e), \nonumber
\end{eqnarray}
because $(x+y)/(X+Y)\leq 2\e$. This relation can be inverted and we also get
\begin{equation}\label{eq1.72}
\sum_{x,y=0}^M\frac{1}{(X+Y)^2}=\sum_{x,y=0}^M\frac{1}{(X+Y+x+y)^2}\left(1+\Or(\e)\right).
\end{equation}
Therefore we have
\begin{equation}\label{eq1.73}
(\ref{eq1.70})= \sum_{x,y=0}^M\frac{\Or(\e)}{(X+Y+x+y)^2}+\sum_{x,y=0}^M\frac{\sin^2\left[t b((X+x)/t)-t b(-(Y+y)/t)\right]}{(X+Y)^2}.
\end{equation}

Now we apply Taylor expansion to the argument in the sine squared. Denote $\kappa_1=t b(X/t)-t b(-Y/t)$, $\theta_1=b'(X/t)$ and $\theta_2=b'(-Y/t)$. Then the argument in the $\sin^2(\cdots)$ is $\kappa_1+\theta_1 x+\theta_2 y+\Or(\e^2)$. The $\e^2$ error term is smaller than the $\Or(\e)$ in (\ref{eq1.73}), thus
\begin{equation}\label{eq1.74}
(\ref{eq1.73})=\sum_{x,y=0}^M\frac{\Or(\e)}{(X+Y+x+y)^2}+
\sum_{x,y=0}^M\frac{\sin^2\left[\kappa_1+\theta_1 x+\theta_2 y\right]}{(X+Y)^2}.
\end{equation}
Since $b$ is smooth and $b'(0)\in (0,\pi)$, in a neighborhood of $0$ we also have $b'\in(0,\pi)$. Thus, for $\e$ small enough, $0<\theta_1,\theta_2<\pi$ uniformly in $t$, because $|Y|/t\leq \e$ and $|X|/t\leq \e$.
The second sum in (\ref{eq1.74}) can be computed explicitly. For $0<\theta_1,\theta_2<\pi$ we have the identity
\begin{eqnarray}\label{eq1.75}
& &\sum_{x,y=0}^M \sin^2\left[\kappa_1+\theta_1 x +\theta_2 y\right] \nonumber \\
&=& \frac{(M+1)^2}{2} -\frac{\cos(2\kappa_1+\theta_1 M +\theta_2 M) \sin(\theta_1(M+1))\sin(\theta_2(M+1))}{2\sin(\theta_1)\sin(\theta_2)}\nonumber \\
&=&\sum_{x,y=0}^M\frac{1}{2}(1+\Or(1/M^2)).
\end{eqnarray}
We replace (\ref{eq1.75}) into (\ref{eq1.74}) and finally obtain
\begin{equation}
\begin{aligned}
&\sum_{x,y=0}^M\frac{\sin^2\left[t b((X+x)/t)-t b(-(Y+y)/t)\right]}{(X+Y+x+y)^2} \\
&= \sum_{x,y=0}^M\frac{1}{2(X+Y+x+y)^2}\left(1+\Or(\e,(\e^4 t)^{-1})\right).
\end{aligned}
\end{equation}
This estimate holds for all the region $I_2$, thus
\begin{equation}\label{eq1.77}
\sum_{(\xi_1,\xi_2)\in I_2} \frac{\sin^2\left[t b(\xi_1/t)-t b(-\xi_2/t)\right]}{(\xi_1+\xi_2)^2} =\sum_{(\xi_1,\xi_2)\in I_2}\frac{1}{2(\xi_1+\xi_2)^2}\left(1+\Or(\e,(\e^4 t)^{-1})\right).
\end{equation}
The estimates of (\ref{eq1.69}) and (\ref{eq1.77}) imply the statement of the Lemma.
\end{proofOF}

\section{Correlations along space-like paths}\label{SectCorrelations}
\begin{center}
\begin{figure}
\psfrag{Tiles}{Facet's types}
\psfrag{Lozenges}{Type of lozenges}
\psfrag{Type}{Type / related angle}
\psfrag{position}{$=(x,n)$ position}
\psfrag{eta}[c]{I / $\pi_\eta$}
\psfrag{tau}[c]{II / $\pi_\tau$}
\psfrag{nu}[c]{III / $\pi_\nu$}
\includegraphics[height=5cm]{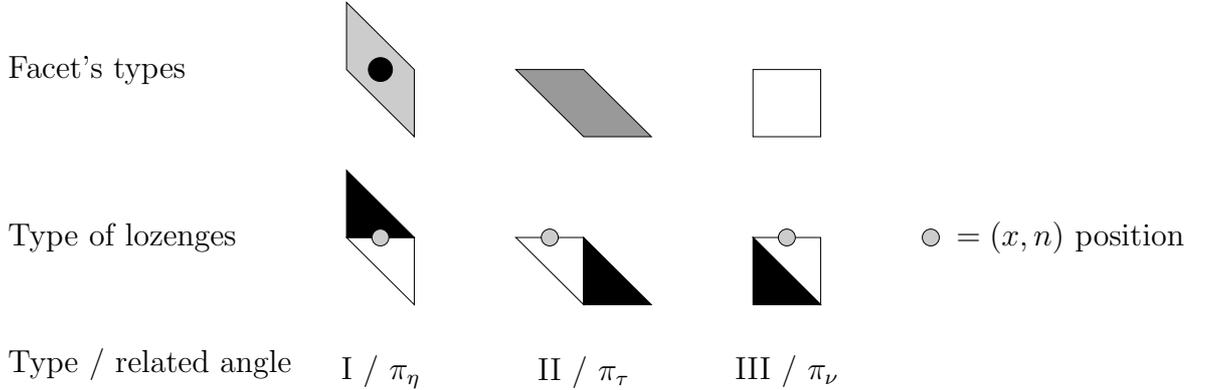}
\caption{Facet's types of Figure~\ref{FigureIntro}, their associated lozenges and angles.}
\label{FigureTilings}
\end{figure}
\end{center}

In this section we present an extension of
Theorem~\ref{ThmDetStructure} to the three types of lozenges (see
Figure~\ref{FigureTilings}). Then we explain the three different
ways of computing height differences. These are then used in the proof
of Theorem~\ref{TheoremCorr}.

\subsection{Joint distribution of the three types of lozenges}

As we saw in the introduction, particle configurations can also be
interpreted as lozenge tiling (see Figure~\ref{FigureIntro}) of a
half-plane. One can draw the corresponding triangular lattice by
associating to the three types of facets three lozenges made by one
black and one white triangle as indicated in
Figure~\ref{FigureTilings}. We define the position of a black /
white triangle to be $(x,n)$-coordinate on the mid-point of its
horizontal side.
\begin{center}\includegraphics[height=2em]{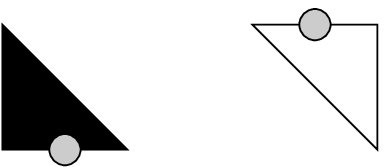}\end{center}
Thus, in our system of coordinates, these positions are pairs of integers $(x,n)$ with
$x\in\Z$, $n\in \{0,1,\ldots\}$ for black and $n\in\{1,2,\ldots\}$
for white triangles. We first state the result in the common way
from tiling point of view, and then we will reformulate it by using
the kernel $K$ defined in (\ref{eqDoubleIntRepr}).

For any pair of black and white triangles with space-time
coordinates $(x,n,t)$ and $(x',n',t')$, define the kernel
\begin{equation}
\widetilde {\cal K}(\B(x,n,t);\W(x',n',t'))=(-1)^{x-x'+n-n'} {\cal K}(x,n,t;x',n',t'),
\end{equation}
where $\cal K$ is the kernel defined in (\ref{StartingKernel}).

\begin{thm}\label{ThmLozengeOriginal}
Consider a finite set of lozenges at time moments $t_1\leq t_2\leq \cdots\leq t_M$, consisting of triangles
\begin{equation}
(b_i,w_i):=(\B(x_i,n_i,t_i),\W(x_i',n_i',t_i)).
\end{equation}
Assume that $n_i\geq n_j$ if $t_i<t_j$. Then
\begin{multline}\label{eqLoz1}
\Pb\big\{\textrm{There is a lozenge }(b_i,w_i) \textrm{ at time }t_i, \\ \textrm{for every }i=1,\ldots,M\big\} = \det\left[\widetilde {\cal K}(b_i,w_j)\right]_{1\leq i,j\leq M}.
\end{multline}
\end{thm}
\begin{proofOF}{Theorem~\ref{ThmLozengeOriginal}}
We prove the statement by induction on the number of lozenges $(b_i,w_i)$ which are not of the form \TypeI. When this number is zero, then the statement is Theorem~\ref{ThmDetStructure}, which is the base of the induction.

Consider any set $\cal S$ of lozenges at any time moments, plus
another lozenge. Then, the l.h.s.\ of (\ref{eqLoz1}) obviously
satisfies
\begin{equation}\label{eqLoz2}
\Pb\left({\cal S} \cup \TypeI\right)+\Pb\left({\cal S}\cup
\TypeII\right)+\Pb\left({\cal S}\cup \TypeIII\right) =
\Pb\left({\cal S}\right),
\end{equation}
where in the l.h.s we either keep the white triangle fixed, or we
keep the black triangle fixed (and we assume that $\cal S$ does not
contain the fixed triangle). Next we verify that the same relation
holds for the r.h.s.\ of (\ref{eqLoz1}).
\medskip

\textbf{Case (a):} the \textbf{fixed} triangle is \textbf{white}. From the explicit formula for the kernel, we get
\begin{equation}\label{eqLoz3}
\begin{aligned}
&\widetilde {\cal K}(\B(x,n,t);\W)+\widetilde {\cal K}(\B(x,n-1,t);\W)+\widetilde {\cal K}(\B(x+1,n-1,t);\W)\\
&=\left\{\begin{array}{ll}
1,&\textrm{if }\W=\W(x,n,t),\\
0,&\textrm{otherwise}.
\end{array}
\right.
\end{aligned}
\end{equation}
This implies relation (\ref{eqLoz2}) for the r.h.s.\ of (\ref{eqLoz1}).

\medskip

\textbf{Case (b):} the \textbf{fixed} triangle is \textbf{black}.
There are two possibilities: (i) the black triangle is on the lower
boundary $\{(x,n,t)\mid n=0\}$, or (ii) it is not on the boundary.
In case (ii), the relevant relation on the kernel is
\begin{equation}\label{eqLoz4}
\begin{aligned}
&\widetilde {\cal K}(\B;\W(x',n',t))+\widetilde {\cal K}(\B;\W(x',n'+1,t))+\widetilde {\cal K}(\B;\W(x'-1,n'+1,t))\\
&=\left\{\begin{array}{ll}
1,&\textrm{if }\B=\B(x',n',t),\\
0,&\textrm{otherwise}.
\end{array}
\right.
\end{aligned}
\end{equation}
In case (i), our assumption $n_i\geq n'_j$ whenever $t_i<t_j'$ implies that we are considering the last time moment, $t_M$. Then, in the formula for the kernel (\ref{StartingKernel}), the first residue term drops out and the second term vanishes on $\W(x',0,t)$ (since at $z=1$ there is no pole anymore). Thus (\ref{eqLoz4}) still gives the needed relation:
\begin{equation}\label{eqLoz5}
\Pb\left({\cal S}\cup \TypeII\right)+\Pb\left({\cal S}\cup
\TypeIII\right) = \Pb\left({\cal S}\right).
\end{equation}

\medskip

With the relation (\ref{eqLoz2}) verified (which, in one case,
degenerates to (\ref{eqLoz5})), let us explain the induction step.
Let us take a lozenge in the set \mbox{$\{(b_i,w_i),1\leq i\leq M\}$} which
is not of the type \TypeI. For example, consider \TypeIII and denote
by $\cal S$ the set of remaining $M-1$ lozenges. Then
\begin{equation}\label{eqLoz6}
\Pb\left({\cal S}\cup \TypeIII\right) = \Pb\left({\cal
S}\right)-\Pb\left({\cal S}\cup \TypeIbis\right)-\Pb\left({\cal
S}\cup \TypeIIbis\right),\quad \textrm{with }\W\textrm{ fixed}.
\end{equation}
So, we have a linear combination of two terms with one less lozenge
of type different from \TypeI, plus the third term with \TypeII
whose black triangle is one position on the right with respect to
the \TypeIII. For this term we use
\begin{equation}\label{eqLoz7}
\Pb\left({\cal S}\cup \TypeII\right) = \Pb\left({\cal
S}\right)-\Pb\left({\cal S}\cup \TypeIter\right)-\Pb\left({\cal
S}\cup \TypeIIIter\right),\quad \textrm{with }\B\textrm{ fixed}.
\end{equation}
So, the third term in (\ref{eqLoz6}) is rewritten as a linear
combination of two terms with one less lozenge of type different
from \TypeI, plus a third term with a lozenge of type \TypeIII, and
this lozenge is one position to the right from the initial \TypeIII\
in the l.h.s. of (\ref{eqLoz6}). This can be continued iteratively.
A similar argument holds for lozenges of type \TypeII\ with
(\ref{eqLoz6}) and (\ref{eqLoz7}) applied in the opposite order.

Thus, we can represent the r.h.s.\ of (\ref{eqLoz1}) as linear
combination of those with fewer lozenges of type \TypeII, \TypeIII,
plus an expression of the same kind but with one of the \TypeII or
\TypeIII\ lozenges far to the right.

We still have to verify that the formula with one more lozenge of
type \TypeII or \TypeIII\ agree when such a lozenge moves to
$+\infty$. Since the determinant in (\ref{eqLoz1}) is invariant with
respect to conjugation, consider the kernel $2^{x-x'} \widetilde
{\cal K}(\B(x,n,t);\W(x',n',t')) $ instead. Then, one verifies that
\begin{equation}
\begin{aligned}
2^{x-x'} \widetilde {\cal K}(\B(x,n,t);\W(x',n',t'))\to 0\quad \textrm{as }x'\to+\infty,\\
2^{x-x'} \widetilde {\cal K}(\B(x,n,t);\W(x',n',t'))\to 0\quad \textrm{as }x\to+\infty,
\end{aligned}
\end{equation}
and also for a lozenge $(b,w)$ far to the right (i.e., when
$x\to+\infty$) we have
\begin{equation}
2^{x-x'} \widetilde {\cal K}(b;w)\to \left\{
\begin{array}{ll}
1,& \textrm{if }(b,w)=\TypeIII,\\
0,& \textrm{if }(b,w)=\TypeII.
\end{array}
\right.
\end{equation}
Therefore, if in r.h.s.\ of (\ref{eqLoz1}) there is one lozenge \TypeII that is far to the right, the determinant tends to zero, which agrees with
\begin{equation}
\Pb\left({\cal S}\cup \TypeII\right)\to 0\quad\textrm{ as
}\TypeII\to+\infty.
\end{equation}
On the other hand, if in r.h.s.\ of (\ref{eqLoz1}) there is one
lozenge \TypeIII that is far to the right, the determinant tends to
the determinant of its minor corresponding to $\cal S$, which is in
agreement with
\begin{equation}
\Pb\left({\cal S}\cup \TypeIII\right)\to \Pb\left({\cal
S}\right)\quad \textrm{ as }\TypeIII\to+\infty.
\end{equation}
This completes the induction step.
\end{proofOF}

In the next section we will describe height function differences as
sum over lozenges of type \TypeI or \TypeII. To each lozenge one can
associate a position. We decided to set the position of a lozenge to
be equal to the position of the \textbf{white} triangle, see
Figure~\ref{FigureTilings}. Now we restate
Theorem~\ref{ThmLozengeOriginal} in a slightly different form.

\begin{thm}\label{ThmLozenges}
For any $N=1,2,\dots$, pick $N$ triples
\begin{equation*}
\varkappa_j=(x_j,n_j,t_j)\in \Z\times\Z_{>0}\times \R_{\geq 0}
\end{equation*}
such that $x_j+n_j\geq 0$ and
\begin{equation}
t_1\leq t_2\leq \dots\leq t_N,\qquad n_1\geq n_2\geq \dots\geq n_N.
\end{equation}
Then
\begin{multline}
\Pb\{\textrm{For each }j=1,\dots,N \textrm{ at } (x_j,n_j,t_j) \textrm{ there is a lozenge} \\
 \textrm{of type }\theta_j\in \{\mathrm{I,II,III}\}\}=\det{[K_\theta(\varkappa_i,\theta_i;\varkappa_j,\theta_j)]}_{i,j=1}^N,
\end{multline}
where
\begin{multline}\label{LozengeKernel}
K_\theta(x_1,n_1,t_1,\theta_1;x_2,n_2,t_2,\theta_2)\\
=\left\{
\begin{array}{ll}
K(x_1+n_1,n_1,t_1;x_2+n_2,n_2,t_2),&\textrm{if }\theta_1=\mathrm{I},\\
-K(x_1+n_1,n_1-1,t_1;x_2+n_2,n_2,t_2),&\textrm{if }\theta_1=\mathrm{II},\\
K(x_1+n_1-1,n_1-1,t_1;x_2+n_2,n_2,t_2),&\textrm{if }\theta_1=\mathrm{III},
\end{array}
\right.
\end{multline}
with $K$ as defined in Section~\ref{SectKernelRepresentations}.
\end{thm}
\begin{proofOF}{Theorem~\ref{ThmLozenges}}
The proof is simple. One just applies the correspondence
\begin{eqnarray}
\textrm{Type I at }(x,n,t) &\Leftrightarrow& (\B(x,n,t),\W(x,n,t))\\
\textrm{Type II at }(x,n,t) &\Leftrightarrow& (\B(x+1,n-1,t),\W(x,n,t))\\
\textrm{Type III at }(x,n,t) &\Leftrightarrow& (\B(x,n-1,t),\W(x,n,t))
\end{eqnarray}
to (\ref{eqLoz1}) and then rewrites $\widetilde {\cal K}$ in terms
of $\cal K$. Then, using the relation (\ref{eq4.1}), we get the
expression in terms of $K$. Finally, one conjugates the kernel by
$(-1)^{x_1-x_2}$ and obtains the desired kernel $K_\theta$.
\end{proofOF}

\subsection{Height differences as time integration of fluxes}\label{SectPaths}
To determine the height function at a position $(m,n)$ at a given
time $t$, one can act in three different ways: \\
(a) Sum along the
$x$-direction:
\begin{equation}\label{eq5.4}
h(m,n,t)=\sum_{x> m}\Id\left(\textrm{lozenge of type }\mathrm{I}\textrm{ at }(x,n,t)\right).
\end{equation}
(b) Sum along the $n$-direction: for $n'>n$,
\begin{equation}
h(m,n,t)=h(m,n',t)+H_{n,n'}(m,t),
\end{equation}
where
\begin{equation}\label{eq5.6}
H_{n,n'}(m,t)=-\sum_{p=n+1}^{n'} \Id\left(\textrm{lozenge of type }\mathrm{II}\textrm{ at }(m,p,t)\right).
\end{equation}
(c) Integrate the current over time: for $t>t'$,
\begin{equation}\label{eq5.8}
h(m,n,t)=h(m,n,t')+J_{t',t}(m,n),
\end{equation}
where $J_{t',t}(m,n)$ is the number of particles ($=$ lozenges of type {\rm I}) which jumped from site $(m,n)$ to site $(m+1,n)$ during the time interval $[t',t]$.

\begin{figure}
\begin{center}
\psfrag{n}{$n$}
\psfrag{t}{$t$}
\psfrag{x}{$x$}
\includegraphics[height=4.5cm]{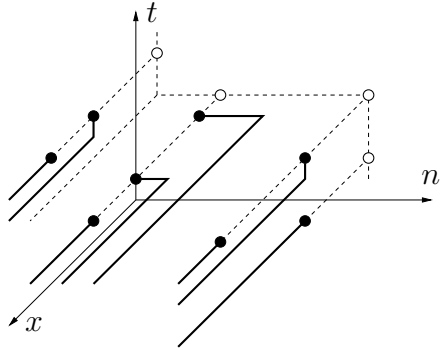}
\caption{The black dots represents the space-time positions where we want to study the height functions. They live on a space-like surface (i.e., for any two points $(x_1,n_1,t_1)$, $(x_2,n_2,t_2)$ on it, either $(n_1,t_1)\prec(n_2,t_2)$ or $(n_2,t_2)\prec(n_1,t_1)$). The white dots represents the projection of the black dots onto the $(n,t)$-plane.}
\label{FigPaths}
\end{center}
\end{figure}

In principle, one could use (a) alone to determine the height.
However, this turns out to be not very practical when dealing with
joint distributions of height functions at different points
$(m_1,n,t),\ldots,(m_K,n,t)$. The reason is that the height
functions are linear functions of lozenges of type {\rm I} but the
same lozenges appears in several of them. The result is a very
tedious computation. This can be avoided by using (b) and (c)
depending on the cases, see Figure~\ref{FigPaths} for an
illustration.

Therefore, the expression
\begin{equation}
\E\bigg(\prod_{k=1}^N\left[h(m_k,n_k,t_k)-\E(h(m_k,n_k,t_k))\right]\bigg)
\end{equation}
can be expressed as a sum of terms of the form
\begin{multline}\label{eqPaths}
\E\bigg(\prod_{k=1}^{M}\left[h(m_k,n_k,t_k)-\E(h(m_k,n_k,t_k))\right] \\
\prod_{\ell=M+1}^{R}\left[H_{n_\ell,n_\ell'}(m_\ell,t_\ell)-\E(H_{n_\ell,n_\ell'}(m_\ell,t_\ell))\right] \\
\prod_{j=R+1}^{N}\left[J_{t_j',t_j}(m_j,n_j)-\E(J_{t_j',t_j}(m_j,n_j))\right]\bigg).
\end{multline}

We now derive a formula for (\ref{eqPaths}).
\begin{lem}\label{LemmaJointMoments}
Assume that the following paths do not intersect and lie on a space-like surface:\\
\begin{equation}
\begin{array}{ll}
\{(x,n_k,t_k) | \, x>m_k\}, & k=1,\ldots,M,\\
\{(m_\ell,p,t_\ell) |\, p=n_\ell+1,\ldots,n_\ell'\}, & \ell=M+1,\ldots,R,\\
\{(m_j,n_j,t)|\, t\in [t_j',t_j]\}, & j=R+1,\ldots,N.
\end{array}
\end{equation}
Then
\begin{multline}\label{eqLemmaJointMoments}
(\ref{eqPaths}) = \sum_{x_1> m_1}\cdots \sum_{x_M> m_M} \sum_{p_{M+1}=n_{M+1}+1}^{n_{M+1}'}\cdots \sum_{p_R=n_R+1}^{n_R'} \\
\int_{t_{R+1}'}^{t_{R+1}}\dx s_{R+1}\cdots \int_{t_N'}^{t_N}\dx s_N \det\left[\begin{array}{ccc} A_{1,1} & A_{1,2} & A_{1,3} \\ A_{2,1} & A_{2,2} & A_{2,3} \\ A_{3,1} & A_{3,2} & A_{3,3} \end{array}\right],
\end{multline}
with the matrix blocks $A_{i,j}$ as follows:
\begin{align}
A_{1,1} &=\left[(1-\delta_{i,j})K(x_i+n_i,n_i,t_i;x_j+n_j,n_j,t_j)\right]_{1\leq i,j \leq M}, \nonumber\displaybreak[0]\\
A_{2,1} &=\left[K(m_i+p_i,p_i-1,t_i;x_j+n_j,n_j,t_j)\right]_{M+1\leq i \leq R, \, 1\leq j \leq M}, \nonumber\displaybreak[0]\\
A_{3,1} &=\left[K(m_i+n_i,n_i,s_i;x_j+n_j,n_j,t_j)\right]_{R+1\leq i \leq N, \, 1\leq j \leq M}, \nonumber\displaybreak[0]\\
A_{1,2} &=\left[K(x_i+n_i,n_i,t_i;m_j+p_j,p_j,t_j)\right]_{1\leq i \leq M, \, M+1\leq j \leq R}, \nonumber\displaybreak[0]\\
A_{2,2} &=\left[(1-\delta_{i,j})K(m_i+p_i,p_i-1,t_i;m_j+p_j,p_j,t_j)\right]_{M+1\leq i,j \leq R}, \displaybreak[0]\\
A_{3,2} &=\left[K(m_i+n_i,n_i,s_i;m_j+p_j,p_j,t_j)\right]_{R+1\leq i \leq N, \, M+1\leq j \leq R}, \nonumber\displaybreak[0]\\
A_{1,3} &=\left[-\partial_{s_j} K(x_i+n_i,n_i,t_i;m_j+n_j,n_j,s_j)\right]_{1\leq i \leq M, \, R+1\leq j \leq N}, \nonumber\displaybreak[0]\\
A_{2,3} &=\left[\partial_{s_j} K(m_i+p_i,p_i-1,t_i;m_j+n_j,n_j,s_j)\right]_{M+1\leq i \leq R,\, R+1\leq j \leq N}. \nonumber\displaybreak[0]\\
A_{3,3} &=\left[-(1-\delta_{i,j})\partial_{s_j} K(m_i+n_i,n_i,s_i;m_j+n_j,n_j,s_j)\right]_{R+1\leq i,j \leq N}. \nonumber
\end{align}
\end{lem}
\begin{proofOF}{Lemma~\ref{LemmaJointMoments}}
Below we prove that
\begin{equation}\label{eqB3.5}
\E\bigg(\prod_{k=1}^{M} h(m_k,n_k,t_k) \prod_{\ell=M+1}^{R} H_{n_\ell,n_\ell'}(m_\ell,t_\ell)
\prod_{j=R+1}^{N}J_{t_j',t_j}(m_j,n_j)\bigg)
\end{equation}
is equal to (\ref{eqLemmaJointMoments}) but without the
$1-\delta_{i,j}$ terms in $A_{1,1}, A_{2,2}$, and $A_{3,3}$. The
fact that the subtraction of the averages is given by putting zeros
on the diagonal is a simple but important property, which was
noticed for example in~\cite{Ken04} (see the proof of Theorem 7.2).

For $N=R$, (\ref{eqB3.5}) is a direct application of
Theorem~\ref{ThmLozenges} to the formulas (\ref{eq5.4}) and
(\ref{eq5.6}). The absence of the minus sign in $A_{2,*}$ is a
consequence of the minus in the definition of the $H$'s in
(\ref{eq5.6}). Next we extend the result when $N>R$, by first
considering $N=R+1$ for clarity. Denote by $\eta(x,n,t,\theta)$ the random variable that is equal to $1$ if there is a type $\theta$ lozenge at $(x,n,t)$ and $0$ otherwise. Recall that lozenges of type {\rm I}
are exactly what we call particles. The flux of particles can be written as
\begin{equation}\label{eqB3.4}
J_{t',t}(m,n)=\lim_{D\to\infty} \sum_{\ell=1}^D \eta(m,n,\tau_{i-1},{\rm I})(1-\eta(m,n,\tau_i,{\rm I}))
\end{equation}
with $\tau_i=t'+i \Delta \tau$, $i=0,\ldots,D$, $\Delta \tau =
(t-t')/D$. The quantity $\eta(m,n,\tau_{i-1},{\rm
I})(1-\eta(m,n,\tau_i,{\rm I}))$ equals 1 iff the site $(m,n)$ was
occupied by a particle at time $\tau_{i-1}$ and empty at time $\tau_i$. Each
particle tries to jump independently with an exponentially waiting
time. Every time a particle moves, it can also push other particles,
but no more than one on each (higher) level $n=const$. In any case,
since on each level there is a finite number of particles, the
probability that a particle has more than one jump during time
$\Delta \tau$ is $\Or(\Delta \tau^2)$. Thus, the limit
$\Delta\tau\to 0$ is straightforward.

To obtain (\ref{eqB3.5}) we have to determine expression at first
order in $\Delta \tau$ of
\begin{equation}\label{eqB3.6}
\E\left(\eta(m,n,\tau_{i-1},{\rm I})(1-\eta(m,n,\tau_i,{\rm
I}))\prod_{j=1}^Q \eta(m_j,n_j,t_j,\theta_j)\right).
\end{equation}
Then, in the $\Delta \tau\to 0$ limit we will get an integral from $t'$ to $t$.

Set $\overline K_{x,n}(t_1;t_2)=\sum_{k=0}^{n-1}\Psi^{n,t_1}_k(x+n)\Phi^{n,t_2}_k(x+n)$.
Remark that in (\ref{eq1.10}),
$\phi^{((n,\tau_i),(n,\tau_{i-1}))}(x,x)=1$. Then, since
$\tau_i>\tau_{i-1}$, from (\ref{eq1.10}) we obtain
\begin{equation}\label{eqB3.7}
(\ref{eqB3.6})=\det\left[\begin{array}{ccc}
\overline K_{m,n}(\tau_{i-1};\tau_{i-1}) & \overline K_{m,n}(\tau_{i-1};\tau_{i}) & K_\theta(m,n,\tau_{i-1},{\rm I};q)\\
1-\overline K_{m,n}(\tau_{i};\tau_{i-1}) & 1-\overline K_{m,n}(\tau_{i};\tau_{i}) & -K_\theta(m,n,\tau_i,{\rm I};q) \\
K_\theta(q;m,n,\tau_{i-1},{\rm I}) & K_\theta(q;m,n,\tau_i,{\rm I}) & K(q,q)
\end{array}\right]
\end{equation}
where by $q$ we denoted the quadruples $(m_j,n_j,t_j,\theta_j)$, for
$j\in \{1,\ldots,Q\}$. The kernel $K_\theta$ is a simple function of
the kernel $K$, see (\ref{LozengeKernel}). The second line is just
one in the diagonal minus the entries of the kernel (c.f.\ complementation principle in the Appendix of~\cite{Boo00}). Written in terms of $\overline K$ it becomes as above, since the $(2,1)$ entry
has a $1$ coming from $\phi$. Next we perform two operations keeping the
determinant invariant:
\begin{eqnarray*}
\textrm{Second row} &\to &\textrm{Second row} + \textrm{First row}\\
\textrm{Second column}& \to& \textrm{Second column} - \textrm{First column}.
\end{eqnarray*}
We get that (\ref{eqB3.7}) is equal to
\begin{eqnarray}
& &\det\left[\begin{array}{ccc}
\overline K_{m,n}(\tau_{i-1};\tau_{i-1}) &
\Delta \tau \partial_2\overline K_{m,n}(\tau_{i-1};\tau_{i-1}) +\Or(\Delta\tau^2)
& K_\theta(m,n,\tau_{i-1},{\rm I};q)\\
1-\Or(\Delta \tau) & \Or(\Delta \tau^2) & \Or(\Delta \tau) \\
K_\theta(q;m,n,\tau_{i-1},{\rm I}) & \Delta \tau \partial_2
K_\theta(q;m,n,\tau_{i-1},{\rm I})+\Or(\Delta\tau^2) & K_\theta(q,q)
\end{array}\right]\nonumber \\
& &\quad=-\Delta \tau \det\left[\begin{array}{cc}
\partial_2\overline K_{m,n}(\tau_{i-1};\tau_{i-1}) & K_\theta(m,n,\tau_{i-1},{\rm I};q)\\
\partial_2 K_\theta(q;m,n,\tau_{i-1},{\rm I}) & K_\theta(q,q)
\end{array}\right] +\Or(\Delta \tau^2)
\end{eqnarray}
where $\partial_2$ means the derivative with respect to $\tau_{i-1}$ in the second entry of the kernel. This formula and (\ref{eqB3.4}) imply
\begin{eqnarray}\label{eq5.11}
& &\E\left(J_{t',t}(m,n)\prod_{j=1}^Q \eta(m_j,n_j,t_j;\theta_j)\right)\\
&=&\int_{t'}^t\dx s
\det\left[\begin{array}{cc}
-\partial_2 K_\theta(m,n,s,{\rm I};m,n,s,{\rm I}) & K_\theta(m,n,s,{\rm I};m_j,n_j,t_j,\theta_j)\\
-\partial_2 K_\theta(m_i,n_i,t_i,\theta_i;m,n,s,{\rm I}) & K_\theta(m_i,n_i,t_i,\theta_i;m_j,n_j,t_j,\theta_j)
\end{array}\right]_{1\leq i,j\leq Q}. \nonumber
\end{eqnarray}
The case of several factors $J$ is obtained by induction. Expressing
$K_\theta$ for $\theta\in\{{\rm I,II}\}$ in terms of $K$ only, and
considering the fact that $H$ is \emph{minus} the sum of lozenges of
type $\rm II$, we obtain the result.
\end{proofOF}

\begin{figure}
\begin{center}
\psfrag{H}{$\mathbbm{H}$}
\psfrag{1}{$1$}
\psfrag{2}{$2$}
\psfrag{3}{$3$}
\psfrag{4}{$4$}
\psfrag{5}{$5$}
\includegraphics[height=4.5cm]{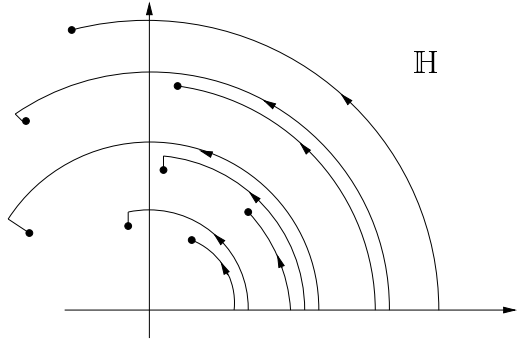}
\caption{The image of paths of Figure~\ref{FigPaths} under the map $\Omega$.}
\label{FigComplexPlane}
\end{center}
\end{figure}

\subsection{Proof of Theorem~\ref{TheoremCorr}}

Consider the expectation
\begin{equation}\label{eqMoment}
\E\bigg(\prod_{k=1}^N\left[h(m_k,n_k,t_k)-\E(h(m_k,n_k,t_k))\right]\bigg).
\end{equation}
Our goal is to determine its limit as $L\to\infty$ under the
macroscopic scaling: $t_k=\tau_k L$, $n_k=[\eta_k L]$,
$m_k=[(\nu_k-\eta_k) L]$, with $\nu_k\in
((\sqrt{\eta_k}-\sqrt{\tau_k})^2,(\sqrt{\eta_k}+\sqrt{\tau_k})^2)$.

The expression (\ref{eqMoment}) is a linear combinations of
expressions from Lemma~\ref{LemmaJointMoments}. The r.h.s.\ of
(\ref{eqLemmaJointMoments}) contains an $N\times N$ determinant; let
us write it as the sum over permutations $\sigma\in {\cal S}_N$ of
terms each of which is $\sgn\sigma$ times the product of matrix
elements $(i,\sigma_i)$, $i=1,\dots,N$.

The contribution of all permutations with fixed points is zero
(because the diagonal matrix elements are zeroes). All other
permutations can be written as unions of several cycles of length
$\ell \geq 2$. The contributions of the permutations with only
cycles of length $2$ lead to the final result, i.e., to prove the theorem
we first need to show that the sum of the contributions of permutations with cycles of
length $\ell \geq 3$ vanishes in the $L\to\infty$ limit.

Consider a cycle of length $\ell\geq 3$ and use the indices
$1,\ldots,\ell$ for the corresponding points $(m_i,n_i,t_i)$. Let us
order them so that
\begin{equation}
\eta_1\geq \eta_2\geq \ldots \geq \eta_\ell,\quad \tau_1\leq \tau_2\leq \ldots\leq \tau_\ell,\quad \textrm{no double points},
\end{equation}
i.e., $(\eta_j,\tau_j)\prec (\eta_{j-1},\tau_{j-1})$.

For an $\ell$-cycle we need to take the product of the kernels (or
their time derivatives depending on the case), and do the summation
over \mbox{$x_k>[(-\eta_k+\nu_k) L]$}, or over $p_\ell\in [[\eta_\ell L]+1,[\eta_\ell' L]]$, or
the integration over $[\tau_i' L,\tau_i L]$ depending on whether in
(\ref{eqLemmaJointMoments}) we have a sum or an integral.

We first collect all the factors related to a fixed index $i$. There are three
 possible cases:\\[0.5em]
\textbf{(a)} The index $i$ is related to a sum over
$[(-\eta_i+\nu_i)L,\infty)$. Then we have to analyze
\begin{equation}\label{eq3.22b}
\sum_{x>[\nu_i L]} K(x,[\eta_i L],\tau_i L;\varkappa_{\sigma_i})
K(\varkappa_{\sigma_i^{-1}};x,[\eta_i L],\tau_i L);
\end{equation}
\textbf{(b)} The index $i$ is related to a sum over $[[\eta_i L]+1,[\eta_i'L]]$. Then we have to analyze:
\begin{multline}\label{eq3.22c}
\sum_{p=[\eta_i L]+1}^{[\eta_i'L]} K([\nu_i L]+p-[\eta_i L],p-1,\tau_i L;\varkappa_{\sigma_i})\\
\times K(\varkappa_{\sigma^{-1}_i};[\nu_i L]+p-[\eta_i L],p,\tau_i
L);
\end{multline}
\textbf{(c)} The index $i$ is related to an integrated variable. We have in
this case
\begin{equation}\label{eq3.23b}
\int_{\tau_i' L}^{\tau_i L} \dx t K([\nu_i L],[\eta_i
L],t;\varkappa_{\sigma_i}) K(\varkappa_{\sigma^{-1}_i};[\nu_i
L]+1,[\eta_i L],t).
\end{equation}
The second kernel has a shift by one in the second $x$-entry. This comes from the identity
\begin{equation}\label{eq5.18}
-\partial_{s'}K(m,n,s;m',n',s')=K(m,n,s;m'+1,n',s'),
\end{equation}
which immediately follows from (\ref{eqDoubleIntRepr}).

We analyze these three expression in the $L\to\infty$ limit using
results of Section~\ref{SectKernelAsympt}. First of all, since
$w_c-z_c$ remains bounded away from zero all along the
integrals/sums, the bounds of Section~\ref{SectKernelAsympt} imply
that the contributions of the error term $\Or(L^{-1/8})$
in (\ref{eqAsymptKernel}) are of the same order, namely
$\Or(L^{-1/8})$. Therefore we can get rid of them immediately and we will not write them explicitly in what follows.

\medskip

\noindent \textbf{Case (a)} We divide the sum in three parts for
which we use \mbox{Propositions~\ref{PropKernelAsymptExp}-\ref{PropBorderEdge2}}.

\medskip
\noindent \textbf{Case (a/1)} Sum in the interval
\begin{equation}
I_1=\{x\in \N, x\geq (\sqrt{\tau_i}+\sqrt{\eta_i})^2L-\ell L^{1/3}\}.
\end{equation}
Then, by Propositions~\ref{PropBorderEdge}-\ref{PropBorderEdge2},
\begin{equation}\label{eq6.19}
\begin{aligned}
&\bigg|\sum_{x\in I_1} K(x,[\eta_i L],\tau_i L;\varkappa_{\sigma_i})
K(\varkappa_{\sigma^{-1}_i};x,[\eta_i L],\tau_i L)\bigg| \\
&\leq \sum_{x\in I_1} \frac{\cte}{L^{2/3}} \exp\left(-2\frac{x-(\sqrt{\tau_i}+\sqrt{\eta_i})^2L}{(\tau_i L)^{1/3}}\right)\times
\textrm{terms in }\varkappa_{\sigma_i},\varkappa_{\sigma^{-1}_i}\\
&\leq \frac{\cte}{L^{1/3}} \times \textrm{terms in
}\varkappa_{\sigma_i},\varkappa_{\sigma^{-1}_i}.
\end{aligned}
\end{equation}
Therefore, as $L\to\infty$, the contribution of this sum goes to
zero.

\medskip
\noindent \textbf{Case (a/2)} Sum in the interval
\begin{equation}
I_2=\{x\in \N, (\sqrt{\tau_i}+\sqrt{\eta_i})^2L-L^{2/3}<x< (\sqrt{\tau_i}+\sqrt{\eta_i})^2L-\ell L^{1/3}\}.
\end{equation}
By Propositions~\ref{PropBorderBulk}-\ref{PropBorderEdge},
\begin{equation}\label{eq6.20}
\begin{aligned}
& \bigg|\sum_{x\in I_2} K(x,[\eta_i L],\tau_i
L;\varkappa_{\sigma_i})
K(\varkappa_{\sigma^{-1}_i};x,[\eta_i L],\tau_i L)\bigg| \\
&\leq \sum_{x\in I_2} \frac{\cte}{L\sqrt{\eta_i\tau_i-\tfrac14(\tau_i+\eta_i-x/L)^2}} \times
\textrm{terms in }\varkappa_{\sigma_i},\varkappa_{\sigma^{-1}_i} \\
&\leq \frac{\cte}{L^{1/6}} \times \textrm{terms in
}\varkappa_{\sigma_i},\varkappa_{\sigma^{-1}_i}.
\end{aligned}
\end{equation}
Therefore, as $L\to\infty$, this contribution is also infinitesimally small.

\medskip

In the following (Cases (a/3), (b), and (c)) we will assume that all the entries $\varkappa_i$'s of the kernel are in $\cal D$ and apply Proposition~\ref{PropKernelAsymptExp} and its Corollary~\ref{CorBulkShifted}. Let us justify it. The variables corresponding to time integration (Case (c)) and sum over the $p$ variables (Case (b)) in (\ref{eqLemmaJointMoments}) are always in $\cal D$. Therefore, the only $\varkappa_i$'s which are not in $\cal D$ correspond to Cases (a/1) and (a/2) above. From Propositions~\ref{PropKernelAsymptExp}-\ref{PropBorderEdge}, the contributions in the $\varkappa_i$ variable are of order
\begin{equation}\label{eq5.44}
\frac{\Or(1)}{L\sqrt{\eta_i\tau_i-\tfrac14(\eta_i+\tau_i-\nu_i)^2}},
\end{equation}
if $\varkappa_i\in{\cal D}$. The sum in Case (a/3) is then bounded by $\Or(1)$ because the sum is over $\Or(L)$ sites and the square-root singularity is integrable. Even simpler is Case (b) where we never come close to the singularity and the sum is over $\Or(L)$ sites. Finally, in Case (c), the integration is over a time span $\Or(L)$. Therefore, the contributions of the terms of Cases (a/3), (b), and (c) are $\Or(1)$, and for every sum reaching the edge we get a factor $\Or(L^{-1/6})$. Thus, in the following we need to determine the asymptotics of Cases (a/3), (b), and (c) in the case where all the entries $\varkappa_i$'s are in $\cal D$.

\medskip

\noindent \textbf{Case (a/3)} Sum in the interval
\begin{equation}
I_3=\{x\in \N, [\nu_i L] < x \leq (\sqrt{\tau_i}+\sqrt{\eta_i})^2L-L^{2/3}\}.
\end{equation}
Define the functions
\begin{equation}
A(\nu,\eta,\tau)=\frac{1}{2\pi
|G''(\Omega(\nu,\eta,\tau))|\sqrt{\nu/\tau}}
\end{equation}
and
\begin{equation}
F(\nu,\eta,\tau)=L\, \Im(G(\Omega(\nu,\eta,\tau))).
\end{equation}
Then, by Proposition~\ref{PropKernelAsymptExp} we have
\begin{equation}\label{eq3.27}
\begin{aligned}
& \sum_{x\in I_3}K(x,[\eta_i L],\tau_i L;\varkappa_{\sigma_i})
K(\varkappa_{\sigma^{-1}_i};x,[\eta_i L],\tau_i L) \\
&=\sum_{x\in I_3} \frac{A(x/L,\eta_i,\tau_i)}{L}\bigg[
\frac{e^{-\I\beta_1(i)}}{\omega(\sigma_i)-\omega(i)}\frac{e^{\I\beta_2(i)}}{\omega(i)-\omega(\sigma^{-1}_i)} \\
& +
\frac{e^{-\I\beta_1(i)}}{\omega(\sigma_i)-\omega(i)}\frac{e^{-\I\beta_2(i)}}{\bar\omega(i)-\omega(\sigma^{-1}_i)}
e^{-2\I F(x/L,\eta_i,\tau_i)} \\
& + \frac{e^{\I\beta_1(i)}}{\omega(\sigma_i)-\bar\omega(i)} \frac{e^{\I\beta_2(i)}}{\omega(i)-\omega(\sigma^{-1}_i)}
e^{2\I F(x/L,\eta_i,\tau_i)} \\
& +\frac{e^{\I\beta_1(i)}}{\omega(\sigma_i)-\bar\omega(i)}
\frac{e^{-\I\beta_2(i)}}{\bar\omega(i)-\omega(\sigma^{-1}_i)} +
\circlearrowleft\bigg] \times \textrm{terms in
}\varkappa_{\sigma_i},\varkappa_{\sigma^{-1}_i}
\end{aligned}
\end{equation}
where we used the notation $\omega(i)=\Omega(\nu_i,\eta_i,\tau_i)$ and $\circlearrowleft$ means the other 12 terms obtained by replacing $\omega(\sigma_i)$ by $\bar \omega(\sigma_i)$ and/or $\omega(\sigma^{-1}_i)$ by $\bar \omega(\sigma^{-1}_i)$.

First we want to show that the terms with $F$ in the exponential are
irrelevant in the $L\to\infty$ limit. For that, we sum over
$N=L^{1/3}$ positions around any $\nu L$ in the bulk. Then, for
$0\leq x \leq L^{1/3}$ it holds
\begin{equation}
F(\nu+x/L,\eta,\tau)=L\gamma(\nu,\eta,\tau)+x\partial_\nu\gamma(\nu,\eta,\tau)+\Or(L^{-1/3}),
\end{equation}
where $\gamma(\nu,\eta,\tau)=L^{-1} F(\nu,\eta,\tau)$. All the other functions ($A$, $\beta_1(i)$, $\beta_2(i)$, and $\omega(i)$) are smooth functions in $\nu_i$, i.e., over an interval $L^{1/3}$
vary only by $\sim L^{-2/3}$. Therefore we have to compute an expression of the form
\begin{equation}\label{eq5.47}
\frac{1}{N}\sum_{x=0}^{N-1} e^{2\I F(\nu+x/L,\eta,\tau)}\phi(\nu+x/L,\eta,\tau)
\end{equation}
where $\phi$ is a smooth function given in term of $A$, $\beta_1(i)$, $\beta_2(i)$, and $\omega(i)$.
Thus
\begin{equation}
(\ref{eq5.47})=\phi(\nu,\eta,\tau) e^{2\I L \gamma(\nu,\eta,\tau)}\frac{1}{N}\sum_{x=0}^{N-1}e^{\I b x} + \Or(L^{-1/3})
\end{equation}
with $b=2\partial_\nu\gamma(\nu,\eta,\tau)$. Then, for $0<b<\pi$, we use
\begin{equation}\label{eq3.28}
\frac{1}{N}\sum_{x=0}^{N-1}e^{\I b x} =\frac{e^{\I b N}-1}{N(e^{\I b}-1)}.
\end{equation}
In our case, $b$ is strictly between $0$ and $\pi$ as soon as we are away from the facet. When we reach the lower facet, $b\to 0$. However, in the sum over $I_3$ we are at least at a distance $L^{2/3}$ from the facet, i.e., $b\geq \cte L^{-1/6}$. Therefore
\begin{equation}\label{eq6.24}
|(\ref{eq3.28})|\leq \cte /(b N)\leq \cte L^{-1/6}.
\end{equation}
Since this holds uniformly in the domain $I_3$, we have shown that
the contribution of the terms where the $\exp(\pm 2\I F)$ is present
is at worst of order $L^{-1/6}$. Therefore the only non-vanishing
terms in (\ref{eq3.27}) are
\begin{eqnarray}
& &\sum_{x\in I_3} \frac{A(x/L,\eta_i,\tau_i)}{L}\bigg[
\frac{e^{-\I\beta_1(i)}}{\omega(\sigma_i)-\omega(i)}\frac{e^{\I\beta_2(i)}}{\omega(i)-\omega(\sigma^{-1}_i)}
\nonumber \\
& &+\frac{e^{\I\beta_1(i)}}{\omega(\sigma_i)-\bar\omega(i)}
\frac{e^{-\I\beta_2(i)}}{\bar\omega(i)-\omega(\sigma^{-1}_i)}+\circlearrowleft\bigg]\times
\textrm{terms in }\varkappa_{\sigma_i},\varkappa_{\sigma^{-1}_i}.
\end{eqnarray}
 All the functions appearing now are smooth and changing over distances $x\sim L$. Thus, defining $x=\nu L$, the sum becomes, up to an error of order $\Or(L^{-1/3})$, the integral
\begin{eqnarray}\label{eq3.32}
& &\int_{\nu_i}^{(\sqrt{\tau_i}+\sqrt{\eta_i})^2}\dx \nu A(\nu,\eta_i,\tau_i)\bigg[
\frac{e^{-\I\beta_1(i)}}{\omega(\sigma_i)-\omega(i)}\frac{e^{\I\beta_2(i)}}{\omega(i)-\omega(\sigma^{-1}_i)} \nonumber \\
& &+\frac{e^{\I\beta_1(i)}}{\omega(\sigma_i)-\bar\omega(i)}
\frac{e^{-\I\beta_2(i)}}{\bar\omega(i)-\omega(\sigma^{-1}_i)}
+\circlearrowleft\bigg]\times \textrm{terms in
}\varkappa_{\sigma_i},\varkappa_{\sigma^{-1}_i}.
\end{eqnarray}
The final step is a change of variable. For the term with
$\omega(i)$, we set \mbox{$z_i^{+}=\omega(i)=\Omega(\nu,\eta_i,\tau_i)$}.
Denote the new integration path by
\mbox{$\Gamma^i_{+}=\{\Omega(\nu,\eta_i,\tau_i),\nu:(\sqrt{\tau_i}+\sqrt{\eta_i})^2\to\nu_i\}$}.
The Jacobian was computed in Proposition~\ref{PropGeom2}, namely
\begin{equation}
\frac{\partial \omega(i)}{\partial\nu}= \frac{\I\omega(i)}{\kappa} =
2\pi\I A\, e^{\I\beta_2(i)}e^{-\I\beta_1(i)}.
\end{equation}
For the term with $\bar\omega(i)$ we set $z_i^{-}=\bar\omega(i)=\overline\Omega(\nu,\eta_i,\tau_i)$ and $\Gamma^i_{-}=\bar \Gamma^i_{+}$.
Then (\ref{eq3.32}) becomes
\begin{equation}\label{eq3.30}
\frac{-1}{2\pi\I}\sum_{\e_i=\pm}\e_i\int_{\Gamma^i_{\e_i}}\dx z^i_{\e_i} \bigg[\frac{1}{z^i_{\e_i}-\omega(\sigma_i)}\frac{1}{\omega(\sigma^{-1}_i)-z^i_{\e_i}}+\circlearrowleft\bigg]\times \textrm{terms in }\sigma_i,\sigma^{-1}_i.
\end{equation}
The factor $-1$ comes from the orientation of $\Gamma_{\e_i}^i$, see Figure~\ref{FigComplexPlane}.

\medskip

\noindent \textbf{Case (b)} We sum in the $n$-direction from $[\eta_i L]+1$ to $[\eta_i' L]$ for some \mbox{$\eta_i'>\eta_i$}. While doing this, we
do not exit the domain $\cal D$ remaining in the bulk. Therefore,
the computations are just a small variation of the sum over $I_3$ of
Case~(a). The minor difference comes from the $-1$ shift in $p$ in
the entries of the first kernel. By changing the variable
$\alpha=p/L$, we then obtain
\begin{equation}\label{eq3.27c}
\begin{aligned}
& \lim_{L\to\infty}\sum_{p=[\eta_i L]+1}^{[\eta_i' L]}
K([\nu_i L]+p-[\eta_i L],p-1,\tau_i L;\varkappa_{\sigma_i}) K(\varkappa_{\sigma^{-1}_i};[\nu_i L]+p-[\eta_i L],p,\tau_i L)\\
&=\int_{\eta_i}^{\eta_i'} \dx \alpha A(\nu_i-\eta_i+\alpha,\alpha,\tau_i)
\bigg[\frac{e^{-\I\beta_1(i)} \omega(i)^{-1}}{\omega(\sigma_i)-\omega(i)}\frac{e^{\I\beta_2(i)}}{\omega(i)-\omega(\sigma^{-1}_i)} \\
&
+\frac{e^{\I\beta_1(i)}\bar\omega(i)^{-1}}{\omega(\sigma_i)-\bar\omega(i)}
\frac{e^{-\I\beta_2(i)}}{\bar\omega(i)-\omega(\sigma^{-1}_i)} +
\circlearrowleft\bigg] \times \textrm{terms in
}\varkappa_{\sigma_i},\varkappa_{\sigma^{-1}_i}
\end{aligned}
\end{equation}
with $\omega(i)=\Omega(\nu_i-\eta_i+\alpha,\alpha,\tau_i)$. For the term with $\omega(i)$, we set $z_i^{+}=\omega(i)$ and denote
the new integration path by
$\Gamma^i_{+}=\{\Omega(\nu_i-\eta_i+\alpha,\alpha,\tau_i),\alpha:\eta_i'\to\eta_i\}$
(we set the orientation of the path as in
Figure~\ref{FigComplexPlane}). By Proposition~\ref{PropGeom2} we get
\begin{equation}
\frac{\partial \omega(i)}{\partial \alpha}=\frac{\I}{\kappa}=2\pi\I A\, e^{\I\beta_2(i)}e^{-\I\beta_1(i)} \omega(i)^{-1}.
\end{equation}
The change of variable for the term with $\bar\omega(i)$ is similar.
The result is the same formula as (\ref{eq3.30}) (of course, with
the new $\Gamma_{\e_i}^i$'s).

\medskip

\noindent \textbf{Case (c)} The last case is when we do an
integration over a time interval. Similarly to Case (b), we do not
have to deal with the edges, since, by assumption, we remain in the
bulk of the system. We need to compute
\begin{equation}\label{eq3.27Time}
\begin{aligned}
&\int_{\tau_i'L}^{\tau_i L}\dx t K([\nu_i L],[\eta_i L],t;\varkappa_{\sigma_i})
K(\varkappa_{\sigma^{-1}_i};[\nu_i L]+1,[\eta_i L],t) \\
=&\int_{\tau_i'}^{\tau_i}\dx \tau A(\nu_i,\eta_i,\tau)\bigg[
\frac{e^{-\I\beta_1(i)}}{\omega(\sigma_i)-\omega(i)}\frac{e^{\I\beta_2(i)}}{\omega(i)-\omega(\sigma^{-1}_i)}(1-\omega(i))\\
&+ \frac{e^{-\I\beta_1(i)}}{\omega(\sigma_i)-\omega(i)}\frac{e^{-\I\beta_2(i)}}{\bar\omega(i)-\omega(\sigma^{-1}_i)}(1-\bar\omega(i))
e^{-2\I F(\nu_i,\eta_i,\tau)} \\
&+ \frac{e^{\I\beta_1(i)}}{\omega(\sigma_i)-\bar\omega(i)} \frac{e^{\I\beta_2(i)}}{\omega(i)-\omega(\sigma^{-1}_i)}(1-\omega(i))
e^{2\I F(\nu_i,\eta_i,\tau)} \\
&+\frac{e^{\I\beta_1(i)}}{\omega(\sigma_i)-\bar\omega(i)}
\frac{e^{-\I\beta_2(i)}}{\bar\omega(i)-\omega(\sigma^{-1}_i)}(1-\bar\omega(i))
+ \circlearrowleft\bigg] \times \textrm{terms in }\varkappa_{\sigma_i},\varkappa_{\sigma^{-1}_i}.
\end{aligned}
\end{equation}
The only rapidly changing function is $F$, which, as for the sum,
makes the contributions of the term with it vanishing small as
$L\to\infty$. We do the same change of variable as above, i.e.,
$z_i^{+}=\omega(i)=\Omega(\nu_i,\eta_i,\tau)$. Denote the new
integration path by
$\Gamma^i_{+}=\{\Omega(\nu_i,\eta_i,\tau),\tau\in
[\tau_i',\tau_i]\}$. The Jacobian is computed in
Proposition~\ref{PropGeom2}, namely
\begin{equation}
\frac{\partial \omega(i)}{\partial \tau}=\frac{-\I\omega(i)(1-\omega(i))}{\kappa}= -2\pi\I A e^{\I\beta_2(i)}e^{-\I\beta_1(i)} (1-\omega(i)).
\end{equation}
Thus, we obtain again (\ref{eq3.30}).

\medskip

Thus, after summing / integrating all the $\ell$ variables, we get
the contribution of the $\ell$-cycle, namely
\begin{eqnarray}\label{eq5.45}
& &\frac{(-1)^\ell}{(2\pi\I)^\ell}\sum_{\e_1,\ldots,\e_\ell=\pm}\prod_{i=1}^\ell \e_i \int_{\Gamma^1_{\e_1}}\dx z_1^{\e_1}\cdots \int_{\Gamma^\ell_{\e_\ell}}\dx z_\ell^{\e_\ell} \prod_{i=1}^\ell\frac{1}{z_{i}^{\e_i}-z_{\sigma_i}^{\e_{\sigma_i}}}, \nonumber\\
&=&\frac{(-1)^\ell}{(2\pi\I)^\ell}\sum_{\e_1,\ldots,\e_\ell=\pm}\prod_{i=1}^\ell \e_i \int_{\Gamma^1_{\e_1}}\dx z_1^{\e_1}\cdots \int_{\Gamma^\ell_{\e_\ell}}\dx z_\ell^{\e_\ell} \prod_{i=1}^\ell \frac{1}{z_{\sigma_i}^{\e_{\sigma_i}}-z_{\sigma_{i-1}}^{\e_{\sigma_{i-1}}}},
\end{eqnarray}
where we set $\sigma_0:=\sigma_\ell$. By Lemma 7.3 in~\cite{Ken04},
which refers back to~\cite{BouPhD},
\begin{equation}
\sum_{\sigma=\ell\textrm{--cycle in }{\cal S}_\ell}\prod_{i=1}^\ell
\frac{1}{Y_{\sigma_i}-Y_{\sigma_{i-1}}}=0,\quad \textrm{ for
}\ell\geq 3.
\end{equation}
Therefore, the sum of (\ref{eq5.45}) over all possible $\ell$-cycles
on the same set of indices gives zero for $\ell\geq 3$.

We have shown that we have a Gaussian type formula (sum over all
couplings) for points macroscopically away. We still need to compute
explicitly the covariance for such points. The covariance is
obtained from (\ref{eq5.45}) for $\ell=2$. We need now to consider
the signature of the permutation, which for a $2$-cycle is $-1$. We
thus obtain a sum of $4$ terms which can be put together into (see
the end of Section 7 in~\cite{Ken04} too)
\begin{eqnarray}\label{eq6.33}
& &\frac{1}{(2\pi\I)^2}\int_{\bar \Omega(\nu_1,\eta_1,\tau_1)}^{\Omega(\nu_1,\eta_1,\tau_1)}\dx z_1 \int_{\bar \Omega(\nu_2,\eta_2,\tau_2)}^{\Omega(\nu_2,\eta_2,\tau_2)}\dx z_2 \frac{1}{(z_1-z_2)^2} \\
&=&
\frac{-1}{4\pi^2}\ln\left(\frac{(\Omega(\nu_1,\eta_1,\tau_1)-\Omega(\nu_2,\eta_2,\tau_2))
(\overline\Omega(\nu_1,\eta_1,\tau_1)-\overline\Omega(\nu_2,\eta_2,\tau_2))}{(\overline\Omega(\nu_1,\eta_1,\tau_1)-\Omega(\nu_2,\eta_2,\tau_2))
(\Omega(\nu_1,\eta_1,\tau_1)-\overline\Omega(\nu_2,\eta_2,\tau_2))}\right).
\nonumber \qed
\end{eqnarray}

\subsection{Short and intermediate distance bounds}

Let us first prove the short distance bound (\ref{eqShortDistance}).
\begin{lem}\label{LemmaShortDistance}
For any $\varkappa_j\in{\cal D}$ and any $\e>0$, we have
\begin{equation}
\E\left(H_L(\varkappa_1)\cdots H_L(\varkappa_N)\right)=\Or(L^\epsilon),\quad L\to\infty.
\end{equation}
\end{lem}
\begin{proofOF}{Lemma~\ref{LemmaShortDistance}}
Theorem~\ref{thmLogfluct} implies, for any integer $m\geq 1$,
\begin{equation}
\E(H_L(\varkappa_j)^{2m})=\Or(\ln(L)^m).
\end{equation}
By Chebyshev inequality,
\begin{equation}
\Pb(|H_L(\varkappa_j)|\geq X\ln(L))=\Or(1/X^{2m}),\quad \Pb(|H_L(\varkappa_j)|\geq Y)=\Or(\ln(L)^m/Y^{2m}).
\end{equation}
The final ingredient is that $|H_L(\varkappa_j)|= \Or(L)$, since on
level $n=L$ we have only $L$ particles. Therefore, for any $Y$, we
can bound
\begin{eqnarray}
|\E(H_L(\varkappa_1)\cdots H_L(\varkappa_N))| &\leq& \Pb(|H_L(\varkappa_1)|\leq Y,\ldots,|H_L(\varkappa_N)|\leq Y) Y^N \nonumber \\
& +& \Pb(\exists j\textrm{ s.t. } |H_L(\varkappa_j)| > Y) \Or(L)^N \nonumber \\
&= & \Or(Y^N)+\Or(L^N\ln(L)^m/Y^{2m}).
\end{eqnarray}
Taking $Y=L^{\e/2}$ and $m\gg 1$ large enough, we obtain
\begin{equation}
|\E(H_L(\varkappa_1)\cdots H_L(\varkappa_N))| = \Or(L^\e),\quad
\textrm{for any given }\e>0.
\end{equation}
\end{proofOF}

Next we give an intermediate distance bound, extending the result of Theorem~\ref{TheoremCorr}.
\begin{lem}\label{LemmaIntermDist}
Consider the setting as in Theorem~\ref{TheoremCorr}. If the points $\Omega_i$'s are not closer than $L^{-1/(8N)}$, then the difference between the expectation $\E(H_L(\varkappa_1)\cdots H_L(\varkappa_N))$ and r.h.s.\ of (\ref{eq1.17}) is $\Or(L^{-1/(12N)})$.
\end{lem}
\begin{proofOF}{Lemma~\ref{LemmaIntermDist}}
It is a small extension of Theorem~\ref{TheoremCorr}. For $N=1$ the
two expressions are identically equal to zero. So, consider $N\geq
2$. We have $|\Omega_i-\Omega_j|\geq L^{-1/16}$, so that the
estimate of (\ref{eqAsymptKernel}) can still be applied in the proof
of Theorem~\ref{TheoremCorr}. All the error terms collected are
$\Or(L^{-1/6})$ (see (\ref{eq6.24})) times at most $N$ factors of
order $1/|\Omega_i-\Omega_j|=\Or(L^{1/(8N)})$. This accounts into an
error $\Or(L^{-1/24})$. Now, since the $\Omega_i$'s are not away of
order one, when one $\Or(L^{-1/8})$ in (\ref{eqAsymptKernel}) is
used, it has to be multiplied by at worst $N-1$ factors of order
$1/|\Omega_i-\Omega_j|=\Or(L^{1/(8N)})$. Therefore the error is at
most $\Or(L^{(N-1)/(8N)} L^{-1/8})=\Or(L^{-1/(8N)})$. Similarly, the
contribution where $\Or(L^{-1/8})$ is used $n$ times is
$\Or(L^{(N-n)/(8N)} L^{-n/8})$, which is maximal at $n=1$. Thus, for
$N\geq 2$, we get all together
$\Or(L^{-1/24})+\Or(L^{-1/(8N)})=\Or(L^{-1/(12N)})$.
\end{proofOF}

\subsection{Gaussian Free Field}\label{SectGFF}
The Gaussian Free Field on $\H$, see e.g.~\cite{She03}, is
a generalized Gaussian process (i.e.\ it is a probability measure on
a suitable class of generalized functions on $\H$) that can be
characterized as follows. If we denote by $\GFF$ the random
generalized function and take any sequence $\{\phi_k\}$ of
(compactly supported) test functions, the pairings
$\{\GFF(\phi_k)\}$ form a sequence of mean 0 normal variables with
covariance matrix
\begin{multline*}
\E(\GFF(\phi_k)\GFF(\phi_l))=\int_\H |\dx z|^2 (\nabla
\phi_k(z),\nabla \phi_l(z))\\=\int_{\H^2} |\dx z_1|^2 |\dx z_2|^2
\phi_k(z_1)\phi_l(z_2) {\cal G}(z_1,z_2),
\end{multline*}
where
\begin{equation}
{\cal G}(z,w)=-\frac 1{2\pi}\ln\left|\frac{z-w}{z-\bar w}\right|
\end{equation}
is the Green function of the Laplacian on $\H$ with Dirichlet boundary conditions.

The value of $\GFF$ at a point cannot be defined. However, one can
think of expectations of products of values of $\GFF$ at different
points as being finite and equal to
\begin{multline}\label{eqGFFmoments}
\E[\GFF(z_1)\cdots\GFF(z_m)]\\
=\left\{
\begin{array}{ll}
0&\textrm{ if }m\textrm{ is odd}, \\
\sum\limits_{\textrm{pairings }\sigma} {\cal
G}(z_{\sigma(1)},z_{\sigma(2)})\cdots{\cal
G}(z_{\sigma(m-1)},z_{\sigma(m)}) &\textrm{ if }m\textrm{ is even}.
\end{array}
\right.
\end{multline}

The justification for the notation is the fact that for any finite
number of test functions,
\begin{equation}\label{GFFmoments}
\E(\GFF(\phi_1)\cdots\GFF(\phi_m))=\int_{\H^m} \prod_{k=1}^m |\dx
z_k|^2 \phi_k(z_k) \E\left[\GFF(z_1)\cdots \GFF(z_m)\right].
\end{equation}
The moments (\ref{GFFmoments}) uniquely determine the Gaussian Free
Field.

To state the convergence results, we consider any (smooth)
space-like surface ${\cal U} \subset \R^3$ in the rounded part of
the surface. Namely $\cal U\subset \cal D$, and for any two triples
$(\nu_i,\eta_i,\tau_i)\in \cal U$, $i=1,2$, $\eta_1\leq \eta_2$
implies $\tau_1\geq \tau_2$.

Clearly, the mapping $\Omega$ restricted to $\cal U$ is a bijection.
Consider any smooth parametrization $u=(u_1,u_2)$ of $\cal U$.
Denote by $\Omega_{\cal U}$ the map from $u$ to $\H$, which is the
composition of the map from $u$ to $(\nu,\eta,\tau)$ and $\Omega$.
Then, for any smooth compactly supported test function $f$ on $\cal
U$, we define
\begin{equation}
\langle f,H_L\rangle:=\int_{\cal U}\dx u f(u) H_L(u),
\end{equation}
where $H_L(u)$ is as in (\ref{HL}). Then
\begin{equation}
\langle f,H_L\rangle = \int_\H |\dx z|^2 J(z) f(\Omega_{\cal U}^{-1}(z)) H_L(\Omega_{\cal U}^{-1}(z)),
\end{equation}
where $J(z)$ is the Jacobian of the change of variables $z\to u$ by
$\Omega_{\cal U}^{-1}$.

\begin{thm}\label{ThmGFF2d}
For any $m\in \N$, and any smooth compactly supported functions
$f_1,\ldots,f_m$ on $\cal U$,
\begin{multline}\label{eqThmGFF2d}
\lim_{L\to\infty}\E\left[\prod_{k=1}^m\langle f_k,H_L\rangle\right]
= \int_{\H^m} \prod_{k=1}^m |\dx z_k|^2 f_k^\H(z_k) \E\left[\GFF(z_1)\cdots \GFF(z_m)\right]
\end{multline}
where $f_i^\H(z):=J(z) f(\Omega_{\cal U}^{-1}(z))$.
\end{thm}

\begin{rem} Since moment convergence to a (multidimensional)
Gaussian implies convergence in distribution, Theorem \ref{ThmGFF2d}
implies that the random vector $(\langle f_k,H_L\rangle)_{k=1}^m$
converges in distribution (and with all moments) to the Gaussian
vector with mean zero and covariance matrix $\Vert\int_\H |\dx z|^2
(\nabla f_k^\H(z),\nabla f_l^\H(z))\Vert_{k,l=1,\dots,m}$.
\end{rem}

\begin{proofOF}{Theorem~\ref{ThmGFF2d}}
We have
\begin{multline}
\E\left[\prod_{k=1}^m\langle f_k,H_L\rangle\right]
= \int_{\H^m} \prod_{k=1}^m |\dx z_k|^2 f_k^\H(z_k)
\E\left[H_L(\Omega_{\cal U}^{-1}(z_1))\cdots H_L(\Omega_{\cal U}^{-1}(z_m))\right].
\end{multline}
Theorem~\ref{TheoremCorr} and Lemma~\ref{LemmaIntermDist} allow us to determine the last expected value as soon as $|z_i-z_j|$ are away at least of order $\delta:=L^{-1/(8m)}$. Denote by
\begin{equation}
{\H}^m_\delta=\{(z_1,\ldots,z_m)\in{\H}^m \textrm{ s.t. }|z_i-z_j|\leq \delta, 1\leq i<j\leq m\}.
\end{equation}
Then, as $L\to\infty$, we have
\begin{equation}\label{eq5.76}
\begin{aligned}
&\int_{\H^m_\delta} \prod_{k=1}^m |\dx z_k|^2 f_k^\H(z_k)
\E\left[H_L(\Omega_{\cal U}^{-1}(z_1))\cdots H_L(\Omega_{\cal U}^{-1}(z_m))\right]\\
=&\int_{\H^m_\delta} \prod_{k=1}^m |\dx z_k|^2 f_k^\H(z_k) \E\left[\GFF(z_1)\cdots \GFF(z_m)\right] +\Or(L^{-1/(12m)}). \end{aligned}
\end{equation}
Then, since the logarithm is integrable around zero (in two but also in one dimension), the $L\to\infty$ limit is simply given by
\begin{equation}
\lim_{L\to\infty} (\ref{eq5.76}) =\int_{\H^m} \prod_{k=1}^m |\dx z_k|^2 f_k^\H(z_k) \E\left[\GFF(z_1)\cdots \GFF(z_m)\right].
\end{equation}

We still need to control the contribution coming from $\H^m \setminus \H^m_\delta$. Using Lemma~\ref{LemmaShortDistance}, this is bounded by
\begin{equation}
\bigg|\int_{\H^m\setminus \H^m_\delta} \prod_{k=1}^m |\dx z_k|^2 f_k^\H(z_k)
\E\left[H_L(\Omega_{\cal U}^{-1}(z_1))\cdots H_L(\Omega_{\cal U}^{-1}(z_m))\right]\bigg|
\leq \cte \delta^2 L^\e
\end{equation}
where $\cte$ depends only the functions $f_1,\ldots,f_m$. Since
$\delta^2=L^{-1/(4m)}$ and $\e>0$ can be chosen smaller than
$1/(4m)$, in the $L\to\infty$ limit this contribution vanishes.
\end{proofOF}

We actually have a stronger result. Indeed the same formula holds
also for smooth functions living on one-dimensional paths. Consider
now any simple path $\gamma$ on $\cal U$ and denote by $s$ a
coordinate on $\gamma$. Denote by $\Omega_\gamma$ the composition of
the map from $s$ to $(\nu,\eta,\tau)$ and $\Omega$, and by
$\gamma_\H\subset \H$ the image of $\gamma$ by $\Omega_\gamma$.
Then, we define
\begin{equation}
\langle f,H_L\rangle_\gamma:=\int_{\gamma}\dx s f(s) H_L(s)
\end{equation}
and we get
\begin{equation}
\langle f,H_L\rangle_\gamma = \int_{\gamma_\H} \dx z J_\gamma(z) f(\Omega_{\gamma}^{-1}(z)) H_L(\Omega_{\gamma}^{-1}(z)),
\end{equation}
with $J_\gamma(z)$ the Jacobian of the change of variables from $z$ back to $s$ by $\Omega_{\gamma}^{-1}$.

\begin{thm}\label{ThmGFF1d}
For any $m\in \N$, consider any smooth functions $f_1,\ldots,f_m$ of compact support on $\gamma$. Then
\begin{multline}
\lim_{L\to\infty}\E\left[\prod_{k=1}^m\langle f_k,H_L\rangle_\gamma\right]
= \int_{\gamma_\H^m} \prod_{k=1}^m |\dx z_k| f_k^\gamma(z_k) \E\left[\GFF(z_1)\cdots \GFF(z_m)\right]
\end{multline}
where $f_i^\gamma(z):=J_\gamma(z) f(\Omega_{\gamma}^{-1}(z))$.
\end{thm}

\begin{proofOF}{Theorem~\ref{ThmGFF1d}}
The strategy is the same as in the proof of Theorem~\ref{ThmGFF2d}. The main difference is that the contribution at small distances will be of order $\delta L^\e$. However, this is fine, since we can choose $\e<1/(8m)$.
\end{proofOF}

\section{Asymptotics analysis}\label{SectAsymptAnalysis}
In this section we do the asymptotic analysis of the functions $q_n$'s at the (upper) edge and at the bulk. These are used to obtain the Gaussian fluctuations in Section~\ref{sectGaussFluct}. Then, we do the asymptotic analysis of the extended kernel in the bulk and provide some bounds at the (upper) edge, needed to prove the Gaussian Free Field correlations in Section~\ref{SectCorrelations}.

\subsection{Asymptotics at the edge}\label{SubsectAsymptEdge}
First we will determine the upper edge asymptotic of $I_{n,t}$ defined in (\ref{eqIn}), for which we apply exactly the same strategy as in previous papers (Lemma~\ref{Lemma1b} and~\ref{Lemma2b} are almost identical to the computations of Propositions~15 and 17 in~\cite{BFS07b}). We first explain the strategy and then give the relevant details.

\begin{lem}[Upper edge]\label{Lemma1b}
Let $n=ct$ and $x=(1+\sqrt{c})^2t+s t^{1/3}$, for any $c>0$. Then,
\begin{equation}\label{eqLem1b}
\lim_{t\to\infty} t^{1/3} I_{n,t}(x) \frac{(-\sqrt{c})^n}{e^{-\sqrt{c}t}(1+\sqrt{c})^x}=\tilde \kappa_2 \Ai(\kappa_2 s),
\end{equation}
uniformly for $s$ in bounded sets, with \mbox{$\kappa_2=c^{1/6} (1+\sqrt{c})^{-2/3}$}, and \mbox{$\tilde \kappa_2=(1+\sqrt{c})^{1/3} c^{-1/3}$}. Here $\Ai(\cdot)$ is the classical Airy function.
\end{lem}

\begin{proofOF}{Lemma~\ref{Lemma1b}}
The strategy is the following. With the replacements $n=ct$ and $x=(1+\sqrt{c})^2t+s t^{1/3}$ in (\ref{eqIn}), we have an integral of the form
\begin{equation}
\frac{1}{2\pi\I}\oint_{\Gamma_0}\dx z e^{t f_0(z)+t^{2/3} f_1(z)+t^{1/3} f_2(z)+f_3(z)}
\end{equation}
for some functions $f_k(z)$, $k=0,1,2,3$. The $s$-dependence is in $f_2(z)$.\\[0.5em]
\textbf{Step 1:} Find a steep descent path\footnote{For an integral
$I=\int_\gamma \dx z e^{t f(z)}$, we say that $\gamma$ is a steep
descent path if (1) $\Re(f(z))$ reaches the maximum at some
$z_0\in\gamma$: $\Re(f(z))< \Re(f(z_0))$ for
$z\in\gamma\setminus\{z_0\}$, and (2) $\Re(f(z))$ is monotone along
$\gamma$ except at its maximum point $z_0$ and, if $\gamma$ is
closed, at a point $z_1$ where the minimum of $\Re(f)$ is reached.}
for the function $f_0(z)$, passing through the double critical point
$z_c$ given by the condition $f_0'(z_c)=f_0''(z_c)=0$. In
particular, the steep descent path will be chosen so that close to
the critical point the descent it the steepest. Then, uniformly for
$s$ in a bounded set, the contribution coming from the integration
path away from a
$\delta$-neighborhood of $z_c$ is of order $\Or(e^{-\mu t})$ with $\mu\sim \delta^3$.\\[0.5em]
\textbf{Step 2:} Consider the contribution of the integration only on \mbox{$|z-z_c|\leq \delta$}, with $\delta$ which can still be chosen small enough, but $t$-independent. In a neighborhood of the critical point, we can use Taylor expansion of the functions $f_0,\ldots,f_3$ and get
\begin{multline}\label{eq444}
\exp(t f_0(z_c)+t^{2/3} f_1(z_c)+t^{1/3} f_2(z_c)+f_3(z_c)) \\
\times\frac{1}{2\pi\I}\int_{\Gamma_0\cap |z-z_c|\leq \delta}\dx z \exp(t \kappa_0 (z-z_c)^3/3+t^{2/3}\kappa_1 (z-z_c)^2+ t^{1/3}\kappa_2 (z-z_c)) \\
\times \exp(\Or(t(z-z_c)^4,t^{2/3}(z-z_c)^3,t^{1/3}(z-z_c)^2,(z-z_c))).
\end{multline}
Remark that we do not have a term $t^{2/3}(z-z_c)$ in the exponential. If such a term remains, than the edge scaling in $x$ is not the right one.\\[0.5em]
\textbf{Step 3:} Estimate the error terms. We do the change of variable \mbox{$t^{1/3}(z-z_c)=w$} and choose $\delta$ small enough, so that the error terms are much smaller than the main ones. Subsequently, taking $t$ large enough, the cubic term dominates all the others. Applying $|e^{y}-1|\leq |y|e^{|y|}$ with $y$ standing for the error term $\Or(\cdots)$, and changing the variable $t^{1/3}(z-z_c)=w$, one sees that the difference between the integral with and without the error term is of order $\Or(t^{-1/3})$.\\[0.5em]
\textbf{Step 4:} For the integral without errors, we also do the change of variable $t^{1/3}(z-z_c)=w$ and then we extend the integration paths to infinity. This accounts for an error of order $\Or(e^{-\mu t})$. The final formula is then
\begin{multline}
t^{-1/3} \exp(t f_0(z_c)+t^{2/3} f_1(z_c)+t^{1/3} f_2(z_c)+f_3(z_c)) \\ \times \left(\frac{\pm 1}{2\pi\I}\int\dx w e^{\kappa_0 w^3/3+\kappa_1 w^2+ \kappa_2 w}+\Or(t^{-1/3},e^{-\mu t})\right),
\end{multline}
where the integral goes from $e^{-\pi\I/3}\infty$ to $e^{\pi\I/3}\infty$ if $\kappa_0>0$ and, in case $\kappa_0<0$ it goes from $e^{-2\pi\I/3}\infty$ to $e^{2\pi\I/3}\infty$. The sign $\pm 1$ depends on the position of the critical point: we have $+1$ if $z_c>0$ and $-1$ if $z_c<0$. Finally, the integral can we rewritten in terms of Airy functions using the following identity
\begin{equation}
\frac{1}{2\pi\I}\int_{e^{-\pi\I/3}\infty}^{e^{\pi\I/3}\infty}e^{az^3/3+bz^2+cz}\dx z =a^{-1/3}\Ai(b^2/a^{4/3}-c/a^{1/3})e^{2b^3/3a^2-bc/a}.
\end{equation}

\vspace{0.5em}
\noindent\textbf{Specialization to our case.} In our specific situation, the critical point is $z_c=-\sqrt{c}$, and the functions $f_0,\ldots,f_3$ are
\begin{eqnarray}
f_0(z)&=&z+(1+\sqrt{c})^2\ln(1-z)-c\ln(z),\nonumber \\
f_1(z)&=&0,\nonumber \\
f_2(z)&=&s\ln(1-z),\nonumber \\
f_3(z)&=&-\ln(z).
\end{eqnarray}
The steep descent path used in the analysis is made of pieces of the two following paths, $\gamma_\rho$ and $\gamma_{\rm loc}$ (see Figure~\ref{FigSteepPath}), given by
\begin{equation}
\gamma_\rho=\{-\rho e^{\I\phi},\phi\in(-\pi,\pi]\},\quad \gamma_{\rm loc}=\{-\sqrt{c}+ e^{-\pi\I/3 \sgn(x)}|x|,x\in [0,\sqrt{c}/2]\}.
\end{equation}
\begin{figure}
\begin{center}
\psfrag{zc}[r]{$-\sqrt{c}$}
\psfrag{grho}{$\gamma_{\rho}$}
\psfrag{gloc}{$\gamma_{\rm loc}$}
\includegraphics[height=5cm]{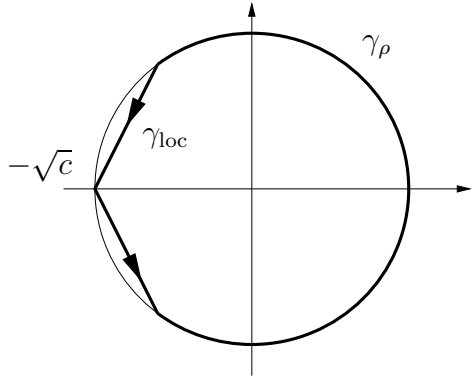}
\caption{The steep descent path used in the asymptotic analysis is the bold one.}
\label{FigSteepPath}
\end{center}
\end{figure}

For $\rho\in (0,\sqrt{c}]$, $\gamma_\rho$ is steep descent path for $f_0$. In fact, we get
\begin{equation}
\frac{\dx \Re(f_0(z=\rho e^{\I\phi}))}{\dx \phi} = -\frac{\rho \sin\phi}{|1-z|^2}(c-\rho^2+2\sqrt{c}-2\rho\cos\phi).
\end{equation}
The last term is minimal for $\phi=0$, where it equals
\begin{equation}\label{eq6.5}
(\sqrt{c}-\rho)(\rho+\sqrt{c}+2)\geq 0,
\end{equation}
for $\rho\in (0,\sqrt{c}]$. $\gamma_\rho$ is a steep descent path
for $f_0$ because the value zero is attained only for
$\rho=\sqrt{c}$ and, in that case, only at one point, $\phi=0$.
However, close to the critical point it is not optimal, because the
steepest descent path leaves $z_c$ at angle $\pm\pi/3$ (there are
rays where $\Im(z-z_c)^3=0$). By symmetry, we need to consider only
$x\geq 0$.
\begin{equation}
\frac{\dx \Re(f_0(z=-\sqrt{c}+e^{-\pi\I/3} x))}{\dx x} = -\frac{x^2 Q(x)}{|z|^2|1-z|^2},
\end{equation}
with $Q(x)=\sqrt{c}(1+\sqrt{c})-x(1+x)/2-\sqrt{c}x$. $Q(0)>0$, and
the computation of the (at most) two zeros of $Q(x)$ shows that none
are in the interval $[0,\sqrt{c}/2]$. Thus $\gamma_{\rm loc}$ is
also a steep descent path for $f_0$. Since this is the steepest
descent path for $f_0$ around the critical point, we choose as path
$\Gamma_0$ in $I_{n,t}(x)$ the one formed by $\gamma_{\rm loc}$
close to the critical point, until it intersects
$\gamma_{\rho=\sqrt{3c/4}}$ ,and then we follow
$\gamma_{\sqrt{3c/4}}$.

The Taylor expansions near the critical point $z_c=-\sqrt{c}$ of the
functions $f_k$ are given by
\begin{eqnarray}\label{eqSeriesf}
f_0(z)&=&f_0(-\sqrt{c})+\tfrac13 \kappa_0 (z+\sqrt{c})^3+\Or((z+\sqrt{c})^4),\quad \kappa_0=\frac{1}{\sqrt{c}(1+\sqrt{c})},\nonumber \\
f_2(z)&=&f_2(-\sqrt{c})+\kappa_2(z+\sqrt{c})+\Or((z+\sqrt{c})^2),\quad \kappa_2=-\frac{s}{1+\sqrt{c}},\nonumber \\
f_3(z)&=&-\ln(-\sqrt{c})+\Or(z+\sqrt{c}).
\end{eqnarray}
Thus in our case we have
\begin{equation}
a=\kappa_0=1/(\sqrt{c}(1+\sqrt{c})),\quad b=0,\quad c=-s/(1+\sqrt{c}),\quad e^{f_3(z_c)}=-1/\sqrt{c}.
\end{equation}
This, together with the relation
\begin{equation}
\exp(t f_0(z_c)+t^{2/3} f_1(z_c)+t^{1/3} f_2(z_c))=\frac{(1+\sqrt{c})^x e^{-\sqrt{c}t}}{(-\sqrt{c})^n}
\end{equation}
proves (\ref{eqLem1b}).
\end{proofOF}

\begin{lem}\label{Lemma2b}
Fix $\ell>0$ and consider the scaling of Lemma~\ref{Lemma1b}. Then
\begin{equation}
\left|t^{1/3} I_{n,t}(x) \frac{(-\sqrt{c})^n}{e^{-\sqrt{c}t}(1+\sqrt{c})^x}\right|\leq \cte e^{-s},
\end{equation}
uniformly for $s\geq -\ell$, where $\cte$ is a constant independent of $t$.
\end{lem}
\begin{proofOF}{Lemma~\ref{Lemma2b}}
For any finite $\tilde\ell$, the bound for $s\in[-\ell,\tilde\ell]$ is a consequence of Lemma~\ref{Lemma1b}. The value of $\tilde\ell$ can be chosen large but independent of $t$. The strategy for $s\geq \tilde\ell$ is just a small modification of the computation made in Lemma~\ref{Lemma1b}, and was already used for example in Proposition 17 of~\cite{BFS07b} and in Proposition 5.3 of~\cite{BF07}. Let us explain it.

In Lemma~\ref{Lemma1b} we have seen that $\gamma_\rho$ is steep descent path for $f_0$, for any $\rho\in(0,\sqrt{c}]$. Set $\tilde s=(s+\ell+\tilde\ell) t^{-2/3}\geq \tilde\ell t^{-2/3}>0$ and $\tilde f_0(z) =f_0(z)+\tilde s \ln(1-z)$. For any $\tilde s\geq 0$, $\gamma_\rho$ is also a steep descent path for $\tilde f_0(z)$. However, for $\tilde s>0$ there are two real critical points for $\tilde f_0$, say at $z_c^\pm$ with $|z_c^+| > |z_c^-|$. For $\tilde s$ small, we have at lowest order in $\tilde s$, $z_c^\pm\simeq -\sqrt{c}\mp \sqrt{\tilde s}\sqrt{\kappa_2/\kappa_0}$, with $\kappa_0$ and $\kappa_2$ given in (\ref{eqSeriesf}). To get the best bound we should pass through $z_c^-$. However, this precision is not needed to get exponential bound and we can choose the integration path passing through
\begin{equation}\label{eq4.7BB}
-\rho=\left\{\begin{array}{ll}
-\sqrt{c}+(\tilde s \kappa_2/\kappa_0)^{1/2},&\textrm{ if }0\leq \tilde s \leq \e,\\
-\sqrt{c}+(\e \kappa_2/\kappa_0)^{1/2},&\textrm{ if }\tilde s \geq \e.
\end{array}\right.
\end{equation}
With this choice, for $\e$ small enough, we have $-\sqrt{c}<-\rho<z_c^-$ and in particular, for small $\tilde s$, $-\rho$ is very close to the position of the critical point. As in Lemma~\ref{Lemma1b}, we use the fact that $\gamma_\rho$ is steep descent to control the contribution away from $|z+\rho|\leq \delta$, while the contribution close to $z=-\rho$ is controlled by the Taylor expansion of $\Re(\tilde f_0(z))$, leading to a Gaussian bound. By choosing $\tilde\ell$ large enough, all the terms coming from $\Re(f_k(z))$, $k=1,2,3$ are dominated by the leading term of $\Re(f_0(z))$. The final result is that
\begin{equation}
\left|t^{1/3} I_{n,t}(x)\frac{(-\sqrt{c})^n}{e^{-\sqrt{c}t}(1+\sqrt{c})^x}\right|\leq \cte Q(\rho),
\end{equation}
with
\begin{equation}
Q(\rho)=\exp(\Re(t (f_0(-\rho)-f_0(z_c))+t^{2/3}
(f_1(-\rho)-f_1(z_c))+t^{1/3} (f_2(-\rho)-f_2(z_c))).
\end{equation}
$Q(\rho$) is decreasing for $-\rho$ from $z_c$ to $z_c^-$ and $-\rho-z_c$ is at most of order $\sqrt{\e}$. Thus, we can easily bound $Q(\rho)$ by using Taylor expansions. Simple computations lead to the desired exponential bound.
\end{proofOF}

To get the needed bound on $q_n$ (see (\ref{eqCharlier})-(\ref{eqQn})) around the edge, we use the bound of Lemma~\ref{Lemma2b} on $I_{n,t}$ which has still to be multiplied by $B_{n,t}(x)$ given in (\ref{eqQn}).

\begin{lem}\label{Lemma3b}
Let $n=ct$, $x=(1+\sqrt{c})^2t+s t^{1/3}$ and fix $\ell>0$. Then
\begin{equation}
\left|q_n(x,t)\right|\leq \cte t^{-1/3} e^{-s},
\end{equation}
for any $s\geq -\ell$, and $\cte$ is a $t$-independent constant.
\end{lem}
\begin{proofOF}{Lemma~\ref{Lemma3b}}
This result follows from Lemma~\ref{Lemma2b} if
\begin{equation}
\tilde B_{n,t}(x)=\Big|B_{n,t}(x)\frac{e^{-\sqrt{c}t}(1+\sqrt{c})^x}{(-\sqrt{c})^{n}}\Big|\leq \cte.
\end{equation}
For the factorials we use Stirling formula, namely
\begin{equation}\label{Stirling}
n!=\sqrt{2\pi n}\left(\frac{n}{e}\right)^{n}e^{f_n},\quad \frac{1}{1+12 n}\leq f_n \leq \frac{1}{12n}.
\end{equation}
We obtain
\begin{equation}
\tilde B_{ct,t}((1+\sqrt{c})^2t)=\left(\frac{c}{(1+\sqrt{c})^2}\right)^{1/4}(1+\Or(1/t)).
\end{equation}
For $x=\xi t$, $\xi\in [(1+\sqrt{c})^2,\infty)$, we compute
\begin{equation}
\frac{\tilde B_{ct,t}(\xi t)}{\tilde B_{ct,t}((1+\sqrt{c})^2t)}=\left(\frac{(1+\sqrt{c})^2}{\xi}\right)^{1/4}(1+\Or(1/t))e^{t h(\xi)},
\end{equation}
with
\begin{equation}\label{eq6.22}
h(\xi)=\tfrac12\xi(1-\ln(\xi)+2\ln(1+\sqrt{c}))-\tfrac12 (1+\sqrt{c})^2.
\end{equation}
Since $h'(\xi)=0$ at $\xi=(1+\sqrt{c})^2$ and $h''(\xi)=-1/(2\xi)<0$, we have \mbox{$e^{t h(x)}\leq 1$}.
\end{proofOF}

\subsection{Asymptotics in the bulk}
In this section we derive a precise expansion of $I_{n,t}(x)$ for $x=\lambda t$, with $\lambda \in ((1-\sqrt{c})^2,(1+\sqrt{c})^2)$. For any fixed $c>0$, set $n=ct$ and $x=\lambda t$. Then
\begin{equation}\label{eqInt}
I_{n,t}(x)=\frac{1}{2\pi\I}\oint_{\Gamma_0}\frac{\dx w}{w} e^{t g(w)},\quad g(w)=G(w|\lambda,c,1),
\end{equation}
see (\ref{eqDefinG}) for the definition of $G$. Recall a few results from Section~\ref{SectGeometry}. For $\lambda\in ((1-\sqrt{c})^2,(1+\sqrt{c})^2)$, $g$ has two complex conjugate critical points, $w_c$ and $\bar{w}_c$, with $w_c=\Omega(\lambda,c,1)$. In particular, $|w_c|=\sqrt{c}$, $|1-w_c|=\sqrt{\lambda}$, and $|g''(w_c)|=\frac{1}{\sqrt{\lambda c}}\sqrt{4c-(1+c-\lambda)^2}$.
When $(\eta,\nu,\tau)=(c,\lambda,1)$, we denote by $\pi_c$ the angle $\pi_\eta$ and by $\pi_\lambda$ the angle $\pi_\nu$. Then
\begin{equation}\label{eqRealImag}
\begin{aligned}
\Re(g(w_c))&=\frac{1+c-\lambda}{2}-\frac{c}{2}\ln(c)+\frac{\lambda}{2}\ln(\lambda), \\
\Im(g(w_c))&=\Im(w_c)-\lambda \pi_c-c\pi_\lambda,\\
\arg(g(w_c))&= -\frac{\pi}{2}+\pi_c-\pi_\lambda.
\end{aligned}
\end{equation}

\begin{lem}\label{Lemma4}
Set $\alpha=\alpha(c,\lambda)=\Im(g(w_c))$ and $\beta=\beta(c,\lambda)=-\tfrac12(\pi_c+\pi_\lambda+\pi/2)$. Then, as $t\to\infty$,
\begin{equation}\label{eqLem4}
I_{ct,t}(\lambda t)=\frac{e^{t\Re(g(w_c))}}{\sqrt{|g''(w_c)|t}} \Bigg[\sqrt{\frac{2}{\pi|w_c|^2}}\cos(t\alpha+\beta)
+\Or(t^{-1/2})\Bigg].
\end{equation}
For any $\e_0>0$, the errors are uniform for $\lambda\in [(1-\sqrt{c})^2+\e_0,(1+\sqrt{c})^2-\e_0]$.
\end{lem}

\begin{rem}
In fact, we prove the bound of (\ref{eqLem4}) with error term
\begin{equation}
\Or(t^{-1/2})+\Or\Big(\sqrt{|g''(w_c)|t} e^{-\cte |g''(w_c)| \delta^2 t}\Big)
\end{equation}
for some $0<\delta\ll |g''(w_c)|$. In Lemma~\ref{Lemma5} we will have to be careful with the second term of the bound, since $g''(w_c)$ goes to zero at the edge.
\end{rem}

\begin{proofOF}{Lemma~\ref{Lemma4}}
The critical points of $g$, the points such that $g'(w)=0$, are $w_c$ and its complex conjugate $\bar w_c$. Close to $w_c$ the Taylor expansion of $g$ has a first relevant term which is quadratic,
\begin{equation}
g(w)=g(w_c)+\tfrac12 g''(w_c)(w-w_c)^2+\Or((w-w_c)^3).
\end{equation}

Now we construct the steep descent path used in the asymptotics. By symmetry we consider only $\Im(w)\geq 0$, the path for $\Im(w)\leq 0$ will be the complex conjugate image of the first one. Let $\gamma_\rho=\{w=\rho e^{\I\phi},\phi\in [0,\pi]\}$, then
\begin{equation}\label{eqSteep}
\frac{\dx}{\dx \phi}(\Re(g(w=\rho e^{\I\phi}))) = \rho\sin(\phi)\left[\frac{\lambda}{|1-w|^2}-1\right].
\end{equation}
This is positive if $|1-w|< \sqrt{\lambda}$, and negative otherwise.

Locally, consider the path $\gamma_{\rm loc}=\{w=w_c+\hat\theta x,x\in [-\delta,\delta]\}$. Then
\begin{equation}
g(w)=g(w_c)+\tfrac12 g''(w_c)\hat \theta^2 x^2+\Or(x^3),
\end{equation}
where we choose
\begin{equation}\label{eqTheta}
\hat \theta=\exp\left(\frac{\I\pi}{2}-\frac{\I}{2}\arg(g''(w_c))\right)
=\exp\left(\frac{3\pi\I}{4}+\frac{\I(\pi_\lambda-\pi_c)}{2}\right).
\end{equation}
For $-\delta<x<0$, the path $\gamma_{\rm loc}$ is closer to $1$ than $\sqrt{\lambda}$, while for $0<x<\delta$ the path $\gamma_{\rm loc}$ is farther from $1$ than $\sqrt{\lambda}$. This is the case since our $\gamma_{\rm loc}$ has an angle between $\pi/4$ and $3\pi/4$ to the tangent to the circle $|1-w|=\sqrt{\lambda}$.

So, the steep descent path used is the following: we extend
$\gamma_{\rm loc}$ by adding two circular arcs of type
$\gamma_\rho$, for adequate $\rho$, which connect to the real axis;
finally we add the complex conjugate image, see
Figure~\ref{FigPathBulk} too.
\begin{figure}
\begin{center}
\psfrag{0}[t]{$0$}
\psfrag{1}[t]{$1$}
\psfrag{zc}[c]{$w_c$}
\psfrag{G0}[l]{$\Gamma_0$}
\psfrag{z}[r]{$\sqrt{c}$}
\psfrag{1-z}{$\sqrt{\lambda}$}
\includegraphics[height=5cm]{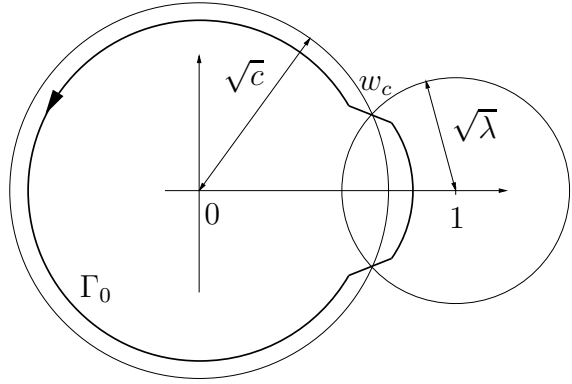}
\caption{Illustration of the steep descent path.}
\label{FigPathBulk}
\end{center}
\end{figure}
In this way, we have a steep descent path. Thus,
\begin{equation}\label{eq1.31}
I_{n,t}(x)=e^{t \Re(g(w_c))}\Or(e^{-\mu t})+2 \Re\left(\frac{1}{2\pi\I}\int_{\gamma_{\rm loc}}\frac{\dx w}{w} e^{t g(w)}\right)
\end{equation}
with $\mu\sim |g''(w_c)| \delta^2$, as soon as $|g''(w_c)|>0$, i.e., as soon as the second order term dominates all higher order terms in the Taylor expansion.

The second term of (\ref{eq1.31}) is given by
\begin{eqnarray}\label{eq1.32}
\frac{1}{2\pi\I}\int_{\gamma_{\rm loc}}\frac{\dx w}{w} e^{t g(w)} &=& \frac{1}{2\pi\I}\int_{-\delta}^\delta \dx x \frac{\hat\theta}{w_c} e^{t g(w_c)} e^{-\tfrac12 t |g''(w_c)| x^2} e^{\Or(tx^3)} \Or(x)\nonumber\\
&=&\frac{1}{2\pi\I} \frac{\hat\theta}{w_c}\int_{-\delta}^\delta \dx x e^{t g(w_c)} e^{-\tfrac12 t |g''(w_c)| x^2} + E_1
\end{eqnarray}
where
\begin{equation}\label{eq1.33}
E_1=\frac{1}{2\pi\I}\frac{\hat\theta}{w_c}\int_{-\delta}^\delta \dx x e^{t g(w_c)} e^{-\tfrac12 t |g''(w_c)| x^2} e^{\Or(tx^3)}\Or(tx^3,x).
\end{equation}
Here we used $|e^{x}-1|\leq |x|e^{|x|}$. Changing variable $y=x\sqrt{t}$, we get that
\begin{eqnarray}\label{eq6.32}
|E_1|&\leq& \cte e^{t \Re(g(w_c))} \frac{1}{t}\int_{-\delta\sqrt{t}}^{\delta \sqrt{t}}\dx y e^{-|g''(w_c)|y^2/2} \Or(y,y^3/\sqrt{t}) e^{\Or(y^3/\sqrt{t})} \nonumber\\
&\leq& \cte \frac{e^{t \Re(g(w_c))}}{t\sqrt{|g''(w_c)|}}
\end{eqnarray}
for $\delta$ small enough, i.e., for $0<\delta\ll |g''(w_c)|$. In this small neighborhood, the quadratic term controls the higher order ones. The final step is to extend the integral on the rest of r.h.s.\ of (\ref{eq1.32}) from $\pm \delta$ to $\pm \infty$. This can be made up to an error $e^{t \Re(g(w_c))}\Or(e^{-\mu t})$ as above.

Resuming we have (counting the contribution from both critical points)
\begin{eqnarray}\label{eq4.29}
I_{n,t}&=&e^{t \Re(g(w_c))}\left[\Or(e^{-\mu t})+\Or(1/(t\sqrt{|g''(w_c)|})\right] \nonumber \\
& & + 2\Re\left(\frac{1}{2\pi\I} \frac{\hat\theta}{w_c}\int_{\R} \dx x e^{t g(w_c)} e^{-\tfrac12 t |g''(w_c)| x^2} \right).
\end{eqnarray}
The error terms are the one indicated in (\ref{eqLem4}), and the
Gaussian integral for the last term gives
\begin{equation}
\frac{2 e^{t\Re(g(w_c))}}{\sqrt{2\pi t \, |w_c|^2\, |g''(w_c)|}} \Re\left(-\I \hat\theta \frac{|w_c|}{w_c} e^{\I t \Im(g(w_c))} \right).
\end{equation}
We then set $\beta=\arg(-\I\hat\theta/w_c)=-\pi/4-(\pi_c+\pi_\lambda)/2$, so that $-\I \hat\theta \frac{|w_c|}{w_c}=e^{\I\beta}$. For $\lambda$ in a compact subset of $((1-\sqrt{c})^2,(1+\sqrt{c})^2)$, $|g''(w_c)|$ is uniformly bounded away from zero and infinity. Thus the Lemma is proven.
\end{proofOF}

The bound of Lemma~\ref{Lemma4} can be easily extended until position away of order $\Or(t^{2/3})$ from the upper edge.
\begin{lem}\label{CorLemma4}
Set $\alpha=\Im(g(w_c))$ and $\beta=-\tfrac12(\pi_c+\pi_\lambda+\pi/2)$. Then, for $\lambda\in[(1-\sqrt{c})^2+\e_0,(1+\sqrt{c})^2-t^{-1/3}]$, for any fixed $\e_0>0$, we have the uniform estimate
\begin{equation}
I_{ct,t}(\lambda t)=\frac{e^{t\Re(g(w_c))}}{\sqrt{|g''(w_c)|t}} \Bigg[\sqrt{\frac{2}{\pi|w_c|^2}}\cos(t\alpha+\beta)
+\Or(t^{-1/2})+\Or\Big(\sqrt{t} e^{-\cte t^{1/3}}\Big)\Bigg].
\end{equation}
\end{lem}
\begin{proofOF}{Lemma~\ref{CorLemma4}}
The analysis of Lemma~\ref{Lemma4} can be made also for this case, with only minor differences. Indeed, for $(1+\sqrt{c})^2-\lambda\sim t^{-1/3}$, we have $|g''(w_c)|\sim t^{-1/6}$ and this time we choose $\delta$ going to zero as $t\to\infty$, setting $\delta=t^{-1/4}$. With this choice, (\ref{eq1.31}) and (\ref{eq6.32}) are still valid because at the border of integration the quadratic term dominates the cubic one. Indeed, with $y=\delta \sqrt{t}=t^{1/4}$, it holds $y^3/\sqrt{t}\sim t^{1/4}\ll t^{1/3}\sim |g''(w_c)| y^2$. Also, the error term coming from steep descent in (\ref{eq1.31}) will vanish as $t\to\infty$ but slower than before, with $\mu t\sim t^{1/3}$.
\end{proofOF}

The results of Lemma~\ref{Lemma4} and Lemma~\ref{CorLemma4} imply the following asymptotics for the functions $q_n$'s.
\begin{lem}\label{Lemma5}
Set $\alpha=\Im(g(w_c))$ and $\beta=-\tfrac12(\pi_c+\pi_\lambda+\pi/2)$ and fix any $\e_0>0$. Then, uniformly in $\lambda\in [(1-\sqrt{c})^2+\e_0,(1+\sqrt{c})^2-t^{-1/3}]$, we have
\begin{equation}\label{eq4.31}
q_{ct}(\lambda t,t)= \frac{1}{\sqrt{\pi}}\frac{t^{-1/2}}{\sqrt[4]{c-\frac{(1+c-\lambda)^2}{4}}} \Bigg[\cos(t\alpha+\beta)+\Or(t^{-1/2})\Bigg].
\end{equation}
\end{lem}
\begin{proofOF}{Lemma~\ref{Lemma5}}
We just have to compute the prefactor $B_{ct,t}(\lambda t) e^{t\Re(g(w_c))}$. We have (\ref{eqRealImag}) and applying Stirling formula for the factorials in $B_{ct,t}(\lambda t)$ we get that
\begin{equation}
B_{ct,t}(\lambda t) e^{t\Re(g(w_c))} = (\lambda/c)^{1/4}(1+\Or(1/t)).
\end{equation}
\end{proofOF}

Now we need to fill the gap between the bulk and the edge. In this region we do not need precise asymptotics, just a bound. Approaching the upper edge, $g''(w_c)$ goes to zero, but then everything can be controlled by the cubic term, because $|g'''(w_c)|\neq 0$ at the edges.

\begin{lem}\label{Lemma6}
For $\e_0>0$ fixed but small enough, and $\ell >0$ large enough, we have the bound
\begin{equation}
\left|q_{ct}(\lambda t,t)\right|\leq \cte \frac{t^{-1/2}}{\sqrt[4]{c-\frac{(1+c-\lambda)^2}{4}}},
\end{equation}
uniformly for $\lambda\in [(1+\sqrt{c})^2-\e_0, (1+\sqrt{c})^2-\ell t^{-2/3}]$.
\end{lem}
\begin{proofOF}{Lemma~\ref{Lemma6}}
Consider the $\e_0$-region close to the upper edge with $\e_0>0$ small enough. We can compute explicitly the direction $\hat \theta$, see (\ref{eqTheta}). It is a continuous function of $\lambda$ and, as $\lambda\uparrow (1+\sqrt{c})^2$, $\hat\theta \uparrow e^{\I 5\pi/4}$ (because $\pi_\lambda\uparrow \pi$ and $\pi_c\downarrow 0$). We need just a bound, so we choose $\hat\theta=e^{\I 5\pi/4}$ and set the local path as
\begin{eqnarray}
\gamma_{\rm loc}&=&\{w=w_c+e^{\I 5\pi/4} x, x\in[-\delta,\Im(w_c)\sqrt{2}]\}.
\end{eqnarray}
The path $\gamma_{\rm loc}$ reaches at $x=\Im(w_c)\sqrt{2}$ the imaginary axis and this is the reason for the upper edge of $\gamma_{\rm loc}$. We have
\begin{equation}
g(w)=g(w_c)+\tfrac12 g''(w_c) (w-w_c)^2+\tfrac16 g'''(w_c) (w-w_c)^3+\Or((w-w_c)^4).
\end{equation}
In a $\delta$-neighborhood of $w_c$, along the direction $\hat\theta$ chosen,
\begin{equation}
\Re(\tfrac12 g''(w_c) (w-w_c)^2)=-\tfrac12 |g''(w_c)| x^2 (1+\Or(\epsilon_0))
\end{equation}
and
\begin{equation}
\Re(\tfrac16 g'''(w_c) (w-w_c)^3)=\tfrac16 |g'''(w_c)| x^3/\sqrt{2} (1+\Or(\sqrt{\epsilon_0})).
\end{equation}
Therefore, for $\e_0$ small enough, the quadratic term helps the convergence. For $x\leq 0$, the cubic term helps the convergence, while for $x\in [0,\Im(w_c)\sqrt{2}]$ we will need to control it by the quadratic term. Thus,
\begin{equation}
I_{n,t}(x)=e^{t \Re(g(w_c))}\Or(e^{-\mu t})+2\Re\left(\frac{1}{2\pi\I}\int_{\gamma_{\rm loc}}\frac{\dx w}{w} e^{t g(w)}\right),
\end{equation}
with $\mu \simeq |g'''(w_c)|\delta^3$, where $g'''(w_c)\to 2/\sqrt{c}(1+\sqrt{c})$ as $\lambda \to (1+\sqrt{c})^2$.

Consider then the contribution coming from the integral over $\gamma_{\rm loc}$. We have
\begin{eqnarray}\label{eq6.47}
\left|\int_{\gamma_{\rm loc}}\frac{\dx w}{w} e^{t g(w)}\right|&\leq & \frac{e^{t\Re(g(w_c))}}{|w_c|}\int_{-\delta}^{\Im(w_c)\sqrt{2}}\dx x \exp\left(-\tfrac12 t |g''(w_c)|x^2\right) \\
&&\times\exp\left(\tfrac16 t |g'''(w_c)|x^3/\sqrt{2}+\Or(x^4 t)\right)(1+\Or(x)),\nonumber
\end{eqnarray}
the last $1/\sqrt{2}$ coming from $\Re(e^{-\I\pi/4})=1/\sqrt{2}$.
A simple verification gives
\begin{equation}\label{eq6.48}
-\tfrac12 t |g''(w_c)|x^2+\tfrac16 t |g'''(w_c)|x^3/\sqrt{2} \leq -\tfrac14 t |g''(w_c)|x^2, \quad 0\leq x \leq \Im(w_c)\sqrt{2}.
\end{equation}
So, for $x\in [0,\Im(w_c)\sqrt{2}]$, the quadratic term is still dominating higher order terms, including the cubic one (the quartic term can be bounded by replacing $\tfrac14$ by $\tfrac16$ in the above estimate).

On the other hand, for $-\delta\leq x\leq 0$, we have that the cubic term is negative and dominates all higher order terms. More precisely, for $\delta$ small enough,
\begin{equation}\label{eq6.49}
\left|\exp\left(\tfrac16 t |g'''(w_c)|x^3/\sqrt{2}+\Or(x^4 t)\right)\right|\leq \exp\left(\tfrac{1}{12} t |g'''(w_c)|x^3\right) \leq 1,
\end{equation}
in the region $x\in [-\delta,0]$.

Using (\ref{eq6.48}) for positive $x$ and (\ref{eq6.49}) for negative $x$, we get
\begin{multline}
(\ref{eq6.47}) \leq \cte e^{t\Re(g(w_c))} \int_{-\delta}^{\Im(w_c)\sqrt{2}}\dx x \exp\left(-\tfrac16 t |g''(w_c)|x^2\right) \\ \leq \cte e^{t\Re(g(w_c))}\frac{1}{\sqrt{|g''(w_c)|t}}.
\end{multline}
Replacing the value of $|g''(w_c)|$ into this expression ends the proof.
\end{proofOF}

\subsection{Asymptotic of the kernel}\label{SectKernelAsympt}
In this section we obtain the precise asymptotics of the extended kernel in the bulk first and a bound to control the behavior starting from the upper edge. Here we use several notations introduced in the Section~\ref{SectGeometry}.

As usual, it is convenient to conjugate the kernel before taking the
limit. For the upper edge (cf. Lemma~\ref{Lemma2b}) set
\begin{equation}
W_{i,u}=\exp\left(-\sqrt{n_i t_i}+x_i\ln(1+\sqrt{n_i/t_i})-n_i \ln(-\sqrt{n_i/t_i})-t_i\right),
\end{equation}
and, in the bulk (see Lemma~\ref{Lemma4}) set
\begin{equation}
W_{i,b}=\exp\left(\tfrac12(t_i+n_i-x_i)-\tfrac12 n_i\ln(n_i/t_i)+\tfrac12 x_i\ln(x_i/t_i)-t_i\right).
\end{equation}
Then, define the conjugation as
\begin{equation}
W_i=\left\{
\begin{array}{ll}
W_{i,b},&\textrm{for }(\sqrt{t_i}-\sqrt{n_i})^2\leq x_i \leq(\sqrt{t_i}+\sqrt{n_i})^2,\\
W_{i,u},&\textrm{for }x_i\geq (\sqrt{t_i}+\sqrt{n_i})^2.
\end{array}
\right.
\end{equation}
Remark that $W_i$ is continuous. Moreover, $|W_{i,u}-W_{i,b}|= \Or(L^{-1/3})$ for \mbox{$|x_i-(\sqrt{t_i}+\sqrt{n_i})^2|=\Or(L^{-1/3})$}. Therefore in such a neighborhood it is actually irrelevant which formula to use.

\begin{prop}\label{PropKernelAsymptExp}
Let us consider two triples $(x_1,n_1,t_1)$ and $(x_2,n_2,t_2)$
parameterized by
\begin{eqnarray}
x_i=[\nu_i L],\quad n_i=[\eta_i L],\quad t_i=\tau_i L.
\end{eqnarray}
Assume that they are in the bulk of the system, namely, that exists $\e_0>0$ such that
\begin{equation}
(\sqrt{\tau_i}-\sqrt{\eta_i})^2+\e_0\leq \nu_i \leq (\sqrt{\tau_i}+\sqrt{\eta_i})^2-L^{-1/3}.
\end{equation}
Denote $z_c=\Omega(\nu_1,\eta_1,\tau_1)$, $w_c=\Omega(\nu_2,\eta_2,\tau_2)$, and assume that these points are not too close: $|z_c-w_c|\geq L^{-1/16}$. Then, the asymptotic expansion
\begin{eqnarray}\label{eqAsymptKernel}
& &(W_1/W_2)K(x_1,n_1,t_1;x_2,n_2,t_2) =
\frac{1}{2\pi L\sqrt{|G''(w_c)|\, |G''(z_c)|}|1-z_c|} \nonumber \\
&\times&\bigg[\frac{1}{w_c-z_c}\frac{e^{\I L \Im(G(w_c))+\I \beta_2}}{e^{\I L \Im(G(z_c))+\I \beta_1}}
+\frac{1}{w_c-\bar z_c}\frac{e^{\I L \Im(G(w_c))+\I \beta_2}}{e^{-\I L \Im(G(z_c))-\I\beta_1}} \\
&&+\frac{1}{\bar w_c-z_c}\frac{e^{-\I L \Im(G(w_c))-\I \beta_2}}{e^{\I L \Im(G(z_c))+\I \beta_1}}
+\frac{1}{\bar w_c-\bar z_c}\frac{e^{-\I L \Im(G(w_c))-\I \beta_2}}{e^{-\I L \Im(G(z_c))-\I \beta_1}}
+\Or(L^{-1/8})\bigg]\nonumber
\end{eqnarray}
holds, with the error uniform in $L$ for $L\geq L_0\gg 1$. The phases $\beta_1$ and $\beta_2$ are given by
\begin{equation}
\beta_1=-\frac{5\pi}{4}-\frac{\pi_{\nu_1}}{2}-\frac{\pi_{\eta_1}}{2},\quad
\beta_2=\frac{3\pi}{4}+\frac{\pi_{\nu_2}}{2}-\frac{\pi_{\eta_2}}{2}.
\end{equation}
\end{prop}

\begin{proofOF}{Proposition~\ref{PropKernelAsymptExp}}
The analysis relies on the double integral representation
(\ref{eqDoubleIntRepr}) of the kernel. The analysis for the cases
$(n_1,t_1)\not\prec (n_2,t_2)$ and $(n_1,t_1)\prec (n_2,t_2)$ are
very similar. Let us explain the first case, corresponding to
$\eta_1>\eta_2$, or $\tau_1<\tau_2$, or
$(\eta_1,\tau_1)=(\eta_2,\tau_2)$. The asymptotics employs several
ingredients already used in Lemma~\ref{Lemma4} and
Lemma~\ref{CorLemma4}. Thus, we introduce the notations
\begin{equation}
c_i=\eta_i/\tau_i\Rightarrow n_i=[c_i t_i],\quad \lambda_i=\nu_i/\tau_i\Rightarrow x_i=[\lambda_i t_i].
\end{equation}
The conjugation factor $e^{t_1-t_2}$ in the kernel representation (\ref{eqDoubleIntRepr}) will not appear in the following computations, since it appears automatically in the factors $W_1/W_2$. Thus, we have to analyze
\begin{equation}\label{eq4.35}
\frac{1}{(2\pi\I)^2}\oint_{\Gamma_0}\dx w \oint_{\Gamma_1}\dx z e^{t_2 g_2(w)-t_1 g_1(z)}\frac{1}{(1-z)(w-z)}
\end{equation}
with $g_i(w)=w+\lambda_i\ln(1-w)-c_i\ln(w) \equiv G(w|\lambda_i,c_i,1)$, $i=1,2$.

The critical points of $g_{2}(w)$ and $g_{1}(z)$ are given by
\begin{equation}
w_c=\Omega(\lambda_2,c_2,1)=\Omega(\nu_2,\eta_2,\tau_2),\quad z_c=\Omega(\lambda_1,c_1,1)=\Omega(\nu_1,\eta_1,\tau_1).
\end{equation}
The integrals over $w$ are, up to the factor $w/(z-w)$, as in
Lemma~\ref{Lemma4}. Therefore, the steep descent path $\Gamma_0$ is
chosen as in Lemma~\ref{Lemma4} and the steep descent path
$\Gamma_1$ is chosen in a similar way. We illustrate these paths if
the critical point is $\zeta_c$, see Figure~\ref{FigSmallS}.
\begin{figure}
\begin{center}
\psfrag{0}[t]{$0$}
\psfrag{1}[t]{$1$}
\psfrag{zc}[c]{$\zeta_c$}
\psfrag{G1}[r]{$\Gamma_1$}
\psfrag{G0}[l]{$\Gamma_0$}
\psfrag{z}[r]{$\sqrt{c}$}
\psfrag{1-z}{$\sqrt{\lambda}$}
\includegraphics[height=5cm]{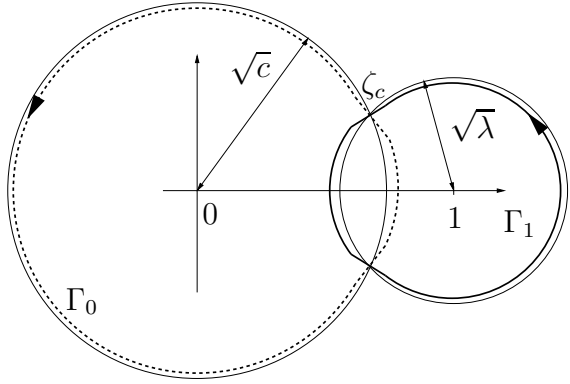}
\caption{Illustration of the steep descent paths.}
\label{FigSmallS}
\end{center}
\end{figure}
In particular, $|w_c|=\sqrt{\eta_2/\tau_2}$ and $|z_c|=\sqrt{\eta_1/\tau_2}$. In our case, we have $|w_c|\leq |z_c|$
and $|w_c-z_c|\geq L^{-1/16}$. The steep descent paths described above actually intersect. Therefore, we have to correct (\ref{eq4.35}) by subtracting the residue at $z=w$, as indicated in Figure~\ref{FigPrincValue}. We call ``main term'' the contribution of the integral with $\Gamma_0$ and $\Gamma_1$ crossing, while we call ``residual term'' the contribution of the residue.

Notice that the integral with the paths $\Gamma_0$ and $\Gamma_1$ crossing is integrable in the usual sense, because the divergence term $1/(w-z)$ is integrable. The contribution of the main term is the following.
\begin{figure}
\begin{center}
\psfrag{(a)}{$(a)$}
\psfrag{(b)}{$(b)$}
\psfrag{0}{$0$}
\psfrag{1}{$1$}
\psfrag{=}[c]{$=$}
\psfrag{-}[c]{$-$}
\psfrag{zc}[t]{$z_c$}
\psfrag{zcbar}[b]{$\bar{z}_c$}
\psfrag{wc}[l]{$w_c$}
\psfrag{wcbar}[l]{$\bar{w}_c$}
\psfrag{G0}{$\Gamma_0$}
\psfrag{C}{$\cal C$}
\psfrag{G1}{$\Gamma_1$}
\psfrag{G1I}{$\Gamma_1$}
\psfrag{G1II}{$\Gamma_1''$}
\includegraphics[height=4cm]{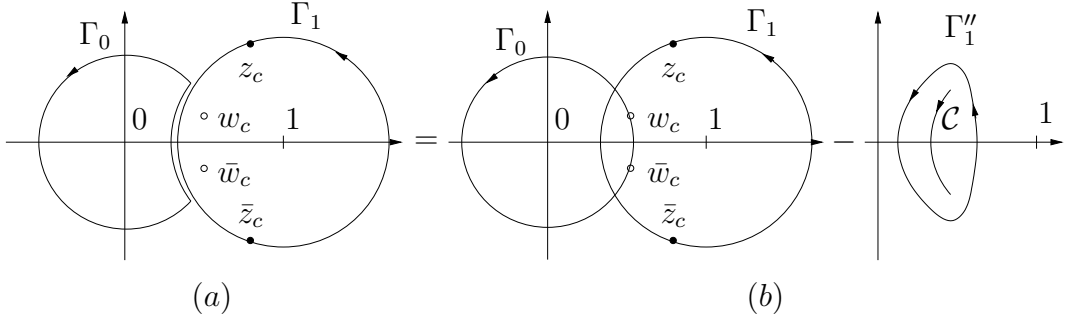}
\caption{The subdivision of the integration (\ref{eq4.35}). We have $|z_c|\geq |w_c|$ and when $|z_c|=|w_c|$, they are not at the same position.}
\label{FigPrincValue}
\end{center}
\end{figure}

Both integrals can be divided as the part in $\mathbbm{H}$ and its complex conjugate. Therefore, in the final expression we get the sum of four terms. Now, we restrict our attention to the integral over the path $\Gamma_0$ and $\Gamma_1$ on $\mathbbm{H}$. The analysis of the integral over $\Gamma_0$ is the same as in Lemma~\ref{Lemma4} except for the missing $1/w_c$ factor and that instead of $2\Re(\cdots)$ we just have $(\cdots)$ in (\ref{eq4.29}). The integral over $\Gamma_1$ is similar.

This time we choose the cutoff for the evaluation of the term with the steep descent path equal to $\delta=L^{-1/4}$. There are two reasons. The first one is that we want to get the expansion valid also for $\nu_i$ up to $L^{-1/3}$ away from the upper edge, compare with Lemma~\ref{CorLemma4}. The second reason is that we have the extra factor $1/(w-z)$. The contributions of the steep descent path do not create problems, since the factor is integrable in the usual sense (just need a bound). However, with our choice of $\delta$, in the contribution of the $\delta$-neighborhoods of $z_c$ and $w_c$ we have
\begin{equation}
1/(w-z)=1/(w_c-z_c)+\Or(\delta/|w_c-z_c|^2)=1/(w_c-z_c)+\Or(L^{-1/8}),
\end{equation}
with $\delta=L^{-1/4}$ and $|w_c-z_c|\geq L^{-1/16}$. In the end, the contribution of the main term is given by
\begin{eqnarray}
& &\frac{1}{w_c-z_c}\frac{e^{t_2\Re(g_2(w_c))}}{\sqrt{2\pi t_2 |g_2''(w_c)|}}\bigg[e^{t_2 \Im(g_2(w_c))}\hat\theta_2(w_c)+\Or\left(L^{-1/2}\right)\bigg]\nonumber \\
&\times &\frac{e^{-t_1\Re(g_1(z_c))}}{\sqrt{2\pi t_1 |g_1''(z_c)|}}\frac{ e^{\I\pi_{c_1}}}{|1-z_c|} \bigg[e^{-t_1 \Im(g_1(z_c))}\hat\theta_1(z_c) + \Or\left(L^{-1/2}\right)\bigg]\nonumber \\
&+&\frac{e^{t_2\Re(g_2(w_c))}}{\sqrt{2\pi t_2 |g_2''(w_c)|}} \frac{e^{-t_1\Re(g_1(z_c))}}{\sqrt{2\pi t_1 |g_1''(z_c)|}}\frac{1}{|1-z_c|} \Or(L^{-1/8}).
\end{eqnarray}
The term $e^{\I\pi_{c_1}}$ is the phase of $1/(1-z_c)$, while
$\hat\theta_i$ are the directions of the steepest descent paths at
the critical points. Explicitly,
\begin{align}
\hat\theta_1(z_c)&=\exp(\I(\pi-\tfrac12 \arg(g''(z_c))))=\exp(\I 3\pi/4+\I(\pi_{\lambda_1}-\pi_{c_1})/2),\\
\hat\theta_2(w_c)&=\exp(\I(\tfrac{\pi}{2}-\tfrac12 \arg(g''(w_c))))=\exp(\I 5\pi/4+\I(\pi_{\lambda_2}-\pi_{c_2})/2). \nonumber
\end{align}
Putting together the four terms (two times two critical points) we get the complete contribution of the main term as
\begin{eqnarray}
& &\frac{e^{t_2\Re(g_2(w_c))-t_1\Re(g_1(z_c))}}{2\pi\sqrt{t_1 t_2 |1-z_c|^2 |g_2''(w_c)| |g_1''(z_c)|}} \bigg[\Or(L^{-1/8})\nonumber \\
&+&\frac{1}{w_c-z_c}\frac{e^{\I t_2 \Im(g_2(w_c))+\I \beta_2}}{e^{\I t_1 \Im(g_1(z_c))+\I \beta_1}}
+\frac{1}{w_c-\bar z_c}\frac{e^{\I t_2 \Im(g_2(w_c))+\I \beta_2}}{e^{-\I t_1 \Im(g_1(z_c))-\I \beta_1}}\nonumber \\
&+&\frac{1}{\bar w_c-z_c}\frac{e^{-\I t_2 \Im(g_2(w_c))-\I \beta_2}}{e^{\I t_1 \Im(g_1(z_c))+\I \beta_1}}
+\frac{1}{\bar w_c-\bar z_c}\frac{e^{-\I t_2 \Im(g_2(w_c))-\I \beta_2}}{e^{-\I t_1 \Im(g_1(z_c))-\I \beta_1}}
\bigg],
\end{eqnarray}
with $\beta_1=-\arg(\hat\theta_1(z_c))-\pi_{c_1}$ and $\beta_2=\arg(\hat\theta_2(w_c))$.
Finally, we replace $g_i(w) t_i=G(w|\nu_i,\eta_i,\tau_i) L$, $\pi_{\lambda_i}=\pi_{\nu_i}$, $\pi_{c_i}=\pi_{\nu_i}$, and
\begin{equation}
e^{t_2\Re(g_2(w_c))-t_1\Re(g_1(z_c))} e^{t_1-t_2}=W_2/W_1
\end{equation}
to get (\ref{eqAsymptKernel}).

The final step is to estimate the contribution of the residual term (the last case of Figure~\ref{FigPrincValue}). It is given by
\begin{equation}\label{eq4.41}
\frac{1}{2\pi\I}\int_{\zeta}^{\bar \zeta} \dx z \frac{e^{(\tau_2-\tau_1) L z} e^{(\eta_1-\eta_2)L\ln(z)}}{(1-z)^{(\nu_1-\nu_2)L+1}},
\end{equation}
where $\zeta$ and $\bar\zeta$ are the two intersection points of the steep descent path $\Gamma_0$ and $\Gamma_1$. Since $\tau_2-\tau_1\geq 0$, $\eta_1-\eta_2\geq 0$, and $|1-z|=\cte$ along the piece of $\Gamma_1$ inside $\Gamma_0$, we have $\Re(z)\leq \Re(\zeta)$ and $\Re(\ln(z))\leq \Re(\ln(\zeta))$. Therefore,
\begin{equation}\label{eq4.42}
|(\ref{eq4.41})|\leq e^{t_2 \Re(g_2(\zeta))-t_1\Re(g_1(\zeta))} \leq e^{t_2\Re(g_2(w_c))-t_1\Re(g_1(z_c))} \Or(e^{-\mu_1 t_1} e^{-\mu_2 t_2}),
\end{equation}
for some positive $\mu_1,\mu_2$ and at least one larger than $L^{-1/8}$. This follows from the fact that either one (or both) critical points are away of order $L^{-1/16}$ from $\zeta$, and $\zeta$ lies on the steep descent paths of $g_2(w)$ and $-g_1(z)$, with local quadratic behavior.
\end{proofOF}

While doing time integration we will also need the following corollary.
\begin{cor}\label{CorBulkShifted}
Consider the same setting of Proposition~\ref{PropKernelAsymptExp}. Then\\
(a) the formula for
$K(x_1,n_1,t_1;x_2+1,n_2,t_2)$ is the same as (\ref{eqAsymptKernel}) but with an extra factor \mbox{$(1-w_c)$}, resp.\ $(1-\bar w_c)$, to the terms with $e^{\I\beta_2}$, resp.\ $e^{-\I\beta_2}$.\\
(b) the formula for
$K(x_1,n_1-1,t_1;x_2,n_2,t_2)$ is the same as (\ref{eqAsymptKernel}) but with an extra factor \mbox{$z_c^{-1}$}, resp.\ $(\bar z_c)^{-1}$, to the terms with $e^{-\I\beta_1}$, resp.\ $e^{\I\beta_1}$.
\end{cor}
\begin{proofOF}{Corollary~\ref{CorBulkShifted}}
The proof is almost identical to the one of Proposition~\ref{PropKernelAsymptExp}. The only difference is that in (\ref{eq4.35}) we have for (a) an extra term $(1-w)$ and for (b) an extra $1/z$.
\end{proofOF}

At this point we have all the needed estimates in the bulk. However,
since our system develops facets, we need to have control at the
upper edge. We will just need some bounds and, since the integrals
are the same as in Section~\ref{SubsectAsymptEdge}, apart from the
factor $1/(w_c-z_c)$, which we will assume bounded away from zero.

\begin{prop}\label{PropBorderBulk}
Consider the setting of Proposition~\ref{PropKernelAsymptExp}, but with one or both of the $\nu_i$ close to the upper edge,
\begin{equation}
(\sqrt{\tau_i}+\sqrt{\eta_i})^2-L^{-1/3}\leq \nu_i \leq (\sqrt{\tau_i}+\sqrt{\eta_i})^2-\ell L^{-2/3}.
\end{equation}
Moreover, assume that $|z_c-w_c|$ is bounded away from zero uniformly in $L$. Then, there exists $\ell$ large enough, such that
\begin{eqnarray}\label{eqAsymptCloseToEdge}
(W_1/W_2)|K(x_1,n_1,t_1;x_2,n_2,t_2)| \leq
\frac{\cte}{L\prod_{i=1}^2\sqrt[4]{\eta_i\tau_i-\tfrac14 (\tau_i+\eta_i-\nu_i)^2}}
\end{eqnarray}
uniformly in $L$ for $L\geq L_0\gg 1$.
\end{prop}
\begin{proofOF}{Proposition~\ref{PropBorderBulk}}
The proof follows the same argument as Lemma~\ref{Lemma6} for the variables which are close to the edge. For the one which is away from the edges, it is a consequence of the analysis Proposition~\ref{PropKernelAsymptExp}.
\end{proofOF}

When one or both positions are at the edge, we need a different bound.
\begin{prop}\label{PropBorderEdge}
Consider the setting of Proposition~\ref{PropKernelAsymptExp}, but now with $\nu_2$ at the edge or in the facet, i.e.,
\begin{equation}
\nu_2 \geq (\sqrt{\tau_2}+\sqrt{\eta_2})^2-\ell L^{-2/3}
\end{equation}
for any fixed $\ell$. Assume $|z_c+\sqrt{\eta_2/\tau_2}|$ is bounded away from zero uniformly in $L$. Then,
\begin{eqnarray}
(W_1/W_2) |K(x_1,n_1,t_1;x_2,n_2,t_2)| &\leq&
\frac{\cte}{\sqrt{L}\sqrt[4]{\eta_1\tau_1-\tfrac14 (\tau_1+\eta_1-\nu_1)^2}} \\
&\times & \frac{1}{L^{1/3}} \exp\left(-\frac{x_2-(\sqrt{\tau_2}+\sqrt{\eta_2})^2L}{(\tau_2 L)^{1/3}}\right),\nonumber
\end{eqnarray}
uniformly in $L$ for $L\geq L_0\gg 1$.
\end{prop}
\begin{proofOF}{Proposition~\ref{PropBorderEdge}}
The proof is obtained along the same lines as Lemmas~\ref{Lemma1b} and~\ref{Lemma2b}. With respect to those cases, the integral has however an extra factor $1/(w-z)$. Since we need just a bound, it can simply be replaced by $1/(w_c-z_c)$ as follows. In Lemma~\ref{Lemma1b} $w_c$ is replaced by $-\sqrt{\eta_2/\tau_2}$, while in Lemma~\ref{Lemma2b}, we need to replace $w_c$ by $\rho$ as given in (\ref{eq4.7BB}). Notice that in the last case we can take $|w_c+\sqrt{\eta_2/\tau_2}|$ as small as desired. The assumption $|z_c+\sqrt{\eta_2/\tau_2}|>0$ uniformly in $L$ ensures then that $1/(w_c-z_c)$ remains bounded as $L\to\infty$.
\end{proofOF}

The last case to consider is when both $\nu_1$ and $\nu_2$ are at the upper edge.
\begin{prop}\label{PropBorderEdge2}
Consider the setting of Proposition~\ref{PropKernelAsymptExp}, but now with $\nu_1$ and $\nu_2$ at the edge or in the facet, i.e., with
\begin{equation}
\nu_i \geq (\sqrt{\tau_i}+\sqrt{\eta_i})^2-\ell L^{-2/3}, \quad i=1,2,
\end{equation}
for any fixed $\ell$. Assume that $|\sqrt{\eta_2/\tau_2}-\sqrt{\eta_1/\tau_1}|$ is bounded away from zero uniformly in $L$.
Then,
\begin{eqnarray}
& &(W_1/W_2) |K(x_1,n_1,t_1;x_2,n_2,t_2)| \leq \cte \\
& &\frac{1}{L^{2/3}} \exp\left(-\frac{x_2-(\sqrt{\tau_2}+\sqrt{\eta_2})^2L}{(\tau_2 L)^{1/3}}\right) \exp\left(-\frac{x_1-(\sqrt{\tau_1}+\sqrt{\eta_1})^2L}{(\tau_1 L)^{1/3}}\right),\nonumber
\end{eqnarray}
uniformly in $L$ for $L\geq L_0\gg 1$.
\end{prop}
\begin{proofOF}{Proposition~\ref{PropBorderEdge2}}
The proof is like Proposition~\ref{PropBorderEdge}. We will have $|w_c+\sqrt{\eta_2/\tau_2}|$ and $|z_c+\sqrt{\eta_1/\tau_1}|$ as small as desired. The assumption $|\sqrt{\eta_2/\tau_2}-\sqrt{\eta_1/\tau_1}|>0$ uniformly in $L$ allows us to easily bound uniformly in $L$ the term $1/(w_c-z_c)$.
\end{proofOF}

\appendix

\section{Determinantal structure of the correlation functions}\label{Appendix}
Let $\X_1,\dots,\X_N$ be finite sets and $c(1),\dots,c(N)$ be arbitrary nonnegative integers. Consider the set
\begin{equation}
\X=(\X_1\sqcup\dots\sqcup\X_1)\sqcup\dots\sqcup(\X_N\sqcup \dots\sqcup\X_N)
\end{equation}
with $c(n)+1$ copies of each $\X_n$. We want to consider a particular form of weight $W(X)$ for any subset $X\subset\X$, which turns out to have determinantal correlations.

To define the weight we need a bit of notations. Let
\begin{equation}
\begin{aligned}
\phi_n(\,\cdot\,,\,\cdot\,):\X_{n-1}\times\X_{n}\to \C,\qquad &n=2,\dots,N, \\
\phi_n(\virt,\,\cdot\,):\X_{n}\to\C,\qquad &n=1,\dots,N,\\
\Psi^N_j(\,\cdot\,):\X_N\to\C,\qquad &j=0,\dots,N-1,
\end{aligned}
\end{equation}
be arbitrary functions on the corresponding sets. Here the symbol
$\virt$ stands for a ``virtual'' variable, which is convenient to
introduce for notational purposes. In applications $\virt$ can
sometimes be replaced by $+\infty$ or $-\infty$. The $\phi_n$ represents the transitions from $\X_{n-1}$ to $\X_n$.

Also, let
\begin{equation}
t_{0}^N\leq\dots\leq t_{c(N)}^N= t_{0}^{N-1}\leq\dots\leq
t_{c(N-1)}^{N-1}=t_0^{N-2}\leq \dots\leq t^2_{c(2)}=t^1_0\leq\dots \leq
t^1_{c(1)}
\end{equation}
be real numbers. In applications, these numbers refer to time moments.
Finally, let
\begin{equation}
{\cal T}_{t_a^n,t_{a-1}^n}(\,\cdot\,,\,\cdot\,):\X_n\times\X_n\to \C,\qquad n=1,\dots,N,\quad a=1,\dots,c(n),
\end{equation}
be arbitrary functions. The ${\cal T}_{t_a^n,t_{a-1}^n}$ represents the transition between two copies of $\X_n$ associated to ``times'' $t^n_{a-1}$ and $t^n_a$.

Then, to any subset $X\subset\X$ assign its weight $W(X)$ as follows. $W(X)$ is zero unless $X$ has exactly $n$ points in each copy of $\X_n$, $n=1,\dots,N$. In the latter case, denote the points of $X$ in the $m$th copy of
$\X_n$ by $x^n_k(t^n_m)$, $k=1,\dots,n$, $m=0,\ldots,c(n)$. Thus,
\begin{equation}
X=\{x^n_k(t^n_m)\mid k=1,\dots,n;\,m=0,\dots,c(n);\, n=1,\dots,N\}.
\end{equation}
Set
\begin{equation}
\begin{aligned}
W(X)= & \prod_{n=1}^{N} \Bigg[\det{\bigl[\phi_n(x_k^{n-1}(t_0^{n-1}),x_l^n(t^n_{c(n)}))\bigr]}_{1\leq k,l\leq n} \\
& \times\prod_{a=1}^{c(n)} \det{\bigl[{\cal T}_{t_a^n,t_{a-1}^n}(x_k^n(t^n_a),x^n_l(t^n_{a-1}))\bigr]}_{1\leq k,l\leq n}
\Bigg]\det{\bigl[\Psi^{N}_{N-l}(x^{N}_k(t_0^{N}))\bigr]}_{1\leq k,l\leq N},
\end{aligned}
\end{equation}
where $x^{n-1}_{n}(\,\cdot\,)=\virt$ for all $n=1,\dots,N$.

In what follows we assume that the partition function of our weights
does not vanish:
\begin{equation}
Z:=\sum_{X\subset \X} W(X)\ne 0.
\end{equation}
Under this assumption, the normalized weights $\widetilde W(X)=W(X)/Z$
define a (generally speaking, complex valued) measure on $2^\X$
of total mass $1$. One can say
that we have a (complex valued) random point process on $\X$, and
its correlation functions are defined accordingly, see e.g.~\cite{RB04}.
We are interested in computing these correlation functions.

Let us introduce the compact notation for the convolution of several transitions. For any $n=1,\dots,N$ and two time moments $t_a^n>t_b^n$ we define
\begin{equation}
{\cal T}_{t_a^n,t_b^n}={\cal T}_{t_a^n,t_{a-1}^n}*{\cal T}_{t_{a-1}^n,t_{a-2}^n}*\cdots *{\cal T}_{t_{b+1}^n,t_b^n},\qquad {{\cal T}}^n = {{\cal T}}_{t_{c(n)}^n,t_0^n},
\end{equation}
where we use the notation $(f*g)(x,y):=\sum_{z}f(x,z)g(z,y).$
For any time moments
$t_{a_1}^{n_1}\ge t_{a_2}^{n_2}$ with $(a_1,n_1)\ne (a_2,n_2)$, we denote the convolution over all the transitions between them by $\phi^{(t_{a_1}^{n_1},t_{a_2}^{n_2})}$:
\begin{equation}
\phi^{(t_{a_1}^{n_1},t_{a_2}^{n_2})}={{\cal T}}_{t^{n_1}_{a_1},t^{n_1}_{0}} * \, \phi_{n_1+1}*{{\cal T}}^{n_1+1} *\cdots*\phi_{n_2}*{{\cal T}}_{t^{n_2}_{c(n_2)},t^{n_2}_{a_2}}.
\end{equation}
If there are no such transitions, i.e., if $t_{a_1}^{n_1}<t_{a_2}^{n_2}$ or $(a_1,n_1)=(a_2,n_2)$, we set
$\phi^{(t_{a_1}^{n_1},t_{a_2}^{n_2})}=0$.

Furthermore, define the matrix $M={\Vert M_{k,l}\Vert}_{k,l=1}^N$ by
\begin{equation}
M_{k,l}=\big(\phi_{k}*{{\cal T}}^k*\cdots *\phi_{N}*{{\cal T}}^{N}*\Psi^{N}_{N-l}\big)(\virt)
\end{equation}
and the vector
\begin{equation}
\Psi^{n,t^n_a}_{n-l}=\phi^{(t^n_a,t^{N}_0)}*\Psi^{N}_{N-l},\qquad l=1,\dots,N.
\end{equation}
The following statement describing the correlation kernel is a part of Theorem 4.2 of~\cite{BF07}.

\begin{thm}\label{ThmPushASEP}
Assume that the matrix $M$ is invertible. Then \mbox{$Z=\det M\neq 0$}, and the (complex valued) random point process on $\X$ defined by the weights $\widetilde W(X)$ is determinantal. Its correlation kernel can be written in the form
\begin{equation}
\begin{aligned}
K(t^{n_1}_{a_1},x_1; t^{n_2}_{a_2},x_2)&= -\phi^{(t^{n_1}_{a_1},t^{n_2}_{a_2})}(x_1,x_2) \\
+& \sum_{k=1}^{N} \sum_{l=1}^{n_2} \Psi^{n_1,t^{n_1}_{a_1}}_{n_1-k}(x_1) [M^{-1}]_{k,l} (\phi_l * \phi^{(t^l_{c(l)},t^{n_2}_{a_2})})(\virt,x_2).
\end{aligned}
\end{equation}
\end{thm}

The proof of Theorem~\ref{ThmPushASEP} given in~\cite{BF07} is based on the
algebraic formalism of~\cite{RB04}. Another proof can be found in Section 4.4 of~\cite{FN08}.
Although we stated Theorem~\ref{ThmPushASEP} for the case when all sets $\X_n$
are finite, one easily extends it to a more general setting. Indeed,
the determinantal formula for the correlation functions is an
algebraic identity, and the limit transition to the case when $\X_n$'s
are allowed to be countably infinite is immediate, under the
assumption that all the sums needed to define the $*$-operations
above are absolutely convergent. Another easy extension (which we do
not need in this paper) is the case when the spaces $\X_j$ become
continuous, and the sums have to be replaced by the corresponding integrals
over these spaces.

\def\bibname{R\lowercase{eferences}}


\end{document}